\documentclass[useAMS,usenatbib]{mn2e}

\usepackage{times,epsfig}
\usepackage{graphics,graphicx}
\usepackage{amsmath}
\usepackage{amsfonts}
\usepackage{epsfig}
\usepackage{url}
\usepackage{morefloats}
\usepackage{tabularx}
\usepackage{color}

\usepackage{mathrsfs}
\usepackage{amssymb}
\usepackage{bm}

\newcommand{\mnras}{MNRAS}

\newcommand{\apj}{ApJ}

\newcommand{\aj}{AJ}

\definecolor{darkred}{rgb}{0.8,0,0}

\def\lsim{\,\lower2truept\hbox{${<\atop\hbox{\raise4truept\hbox{$\sim$}}}$}\,}
\def\gsim{\,\lower2truept\hbox{${> \atop\hbox{\raise4truept\hbox{$\sim$}}}$}\,}
\def\simlt{\mathrel{\rlap{\lower 3pt\hbox{$\sim$}}
        \raise 2.0pt\hbox{$<$}}}
\def\simgt{\mathrel{\rlap{\lower 3pt\hbox{$\sim$}}
        \raise 2.0pt\hbox{$>$}}}

\def\be{\begin{equation}}
\def\ee{\end{equation}}
\def\ba{\begin{eqnarray}}
\def\ea{\end{eqnarray}}
\def\f{\frac}
\def\l{\left}
\def\r{\right}

\def\hksqrt{\mathpalette\DHLhksqrt}
\def\DHLhksqrt#1#2{\setbox0=\hbox{$#1\sqrt{#2\,}$}\dimen0=\ht0
\advance\dimen0-0.2\ht0
\setbox2=\hbox{\vrule height\ht0 depth -\dimen0}%
{\box0\lower0.4pt\box2}}

\title[Cosmological Measurements with Radio Surveys]{Cosmological Measurements with Forthcoming Radio Continuum Surveys}
\author[A. Raccanelli et al.]
{\parbox[t]{\textwidth}
{Alvise Raccanelli$^{1}$\thanks{e-mail: alvise.raccanelli@port.ac.uk (AR)}, 
Gong-Bo Zhao$^{1}$, 
David J. Bacon$^{1}$, 
Matt J. Jarvis$^{2,3}$, 
Will J. Percival$^{1}$, 
Ray P. Norris$^{4}$, 
Huub R\"{o}ttgering$^{5}$, 
Filipe B. Abdalla$^{6}$, 
Catherine M. Cress$^{3,7}$, 
Jean-Claude Kubwimana$^{8}$, 
Sam Lindsay$^{2}$,
Robert C. Nichol$^{1}$, 
Mario G. Santos$^{9}$, 
Dominik J. Schwarz$^{10}$}
\vspace*{8pt}\ \\
$^{1}$ Institute of Cosmology {\fontfamily{ppl}\selectfont \&} Gravitation, University of Portsmouth, Dennis Sciama building, Portsmouth, P01 3FX, UK\\
$^{2}$ Centre for Astrophysics Research, STRI, University of Hertfordshire, Hatfield, AL10 9AB, UK \\
$^{3}$ Physics Department, University of the Western Cape, Bellville 7535, South Africa \\
$^{4}$ CSIRO Astronomy \& Space Science, PO Box 76, Epping, NSW 1710, Australia \\
$^{5}$ Leiden Observatory, Leiden University, P.O. Box 9513, NL-2300 RA Leiden, The Netherlands \\
$^{6}$ Department of Physics and Astronomy, University College London, Gower Street, London, WC1E 6BT, UK \\
$^{7}$ Centre for High Performance Computing, 15 Lower Hope St, Cape Town 7700, South Africa \\
$^{8}$ Institut de Recherche en Astrophysique et Plan\'{e}tologie, 9, Av du Colonel Roche, BP 44346, 31028, Toulouse, France \\
$^{9}$ CENTRA, Instituto Superior Tecnico, Technical University of Lisbon, Lisboa 1049-001, Portugal \\
$^{10}$ Fakult\"{a}t f\"{u}r Physik, Universit\"{a}t Bielefeld, Postfach 100131, 33501 Bielefeld, Germany }
\date{}
\begin{document}
\maketitle
\begin{abstract}
We present forecasts for constraints on cosmological models which can be obtained by forthcoming radio continuum surveys: the wide surveys with the LOw Frequency ARray (LOFAR), Australian Square Kilometre Array Pathfinder (ASKAP) and the Westerbork Observations of the Deep APERTIF Northern sky (WODAN). We use simulated catalogues appropriate to the planned surveys to predict measurements obtained with the source auto-correlation, the cross-correlation between radio sources and CMB maps (the Integrated Sachs-Wolfe effect), the cross-correlation of radio sources with foreground objects due to cosmic magnification, and a joint analysis together with the CMB power spectrum and supernovae.
We show that near future radio surveys will bring complementary measurements to other experiments, probing different cosmological volumes, and having different systematics. 
Our results show that the unprecedented sky coverage of these surveys combined should provide the most significant measurement yet of the Integrated Sachs-Wolfe effect. 
In addition, we show that using the ISW effect will significantly tighten constraints on modified gravity parameters, while the best measurements of dark energy models will come from galaxy auto-correlation function analyses.
Using the combination of EMU and WODAN to provide a full sky survey, it will be possible to measure the dark energy parameters with an uncertainty of \{$\sigma (w_0) = 0.05$, $\sigma (w_a) = 0.12$\} and the modified gravity parameters \{$\sigma (\eta_0) = 0.10$, $\sigma (\mu_0) = 0.05$\}, assuming Planck CMB+SN(current data) priors.
Finally, we show that radio surveys would detect a primordial non-Gaussianity of $f_{\rm NL}$ = 8 at 1-$\sigma$ and we briefly discuss other promising probes.
\end{abstract}

\begin{keywords}
large-scale structure of the universe --- cosmological parameters --- cosmology: observations --- radio continuum: galaxies.
\end{keywords}

%%%%%%%%%%%%%%%%%%%%%%%%%%%%%%%%%%%%%%%%%%%%%%%%%%%%%%%%%%%%%%%%%%
%%%%%%%%%%%%%%%%%%%%%%%%%%%%%%%%%%%%%%%%%%%%%%%%%%%%%%%%%%%%%%%%%%
%%%%%%%%%%%%%%%%%%%%%%%%%%		Introduction		%%%%%%%%%%%%%%%%%%%%%%%%%
%%%%%%%%%%%%%%%%%%%%%%%%%%%%%%%%%%%%%%%%%%%%%%%%%%%%%%%%%%%%%%%%%%
%%%%%%%%%%%%%%%%%%%%%%%%%%%%%%%%%%%%%%%%%%%%%%%%%%%%%%%%%%%%%%%%%%

\section{Introduction}
Radio surveys for cosmology are entering a new phase with the construction of the LOw Frequency ARray (LOFAR, \citealt{rottgering03}), the Australian Square Kilometre Array Pathfinder (ASKAP, \citealt{Johnston08}) and APERTIF, the new Phased Array Feed receiver system for the Westerbork Synthesis Radio Telescope (WSRT, \citealt{oosterloo10}). In each case, the increased sensitivity available, together with a very wide sky coverage, will allow certain cosmological statistics to be measured with substantial accuracy. 
Several studies in the past have concentrated on the cosmological constraints that can be determined from the large redshift surveys using the H{\sc i} 21-cm emission line (e.g. \citealt{abdalla05}; \citealt{abdalla10}). 
However, little attention has been paid to the information that can be gleaned from the large radio continuum surveys which will, in many respects, be much easier to interpret than the H{\sc i} surveys and allow us to push out to much higher redshifts. In this paper we consider three experiments using the deep continuum observations, the auto-correlation of radio sources, the cross-correlation of radio sources with the Cosmic Microwave Background (the late Integrated Sachs-Wolfe effect), and cross-correlation of radio sources with foreground objects (magnification bias). The level of the accuracy of these measurements, and the relative significance of the various potential probes, are the key issues which we wish to address in this paper.

One of the goals of these measurements will be to measure the cosmological parameters of particular current interest. Among the biggest challenges in cosmology is to understand if the standard $\Lambda$ Cold Dark Matter model, and its General Relativity context,  is correct, or if we need a different cosmological model and/or gravitational theory, with the related important implications for fundamental physics.
We will therefore present forecasts of the constraints on cosmological models and gravitational parameters that will be possible to obtain with the LOFAR, ASKAP and WSRT radio telescopes, in isolation and together.

There are many major optical and near infra-red galaxy surveys (e.g. the Baryon Oscillation Spectroscopic Survey (BOSS, \citealt{boss}), BigBOSS (\citealt{bigboss}), the Dark Energy Survey\footnote{http://www.darkenergysurvey.org/} (DES), the Panoramic Survey Telescope And Rapid Response System\footnote{http://pan-starrs.ifa.hawaii.edu/public/} (Pan-STARRS), Euclid \citep{euclid}, the Large Synoptic Survey Telescope\footnote{http://www.lsst.org/lsst} (LSST)) which aim to improve the
precision of cosmological parameter measurements during this decade. One of the goals of this paper is to discover whether there are also significant and complementary opportunities for improvement of cosmological constraints  by forthcoming radio continuum surveys. These surveys have a niche because of their large sky coverage, high median redshift and number of objects observed.

The paper is organised as follows. 
In Section 2 we will describe the next generation of radio surveys, given by LOFAR, EMU and WODAN. 
In Section 3, we will discuss the predictions for source densities and bias as a function of redshift for each survey and for different source populations. 
In Section 4 we will 
present the cosmological probes we will use, and in 
Section 5 we show our predicted cosmological measurements.
In Section 6 we will describe the methodology used to predict the resulting constraints on dark energy and modified gravity models, and in 
Section 7 we present our results. 
In Section 8 we present our conclusions and summarise why LOFAR, EMU and WODAN will be important for cosmology.

\section{Forthcoming Radio Surveys}
In this section we introduce the three large radio surveys which we will focus on in this paper: LOFAR, EMU and WODAN.
We do not consider the surveys to be conducted with the South African SKA Precursor Telescope (MeerKAT, \citealt{jonas09}) as the parameter space probed by MeerKAT is towards much deeper and narrower surveys which are more adept to studying galaxy formation and evolution. 
In each case we will discuss the properties of the surveys, and their expected timescales for observation. A summary of the survey properties is shown in Table \ref{tab:surveys}.

\subsection{LOFAR}
\label{sec:lofar}
LOFAR (the LOw Frequency ARray for radio astronomy, \citealt{rottgering03}) is a multi-national telescope that has stations spanning Europe. The core of LOFAR is situated in the north-east of the Netherlands, with stations on longer baselines both within the Netherlands and across to Germany, UK, France and Sweden. Other stations may also be added throughout the rest of Europe in the coming years.

Each LOFAR station operates at two broad frequency ranges, the high band operating at 120~$<$~$\nu$~$<$~240~MHz and the low band which operates at 10~$<$~$\nu$~$<$~80~MHz. The bulk of the early operations of LOFAR will be dedicated to a number of Key Science Projects (KSPs): Solar Physics and Space Weather, Transients, Cosmic Magnetism, the Epoch of Reionization, Cosmic Rays and Continuum Surveys.
It is the last of these which is pertinent to the aims of this paper.

The Continuum Surveys KSP will explore the bulk of the northern sky at low-radio frequencies. Low-frequency radio observations are ideally suited for carrying out sensitive surveys of the extragalactic sky, firstly because the low-frequency ensures a large instantaneous field of view, allowing an increased survey speed compared to similar telescopes operating at higher frequencies. Second, the bulk of radio emission detected from extragalactic sources is due to synchrotron radiation and therefore increases towards the lower frequencies, although at the very lowest frequencies one might expect a turnover to occur due to synchrotron self absorption.

The LOFAR continuum surveys \citep{rottgering10} follow the usual strategy of a ``wedding cake'' tiered survey. The design of these has focused on addressing the original key science topics within the continuum surveys, namely tracing the formation of massive galaxies, clusters and black holes using high-redshift radio sources, measuring the star-formation history of the Universe through radio emission and tracing intracluster magnetic fields using diffuse radio emission. However, as we demonstrate, these surveys will also provide key data which can be used to constrain the cosmology and gravitational physics in our Universe.

For the purposes of this paper we concentrate solely on the 120~MHz surveys from LOFAR, as these are the most sensitive for our science, i.e. wide-field and highly sensitive. 
The tiers of the LOFAR survey are as follows: the large-area, ``Tier-1" survey will survey the whole of the northern sky down to an expected rms flux-density at 120~MHz of $S_{\rm 120~MHz} = 0.1$~mJy. 

The LOFAR Tier-2 survey will survey to deeper levels over a smaller area.  The baseline strategy is to survey around 550~square degrees at 120~MHz to an rms flux-density of $S_{\rm 120~MHz} = 25\mu$Jy. 

The Tier-3 survey is not considered in this paper due to the relatively small area it will survey ($\sim 70$~square degrees) at 150~MHz to $\sim 6\mu$Jy rms.

We also consider what results could be achieved with the LOFAR commissioning survey. Although this survey is still being fully defined, we take a shallow survey covering the whole northern hemisphere at 150~MHz and use a $10\sigma$ limit of 7~mJy. This allows us to examine what can be achieved with a very conservative survey.

If we assume only the stations situated in the Netherlands are used in carrying out the large area surveys, then the resolution at 120~MHz will be $\sim 6$~arcsec and around $\sim 5$~arcsec at 150~MHz, i.e. very similar to the Faint Images of the Radio Sky at 21 centimetres (FIRST) survey \citep{Becker95}.

\subsection{EMU}
\label{sec:emu}
EMU (Evolutionary Map of the Universe, \citealt{Norris11}) is an all-sky continuum survey planned for the new Australian SKA Pathfinder (ASKAP) \citep{Johnston08} telescope under construction on the Australian candidate SKA site in Western Australia. EMU is one of the two key projects (the other is the WALLABY all-sky HI survey) which are primarily driving the ASKAP design. At its completion,  expected to be in late 2012, ASKAP will consist of 36 12-metre antennas spread over a region 6 km in diameter. Although the collecting area  is no larger than many existing radio telescopes, the phased- array feed at the focus of each antenna  provides about 100 dual-polarisation  pixels, giving ASKAP a thirty square degree of instantaneous field of view. This enables it to survey the sky some thirty times faster than existing radio telescopes at similar frequencies. 

The primary goal of EMU is to make a deep ($10\,\mu$Jy rms) radio continuum survey of the entire Southern Sky, extending as far North as $+30\,\deg$. EMU will cover the same area (75\% of the sky)  as NVSS \citep{condon98}, but will be 45 times more sensitive, and will have an angular resolution (10 arcsec) five times better. It will also have higher sensitivity to extended structures. EMU is expected to begin in late 2012
%\footnote{http://www.atnf.csiro.au/people/rnorris/emu/status/index.htm} 
and it will generate a catalogue of radio sources 38 times greater than NVSS; all
radio data from the EMU survey will be placed in the public domain as soon as the data quality has been checked.

\subsection{WODAN}
\label{sec:wodan}
WODAN (the Westerbork Observations of the Deep Apertif Northern sky survey) is planned to chart the entire northern sky above Dec $> 30^{\circ}$ down to a proposed  rms flux density at 1.4~GHz of $S_{\rm 1.4~GHz} = 10\mu$Jy/beam \citep{rottgering11}. It will be able to do this because of the new phased array feeds (APERTIF) being put on the Westerbork Synthesis Radio Telescope (WSRT, \citealt{oosterloo10}). The phased array feeds will open up the field-of-view of the WSRT to around 8 square degrees allowing very high survey speeds. Such a survey is extremely complementary to the proposed LOFAR Tier-1 survey and will allow source spectral indices to be measured down to very faint levels. Although APERTIF increases the field-of-view of the WSRT considerably it will remain a relatively low-resolution survey instrument, with the resolution limited to the distribution of the WSRT antennae; as such the resolution will be around $\sim 15$~arcsec. However, this resolution is generally not a problem for the experiments we discuss in this paper. The current schedule for the commencement APERTIF surveys is 2013.

\begin{center}
  \begin{table}
	\begin{tabular}{ |l|l|l|l|l|l| }
	  \hline
	  Survey & Area & Frequency& N$_{gal}$ & Mean z & Median z\\
	  \hline
	  LOFAR MS$^3$ & $2\pi$  & 150~MHz & $1.0\times10^{6}$ & 1.6  & 1.3\\
	  \hline
	  LOFAR Tier1  & $2\pi$  & 120~MHz  & $6.5\times10^{6}$ & 1.8 & 1.1  \\
	  \hline
	  EMU  & $3\pi$  & 1400~MHz & $2.2\times10^{7}$ & 1.7 & 1.1 \\
	  \hline
	  WODAN  & $1\pi$  & 1400~MHz  & $7.3\times10^{6}$ & 1.7  & 1.1 \\
	  \hline
	\end{tabular}
\caption{Parameters of the surveys considered. We use the $10\sigma$ flux-density limit for each survey. Total number of radio sources N$_{\rm gal}$, mean and median redshifts calculated using our number density models in Section 3. }
\label{tab:surveys}
   \end{table}
\end{center}

\section{Source Population Models}

In this section we describe our models for source populations for LOFAR, EMU and WODAN surveys; in particular the number density of different source populations as a function of redshift, and the bias of different source populations as a function of redshift. These are required in order to make predictions for cosmological probes such as the auto correlation function, the ISW effect and magnification bias.

\subsection{Number densities}
\label{sec:nz}
We use empirical simulations to predict the number density of radio sources per redshift interval for the envisaged all-hemisphere LOFAR survey, the WODAN survey, and the ASKAP-EMU 3$\pi$ steradian survey. The combination of these surveys will provide complete coverage of 4$\pi$ steradians of the sky; however, the different observing frequencies and depths means that they will produce distinct redshift distributions, which need to be understood in order to use the combination for cosmological constraints.
Throughout this paper we assume that no redshift information is available for individual radio sources.

We use the simulations of  \citet{wilman08, wilman10}, developed for predictions for the Square Kilometre Array continuum survey. These simulations provide specific prescriptions for the redshift evolution of the various populations which dominate the radio source counts: powerful active galactic nuclei at bright fluxes, down to the less luminous radio-quiet AGN, starburst and star-forming galaxies. The simulations cover five different radio frequencies -- 150, 610, 1400, 4860 and 18000~MHz. We use the update of the simulated catalogue \citep{wilman10}, which has been adjusted to incorporate results from mid- and far-infrared data to provide a better estimate of the starburst and star-forming galaxy populations.

The N(z) from these simulations should in principle be modified by Redshift-Space Distortions \citep{rassat09} and magnification bias \citep{loverde07}, which are not included. However these corrections are small and will not affect our results.

Catalogues are generated from the $S^{3}$ database\footnote{http://s-cubed.physics.ox.ac.uk} corresponding to the radio flux-density limits of the proposed LOFAR, EMU and WODAN surveys. As described in Section 2, we assume the depth of the LOFAR survey over the whole hemisphere to be uniform across the sky down to a rms flux-density of $0.1$~mJy as given in the LOFAR Surveys document \cite{morganti10}. For the purposes of this paper we use the 151~MHz data from the $S^3$ data base and extrapolate to 120~MHz using the spectral index determined between 610~MHz and 151~MHz to predict the number density distribution. 
As the simulations include a spectral curvature term, this means that the spectral index between 610 and 151~MHz is generally flatter than the canonical $\alpha \sim 0.7$ and thus leads to a decrease in the number of sources one would expect based on using an $\alpha \sim 0.7$. We adopt these numbers as a conservative approach, however we note that if the radio spectra do not flatten significantly to low frequencies then the constraints from LOFAR Tier 1 will be similar to those of the EMU survey, albeit over $2\,\pi$ ~sr rather than $3\,\pi$~sr.
We then apply a cut to the simulated data and retain only those
sources with an integrated flux density larger then 10 times the rms
noise in the map. Note that for extended sources the nominal sensitivity to
peak flux densities is then less than 10 $\sigma$. However, this definition 
ensures that the virtually all extended sources will be detected.
It is also conservative in the sense that it still has to be proven
that all these new instruments can reach their theoretical thermal
noise levels.

For the EMU and WODAN surveys we again use the \citet{wilman08, wilman10} simulations, this time at the 1.4~GHz frequency. EMU will survey $\sim 75$\% of the sky down to an rms flux-density limit of $10\mu$~Jy, while WODAN will survey 10,000 sq. deg. down to an rms flux-density limit of $10\mu$~Jy; we again extract a catalogue from the $S^3$ database down to this limit and apply cuts at 10~$\sigma$ signal-to-noise.
Note that \cite{Norris11} assume a less conservative 5~$\sigma$ threshold and therefore obtain stronger constraints.

\begin{figure*}
\begin{center}
\epsfig{file=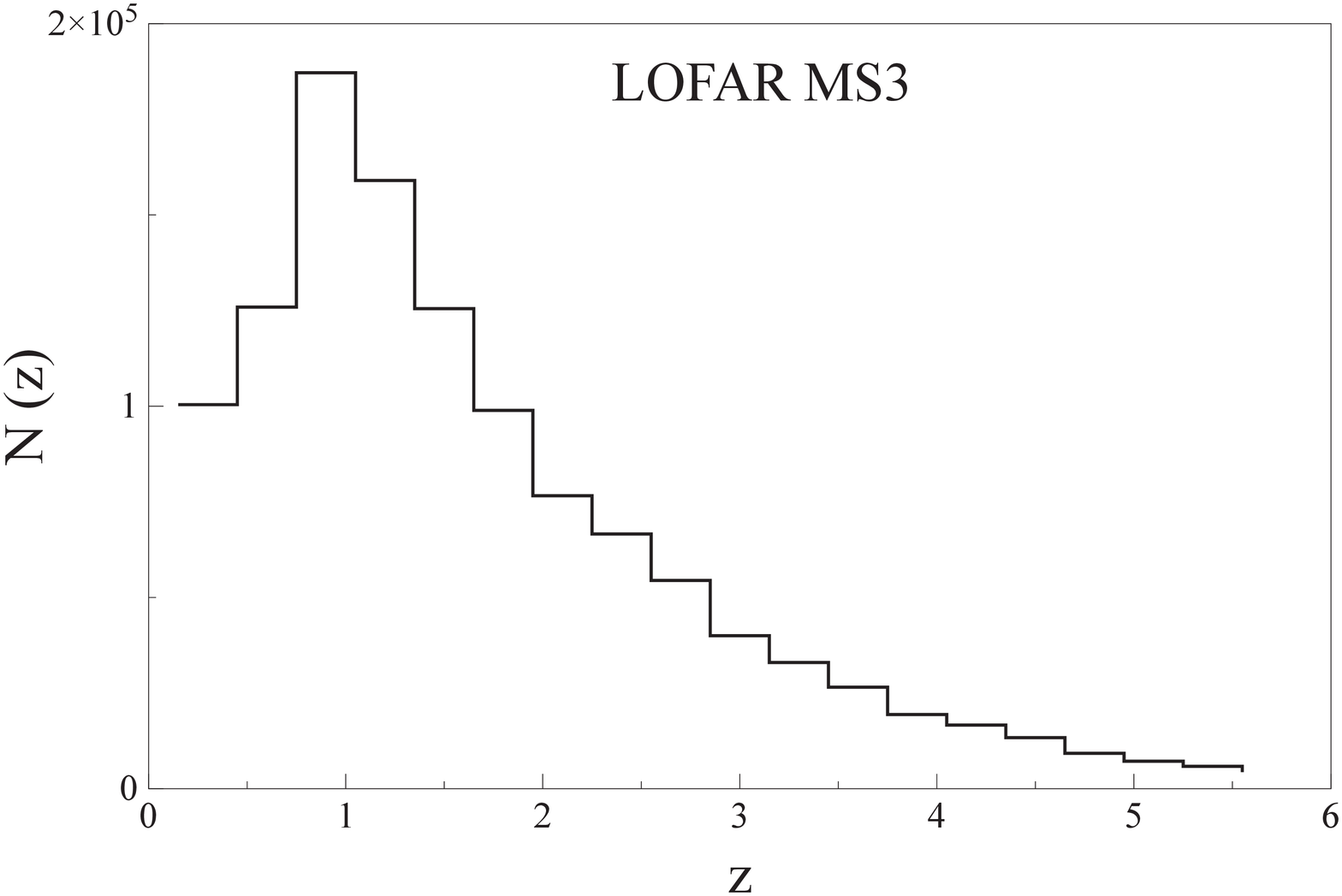, width=0.49\linewidth}
\epsfig{file=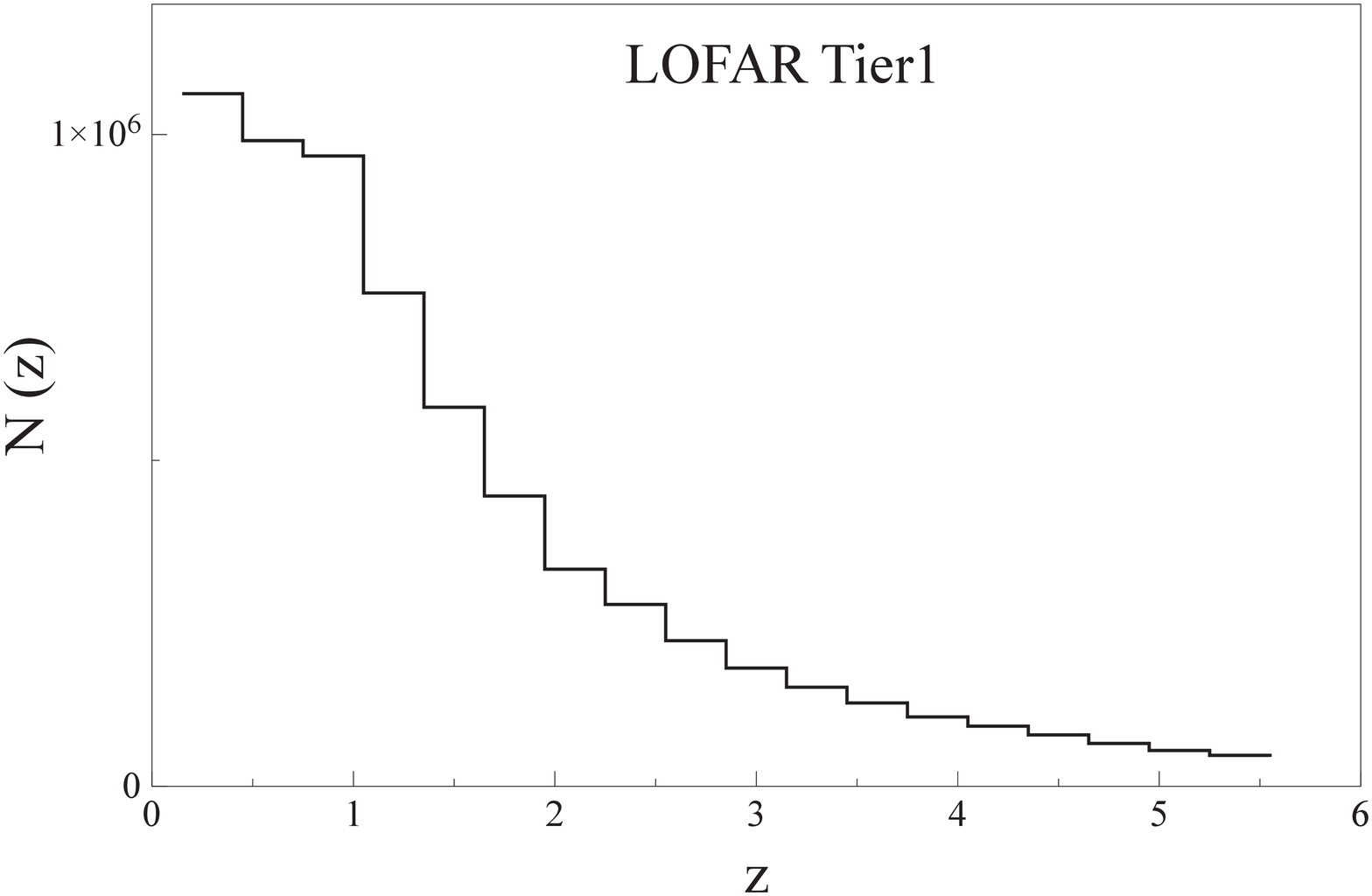, width=0.49\linewidth}
\epsfig{file=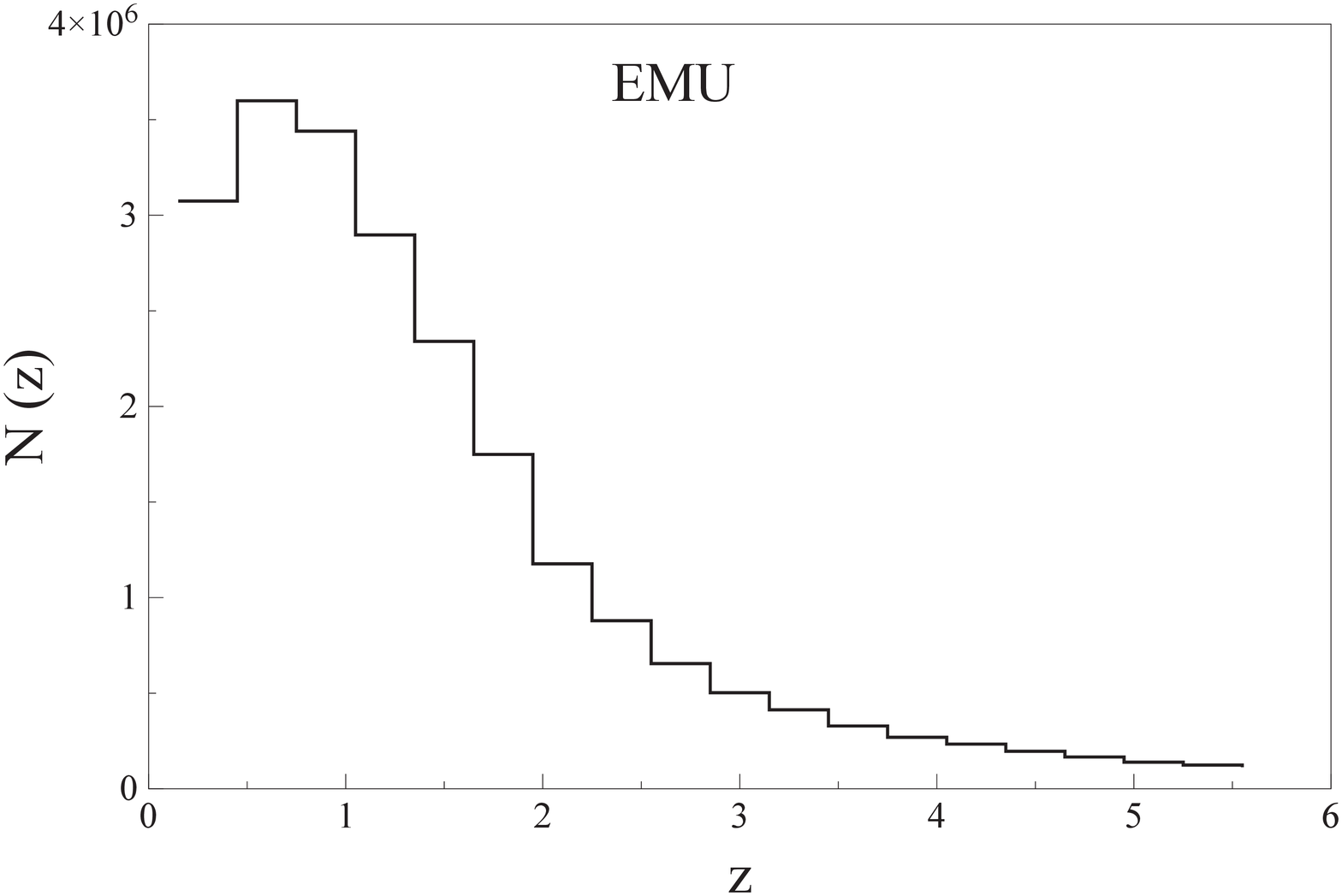, width=0.49\linewidth}
\epsfig{file=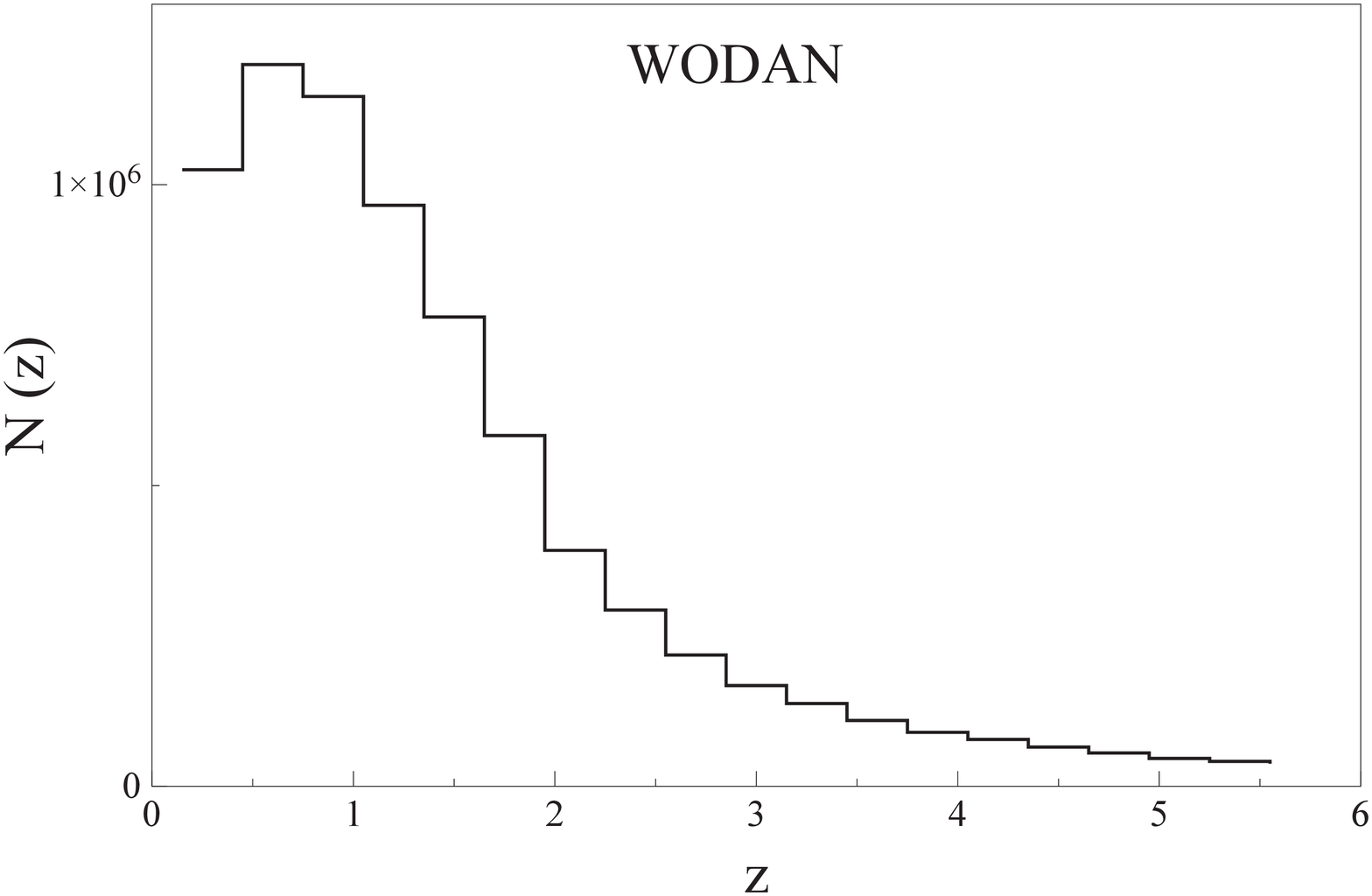, width=0.49\linewidth}
\caption{Redshift distributions found for the different radio surveys: LOFAR MS$^3$ and Tier 1, EMU and WODAN. All source types are included in these overall redshift distributions; on vertical axes are the number of sources per bin of width $\Delta z = 0.3$.}
\label{fig:Nz_tot}
\end{center}
\end{figure*}

\begin{figure*}
\begin{center}
\epsfig{file=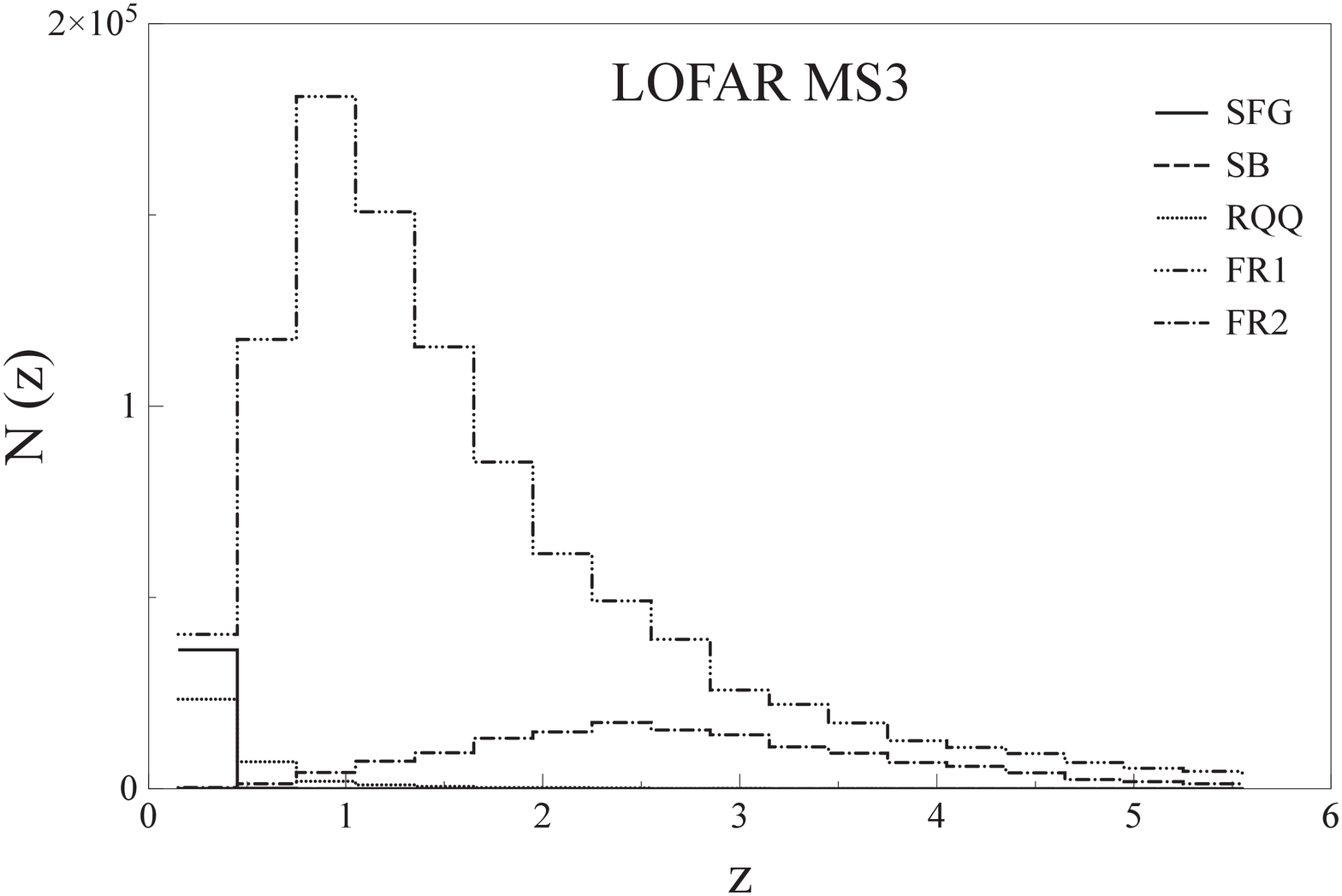, width=0.49\linewidth}
\epsfig{file=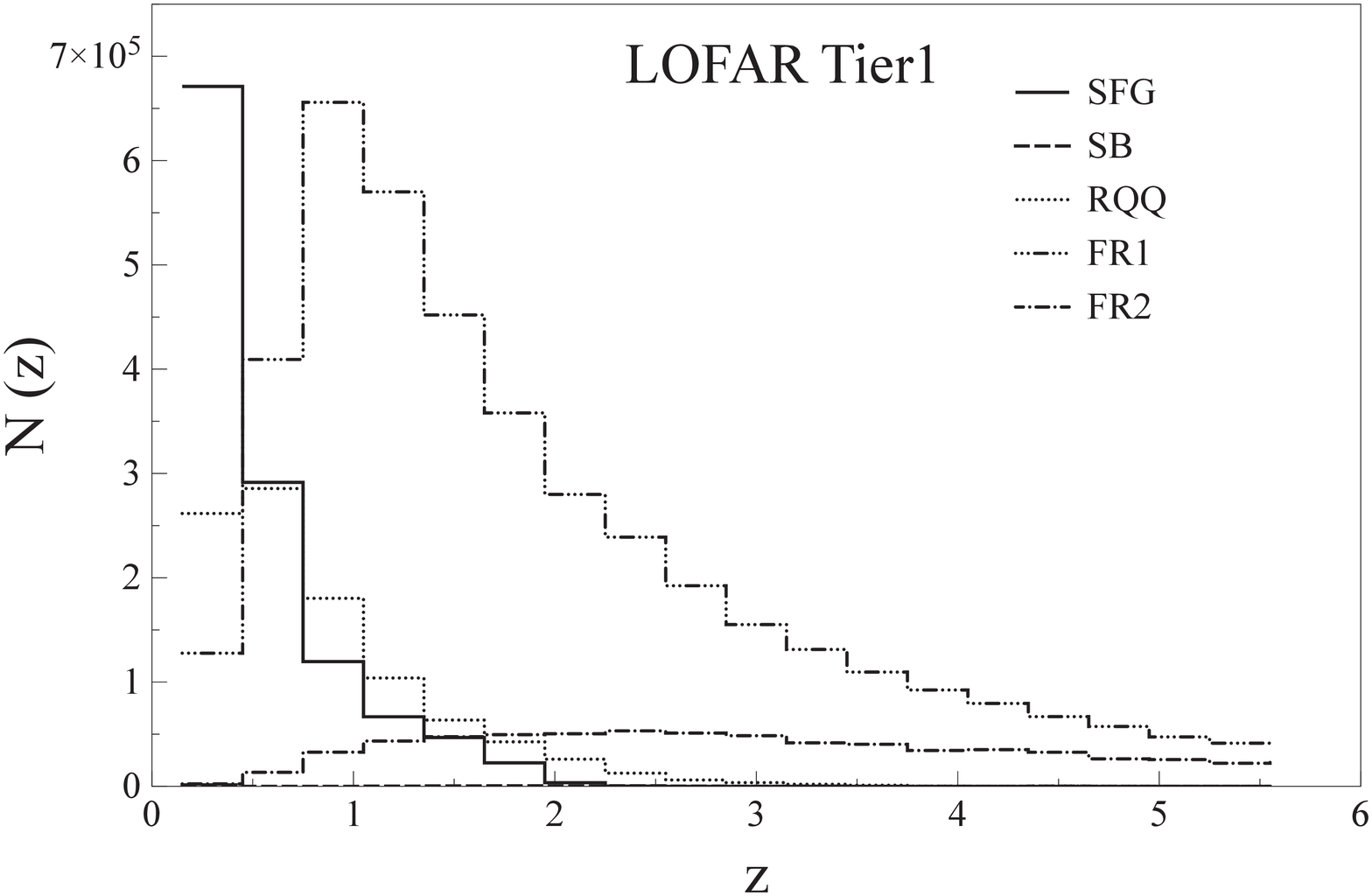, width=0.49\linewidth}
\epsfig{file=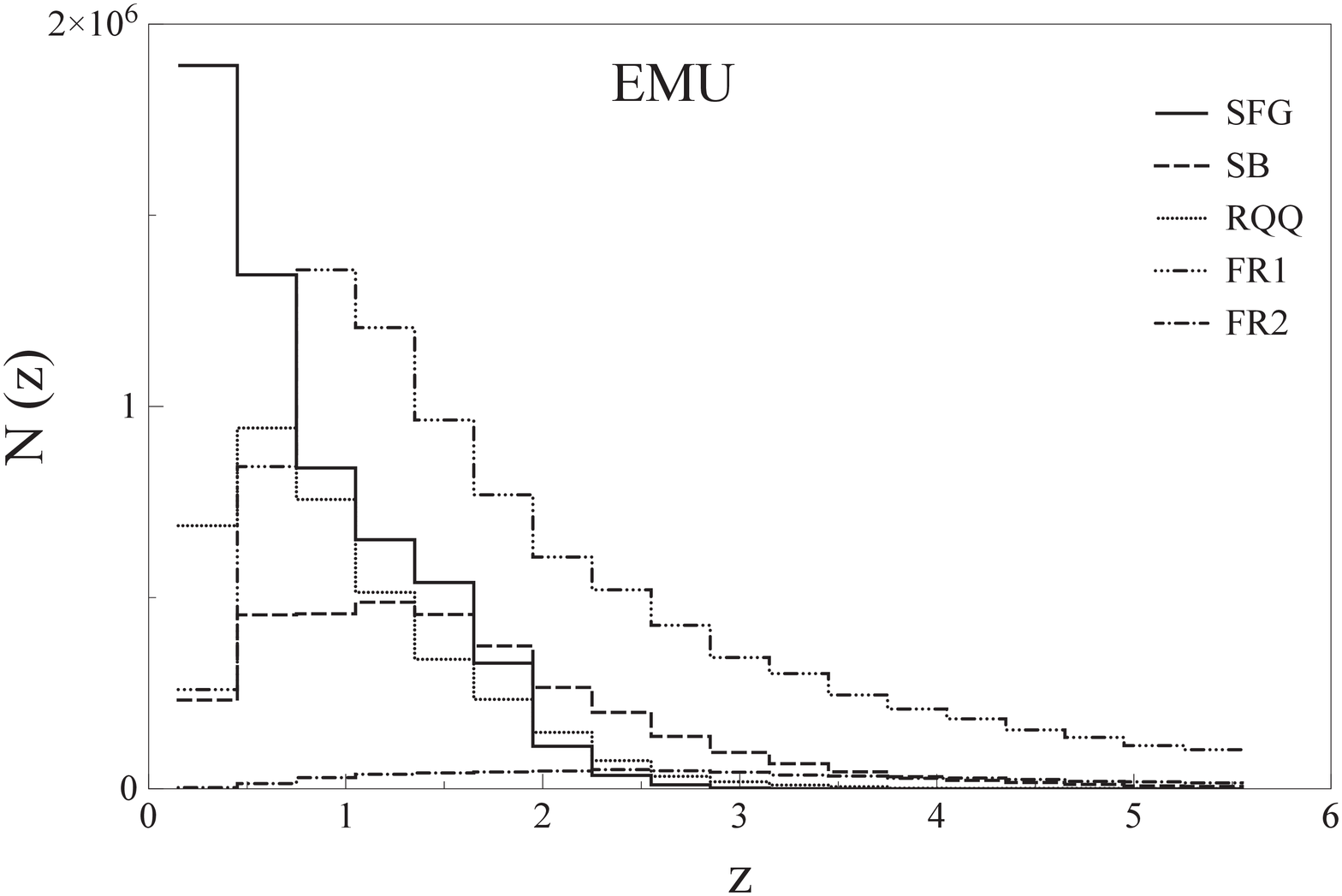, width=0.49\linewidth}
\epsfig{file=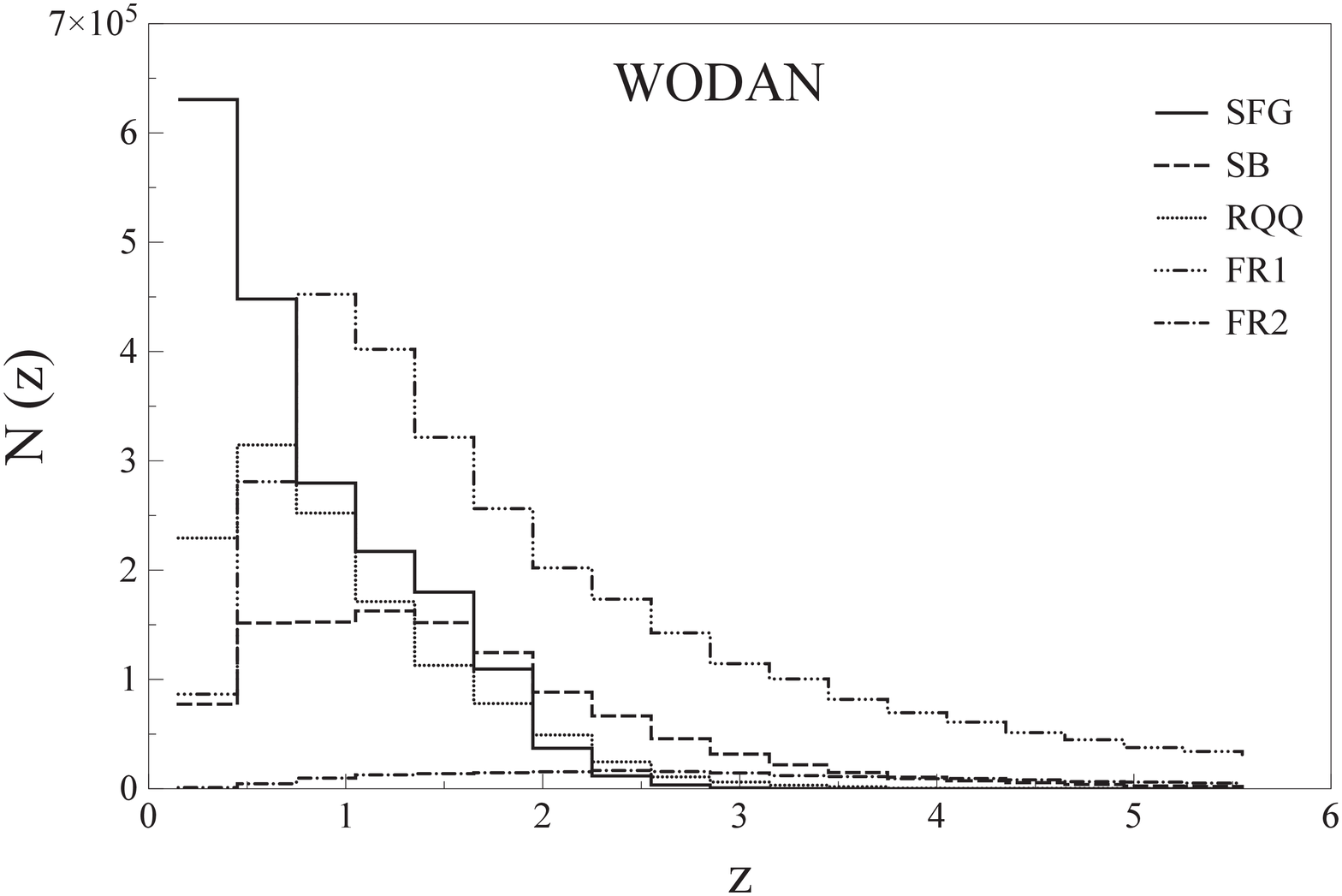, width=0.49\linewidth}
\caption{Redshift distributions of LOFAR, EMU and WODAN surveys, for different source types: Star-Forming Galaxies, Star Burst, Radio Quiet Quasars, FR1 and FR2 sources; on vertical axes are the number of sources per bin of width $\Delta z = 0.3$.}
\label{fig:Nz_pop}
\end{center}
\end{figure*}

In Figures \ref{fig:Nz_tot} and \ref{fig:Nz_pop} we show the resulting redshift distributions adopted for the different surveys; in Figure \ref{fig:Nz_tot} we display the total number of radio sources for the LOFAR (MS$^3$ and Tier1), EMU and WODAN surveys, while in Figure \ref{fig:Nz_pop} we display the number of sources for the different source types (Star Forming Galaxies, StarBursts, Radio Quiet Quasars, Fanaroff-Riley1 and Fanaroff-Riley2; see \cite{wilman08} for details of how each of these is defined) within the surveys.

In addition to predictions for the four surveys considered, we will also consider a combination of EMU and WODAN, given that they will span a similar range of frequencies and depth (see Tab. \ref{tab:surveys}), in order to have a complete full sky catalogue covering both hemispheres.

\subsection{Galaxy Bias}
\label{sec:bias}
\begin{figure}
\epsfig{file=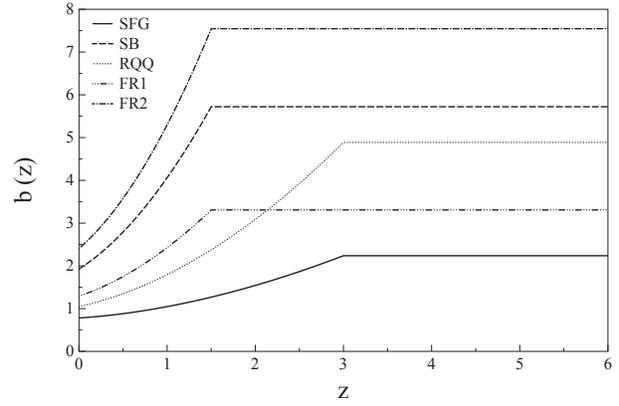, width=0.98\linewidth}
\caption{{Bias as a function of redshift for the different source types, as calculated for our simulated catalogues in accordance with \citet{wilman08}.}}
\label{fig:bias}
\end{figure}

As we will often be using radio sources as a probe of large-scale structure, it is necessary to model how biased the sources are in relation to the underlying structures. On large scales we assume that the two-point correlation function can be written (\citealt{matarrese97}, \citealt{moscardini98}) as:
\begin{equation}
\xi(r,z)=b^{2}(M_{\rm eff},z)\xi_{\rm DM}(r,z)\ ,
\end{equation}
where $M_{\rm eff}$ represents the effective mass of dark matter halos in which sources reside and $\xi_{DM}$ is the correlation function of dark matter. We derive a model of the bias using the peak-background split formalism (\citealt{cole89}, \citealt{mo96}), following the prescription of \citet{sheth99}; in this context  the mass function of halos, altered from \citet{press74}, is given by:
\begin{align}
\label{eq:mf}
\bar{n}(z,M) = & \hksqrt{\frac{2qA^2}{\pi}} \frac{3 H_0^2 \Omega_{0m}}{8
\pi G}\frac{\delta_c}{MD(z)\sigma_M} \cdot \\ \nonumber
& \cdot \left[1+\left(\frac{D(z)\sigma_M}{\hksqrt{q}\delta_c}\right)^{2p} \right]\left|\frac{d\ln \sigma_M}{d \ln M}\right| \cdot \\ \nonumber
& \cdot \exp \left[ -\frac{q\delta_c^2}{2D^2(z)\sigma_M^2} \right] ,
\end{align}
where $\sigma_M^2$ is the mass variance on scale $M$, $\delta_c$ is the critical overdensity for
the spherical collapse, $D(z)$ the linear growth factor of density fluctuations, and $q$ and $p$ are parameters to be fitted with simulations (\citealt{sheth99}); we can describe the bias by:
\begin{align}
\label{eq:bias}
b(M,z) = & 1 + \frac{1}{\delta_c} \left[
\frac{q\delta_c^2}{D^2(z)\sigma_M^2} -1 \right] + \\ \nonumber
& + \frac{2p}{\delta_c} \left(
\frac{1}{1+(\hksqrt{q}\delta_c/[D(z)\sigma_M])^{2p}} \right).
\end{align}
For the purposes of this paper we use the bias in the $S^3$ simulation for each galaxy population, which is computed using the formalism of Eq.~(\ref{eq:mf}), (\ref{eq:bias}) separately for each galaxy population, where each population is assigned a dark matter halo mass. This dark matter halo mass is chosen to reflect the large-scale clustering found by observations. Note that for most of Fig.~\ref{fig:bias},
there are simply no observational measurements available at present, so large uncertainties in bias remain.

The $S^3$ simulation provides us with a source catalogue with the sources identified by type, i.e. starburst, FRII-type radio galaxy etc. Each of these has a different prescription for the bias, as described in \citet{wilman08}.
With this framework, one finds that the increasing bias $b(z)$ with redshift would lead to excessively strong clustering at high redshift, therefore the bias for each population is held constant above a certain cut-off redshift, as described by \citet{wilman08}.
The resulting redshift dependence of the bias we use for the different source types is shown in Fig.~\ref{fig:bias}. While this bias evolution is indicative of that expected, the exact behaviour is not yet well known; to allow for this uncertainty, we will marginalise over the overall bias amplitude, and discuss remaining uncertainties in Section~\ref{sec:uncert}.

%%%%%%%%%%%%%%%%%%%%%%%%%%%%%%%%%%%%%%%%%%%%%%%%%%%%%%%%%%%%%%%%%%
%%%%%%%%%%%%%%%%%%%%%%%%%%%%%%%%%%%%%%%%%%%%%%%%%%%%%%%%%%%%%%%%%%
%%%%%%%%%%%%%%%%%%%%%%	The Auto-Correlation Function		%%%%%%%%%%%%%%%%%%%%%%
%%%%%%%%%%%%%%%%%%%%%%%%%%%%%%%%%%%%%%%%%%%%%%%%%%%%%%%%%%%%%%%%%%
%%%%%%%%%%%%%%%%%%%%%%%%%%%%%%%%%%%%%%%%%%%%%%%%%%%%%%%%%%%%%%%%%%

\section{Cosmological Probes}
\label{sec:cosmoprobes}

In this section, we will describe several cosmological probes which one can measure with the forthcoming radio surveys, in combination with surveys at other wavelengths. Here we describe the necessary framework for calculating the accuracy with which we can measure these probes; in the next section we will describe the signal-to-noise of measurements with the specific planned surveys.

\subsection{The Auto-Correlation Function}
\label{sec:acf}
The first cosmological probe we can examine with the radio surveys is the two-point correlation function of source positions, which is
a measure of the degree of clustering in either the spatial, $\xi(r)$, or the angular, $w(\theta)$, distribution of sources. For the current radio surveys, where individual redshifts will be unknown, there will be little radial information, therefore it will be appropriate to only study the angular two-point correlation function, $w(\theta)$,
which is defined as the excess probability of finding a radio source at an angular distance $\theta$ from another given radio source (\citealt{peebles80}):
\begin{equation}
\delta P = n[1+w(\theta)]\delta\Omega,
\label{eq:w_est}
\end{equation}
where $\delta P$ is the probability, $n$ is the mean surface density and $\delta\Omega$ a surface area element.

The angular two-point correlation function of a given sample of objects can be computed
using one of the many estimators that have been proposed (e.g. \citealt{hamilton93}, \citealt{landy93}). 

Its Fourier transform, the angular power spectrum, can be calculated from the underlying 3D matter power spectrum using:
\begin{equation}
\label{eq:Clgg}
C_{\ell}^{gg} = \langle a_{\ell m}^g a_{\ell m}^{g*} \rangle = 4 \pi
\int \frac{dk}{k} \Delta^2(k) [W_{\ell}^g(k)]^2,
\end{equation}
where $W_{\ell}^g$ is the radio source distribution window function described below in Eq.~\ref{eq:flg},
and $\Delta^2(k)$ is the logarithmic matter power spectrum today, and $a_{\ell m}$ are the spherical harmonics coefficients,
assumed to be standard gaussian random variables.

\subsubsection{The radio source window function}
\label{sec:wlg}
Radio source counts are a biased tracer of the underlying matter distribution, and thus the projected number density of radio sources per steradian is related to the matter distribution via:
\begin{equation}
n(z,\hat{n}) dzd\Omega = \frac{dN}{dz} [1+b(z)\delta(z,\hat{n})] dzd\Omega .
\end{equation}
The window function can then be written as (see e.g. \citealt{raccanelli08, giannantonio08a}):
\begin{equation}
\label{eq:flg}
W_{\ell}^g(k) = \int \frac{dN}{dz} b(z) D(z) j_{\ell}[ck\eta(z)] dz,
\end{equation}
where  $(dN/dz)dz$ is the mean number of sources per steradian with redshift
$z$ within $dz$, brighter than the flux limit, $b(z)$ is the bias factor relating
the source overdensity to the mass overdensity, assumed to be scale-independent,
$D(z)$ is the linear growth factor of mass fluctuations, $j_{\ell}(x)$ is the spherical
Bessel function of order $\ell$, and $\eta(z)$ is the conformal look-back time.

%%%%%%%%%%%%%%%%%%%%%%%%%%%%%%%%%%%%%%%%%%%%%%%%%%%%%%%%%%%%%%%%%%
%%%%%%%%%%%%%%%%%%%%%%%%%%%%	non-Gaussianity		%%%%%%%%%%%%%%%%%%%%%%%%%
%%%%%%%%%%%%%%%%%%%%%%%%%%%%%%%%%%%%%%%%%%%%%%%%%%%%%%%%%%%%%%%%%%

\subsubsection{Non Gaussian clustering}
\label{sec:ng}
The amplitude and shape of clustering on large scales, described by the ACF, can provide important cosmological information. For example, a unique way to test aspects of inflationary theories is given by measuring the statistics of the initial conditions of cosmological perturbations.
An important goal for forthcoming cosmological experiments is to test whether 
initial conditions of the probability distribution function of cosmological perturbations deviate from gaussianity; 
this can be done using the CMB (\citealt{bartolo04}, \citealt{komatsu10} and references therein) or 
the large-scale structure of the Universe (\citealt{matarrese2000}, \citealt{dalal08}, \citealt{slosar08}, \citealt{desjacques10}, \citealt{xia10}).
Deviations from Gaussian initial conditions can be parametrized by the
dimensionless parameter $f_{\rm NL}$:
\begin{equation}
\label{eq:fnl}
\Phi_{\rm NG}=\phi+f_{\rm NL}\left(\phi^2-\langle\phi^2\rangle\right) ,
\end{equation}
where $\Phi$ denotes Bardeen's gauge-invariant potential, which, on sub-Hubble scales reduces to the usual Newtonian peculiar gravitational potential.
Here $\phi$ is a Gaussian random field, and the second term, when $f_{\rm NL}$ is not zero, gives the deviation from gaussianity;
in this paper we refer to the so-called ``local type" $f_{\rm NL}$ and we use the LSS convention (as opposed to the CMB one, where $f_{\rm NL}^{\rm LSS} \sim 1.3 f_{\rm NL}^{\rm CMB}$, \citep{xia10}).

One method for constraining non-Gaussianity from large-scale structure surveys exploits 
the fact that a positive $f_{\rm NL}$ corresponds to positive skewness of the density probability 
distribution, and hence an increased number of massive objects (\citealt{matarrese2000}, \citealt{dalal08}, \citealt{desjacques10}).

In particular, a non-zero $f_{\rm NL}$ in Eq.~(\ref{eq:fnl}) introduces 
a scale-dependent modification of the large-scale halo bias, so that the difference from the usual Gaussian bias, is: 
\begin{equation}
\label{eq:ng-bias}
\Delta b(z, k) = [b_{\rm G}(z)-1] f_{\rm NL}\delta_{\rm ec} \frac{3 \Omega_{0m}H_0^2}{c^2k^2T(k)D(z)}, 
\end{equation}
where $b_{\rm G}(z)$ is the usual bias calculated assuming gaussian initial conditions, assumed to be scale-independent, $D(z)$ is the linear growth factor and $\delta_{\rm ec}$ is the critical value of the matter overdensity for ellipsoidal collapse, $\delta_{\rm ec}=\delta_{\rm c}\hksqrt{q}$.

%%%%%%%%%%%%%%%%%%%%%%%%%%%%%%%%%%%%%%%%%%%%%%%%%%%%%%%%%%%%%%%%%%
%%%%%%%%%%%%%%%%%%%%%%%%%%%%%%%%%%%%%%%%%%%%%%%%%%%%%%%%%%%%%%%%%%
%%%%%%%%%%%%%%%%%%%%%%	The Cross-Correlation Function		%%%%%%%%%%%%%%%%%%%%%%
%%%%%%%%%%%%%%%%%%%%%%%%%%%%%%%%%%%%%%%%%%%%%%%%%%%%%%%%%%%%%%%%%%
%%%%%%%%%%%%%%%%%%%%%%%%%%%%%%%%%%%%%%%%%%%%%%%%%%%%%%%%%%%%%%%%%%

\subsection{The Integrated Sachs-Wolfe Effect} 
%\subsection{The Cross-Correlation Function}
\label{sec:isw}

In addition to making an auto-correlation of source positions, it is possible to cross-correlate the radio source distribution with CMB temperature maps, in order to detect the so-called Integrated Sachs-Wolfe (ISW) effect (\citealt{sachs67}).
Travelling from the last scattering surface to us, CMB photons pass through gravitational potential wells of intervening matter. In an Einstein-de Sitter universe, the blueshift of a photon falling into a well is cancelled by the redshift as it climbs out.
However, in a universe with a dark energy component or modification to General Relativity, the local gravitational potential $\Phi$ varies with time, so potential wells are stretched while photons are traversing the well; this leads to a net blue-shift of the photons, and equivalently to a net change in photon temperature, which accumulates along the photon path, and is proportional to the time variation of the gravitational potential.

The integrated Sachs-Wolfe effect only contributes to the low $\ell$ multipoles of the CMB fluctuations, and is smaller than the primary CMB anisotropies even at those $\ell$. Thus, to make the effect detectable, we have to cross-correlate CMB maps with tracers of large scale structure (\citealt{crittenden96}) such as radio sources, since the source density traces the potential wells.
If the evolution of potentials is modified by dark energy we should observe a correlation between CMB temperature anisotropies and the source distribution; for this reason, ISW measurements will provide a signature for dark energy or modified gravity.
The WMAP data have been cross correlated with a variety of radio, IR, optical, and X-ray surveys (e.g. \citealt{giannantonio06, giannantonio08a}, \citealt{pietrobon06}, \citealt{raccanelli08}; see \citealt{dupe} for a review of recent results and more references) to look for evidence of a decay of the gravitational potential due to the influence of dark energy.

We can write the cross-correlation power spectrum between the surface density
fluctuations of radio sources and CMB temperature fluctuations as:
\begin{equation}
\label{eq:ClgT}
C_{\ell}^{gT} = \langle a_{\ell m}^g a_{\ell m}^{T*} \rangle = 4 \pi
\int \frac{dk}{k} \Delta^2(k) W_{\ell}^g(k)
W_{\ell}^T(k),
\end{equation}
where $W_{\ell}^g$ and $W_{\ell}^T$ are the radio source and CMB window functions, respectively,
and  $\Delta^2(k)$ is the logarithmic matter power spectrum today.

The cross-correlation function as a function of the angular separation $\theta$
is then obtained as:
\begin{equation}
\label{eq:cthetagT}
C^{gT}(\theta) = \sum_{\ell} \frac{2\ell+1}{4 \pi}
C_{\ell}^{gT}L_{\ell}(\cos \theta),
\end{equation}
where $L_{\ell}$ are the Legendre polynomials of order $\ell$ \citep{legendre}.

\subsubsection{The ISW window function}
\label{sec:wlt}
In the Newtonian gauge, scalar metric perturbations are specified by the gauge-invariant potentials $\Psi$ and $\Phi$:
\begin{equation}
ds^2 = -  a^2 (\tau) [ (1+2\Psi) d\tau^2 - (1-2\Phi) d \vec{x}^2].
\end{equation}
The temperature anisotropies due to the ISW effect are expressed by an integral
over the conformal lookback time from today ($\eta=0$) to the CMB decoupling surface $\eta_{dec}$:
\begin{align}
\label{eq:deltat-isw}
\Theta_{ISW}= \frac{\delta T}{T} = -\frac{1}{c^2} \int_0^{\eta_{dec}} (\dot{\Phi} + \dot{\Psi}) d \eta,
\end{align}
where $\tau$ is the conformal time, the dot represents a conformal time derivative and the integral is calculated along
the line of sight of the photon.
In the absence of anisotropic stress, the momentum constraint in GR fixes $\Phi = - \Psi$, so the ISW modification of the temperature of the CMB in GR becomes:
\begin{align}
\Theta_{ISW}= \frac{\delta T}{T} = -\frac{2}{c^2} \int_0^{\eta_{dec}} \frac{\partial \Phi}{\partial \eta} d \eta.
\end{align}

The local gravitational potential is related to the matter distribution via the Poisson equation:
\begin{align}
\label{eq:poisson}
\nabla^2 \Phi = 4 \pi G a^2 \varrho_m \delta_m,
\end{align}
where the gradient is taken with respect to comoving coordinates; taking the
Fourier transform we have:
\begin{align}
\label{eq:gravpot}
\Phi(k,\eta)= - \frac{3}{2} \Omega_{0m} \left(\frac{H_0}{ck}\right)^2
g(\eta) \delta(k),
\end{align}
where $H_0$ is the Hubble constant, $g(\eta) \equiv D(\eta)/a(\eta)$ is the linear
growth suppression factor and $\delta(k)$ is the mass overdensity field.

Combining Eq.~\ref{eq:poisson} and \ref{eq:gravpot}, the window function for the ISW effect can be written as:
\begin{equation}
\label{eq:flT}
W_{\ell}^T(k) = 3 \Omega_{0m} \left(\frac{H_0}{ck}\right)^2 \int \frac{\partial \Phi}{\partial z} j_{\ell}[ck\eta(z)] dz,
\end{equation}
where $\Phi(z)$ is the Newtonian gravitational potential. \\

%%%%%%%%%%%%%%%%%%%%%%%%%%%%%%%%%%%%%%%%%%%%%%%%%%%%%%%%%%%%%%%%%%
%%%%%%%%%%%%%%%%%%%%%%%%%%%%%%%%%%%%%%%%%%%%%%%%%%%%%%%%%%%%%%%%%%
%%%%%%%%%%%%%%%%%%%%%%	Magnification bias		%%%%%%%%%%%%%%%%%%%%%%
%%%%%%%%%%%%%%%%%%%%%%%%%%%%%%%%%%%%%%%%%%%%%%%%%%%%%%%%%%%%%%%%%%
%%%%%%%%%%%%%%%%%%%%%%%%%%%%%%%%%%%%%%%%%%%%%%%%%%%%%%%%%%%%%%%%%%

\subsection{Magnification bias}
\label{sec:cosmag}
Light rays are deflected by large scale structures along the line of sight, which therefore  systematically introduce distortions in the observed images of distant sources; this is the phenomenon of gravitational lensing. The sources behind a lens are magnified in size, while surface brightness is conserved; this leads to an increase in the total observed luminosity of a source.

Observationally we can detect the effects of magnification by cross-correlating two galaxy surveys with disjoint redshift distributions; in this paper, we consider the possibility of using an optical survey such as SDSS-II \citep{sdss-ii} or DES\footnote{http://www.darkenergysurvey.org/} for our low redshift ``lens'' sample (Pan-STARRS will also be available on these timescales and could also be used), which will serve as the foreground which magnifies the background radio distribution; we will discuss this further in Section~\ref{sec:cosmag-pred}.

This ``cosmic magnification" effect was first detected by \cite{scranton}, who cross-corelated foreground SDSS galaxies with SDSS quasars. More recently, \cite{hildebrandt} have detected the effect in samples of normal galaxies in the Canada-France-Hawaii-Telescope Legacy Survey, \cite{wang11} have detected the effect at longer wavelengths using Herschel, while \cite{menard} have built on the SDSS analysis by constraining galaxy-mass and galaxy-dust correlation functions.

The effect can be described in detail as follows. At position $\vec{\varphi}$, we can relate the behaviour of unlensed sources with
number density $N_0(m)dm$ within a magnitude range $[m,m+dm]$, to that of lensed
sources with number density $N(m,\vec{\varphi})dm$.
There are two competing effects in this relationship, namely the flux increase due to magnification
of distant faint sources, which increases the number density of observed images above a certain magnitude threshold; and counteracting this,
the dilution of the number density due to the stretching of the solid angle by lensing. If the source fluxes have a distribution with a power law slope given by:
\begin{equation}
\label{eq:alpha}
\alpha(m) = 2.5 \frac{\partial[\log N_0(m)]}{\partial m} , 
\end{equation}
one can obtain (\citealt{bartelmann01}):
\begin{equation}
N(m,\vec{\varphi})dm=\mu^{\alpha(m)-1}N_0(m)dm,
\end{equation}
where the magnification $\mu$ is:
\begin{equation}\label{mag}
\mu=\frac{1}{\vert(1-\kappa)^2-\vert\gamma\vert^{2}\vert},
\end{equation}
where the convergence $\kappa$ and the shear $\gamma$ are two further lensing distortions. In the weak lensing regime it is possible to Taylor expand the last
equality in Eq.~(\ref{mag}) to obtain:
\begin{equation}
\mu(\vec{\varphi})\simeq1+2\kappa(\vec{\varphi}).
\end{equation}
We therefore see that the magnification is closely related to the convergence $\kappa$, which is related to the matter overdensity via a line-of-sight integral (\citealt{bartelmann01}):
\begin{equation}
 \kappa(\vec{\varphi})=\frac{3\Omega_{0m}H_0^2}{2c^2}\int_0^{w_{H}}dw W(w)f_K(w)\frac{\delta(f_K(w)\vec{\varphi},w)}{a(w)},
\end{equation}
with $\vec{\varphi}$ being the angular position on the sky, $w_H$ the horizon distance, $w(z)$
the comoving radial distance, $f_K(w)$ the angular diameter comoving distance, $a$ the scale factor, and a quantity $W$ involving the redshift distribution and geometry:
\begin{equation}
 W(w)=\int_w^{w_H}dw'Z_{w}(w')\frac{f_K(w'-w)}{f_K(w')},
 \label{eq:W}
\end{equation}
for which $Z_{w}(w)dw$ is the source redshift distribution.

Because of the magnification bias effect described, we can obtain cosmological constraints by cross-correlating foreground and background objects, and hence investigating how clustered lensed background sources appear to be around foreground sources, compared to a random distribution. The most common
estimator of the angular two point correlation (also adopted in this work) is given by:
\begin{equation}
\xi_{SL}=[\bar{N_S}\bar{N_L}]^{-1}\overline{[N_S(\vec{\varphi})-\bar{N_S}][N_L(\vec{\varphi}+\vec{\phi})-\bar{N_L}]} ,
\end{equation}
where in our case $S$ and $L$ indexes denote background sources and foreground
lenses and overbarred quantities correspond to averaged quantities. 

Large scale structures only slightly magnify or demagnify sources, so we can write:
\begin{equation}
\mu^{\alpha-1}=(1+\delta\mu)^{\alpha-1}\simeq 1+(\alpha-1)\delta\mu ,
\label{eq:cosmag-mu}
\end{equation}
leading to an over/under-density in background sources:
\begin{equation}
\frac{N_S(\vec{\varphi})-\bar{N_S}}{\bar{N_S}}\simeq(\alpha-1)\delta\mu(\vec{\varphi}).
\end{equation}
Assuming that foreground sources have bias $b_L$, the number density can be related to the underlying matter density
contrast by:
\begin{equation}
\frac{N_L-\bar{N_L}}{\bar{N_L}}=b_L\delta(\vec{\varphi}) ,
\end{equation}
Then as a consequence (c.f. \citealt{bartelmann01}), $\xi_{SL}(\vec{\varphi})$ is related to the theoretical magnification density contrast
2-point correlation function $\xi_{\mu\delta}(\vec{\varphi})$ via:
\begin{equation}
\xi_{SL}(\vec{\varphi})=(\alpha-1)b_L(\vec{\varphi})\xi_{\mu\delta}(\vec{\varphi}) ,
\label{eq:mag-corr}
\end{equation}
with
\begin{align}
\label{eq:xi}
\xi_{\mu\delta}(\vec{\varphi})=\frac{3H_0^2\Omega_{0}}{2\pi c^2}\int dw
\frac{W(w)G_f(w)}{a(w)f_K^2(w)} \int k dk P_\delta(k) J_0(k\vec{\varphi}),
\end{align}
where $P_\delta(k)$ is the matter power spectrum, $G_f(w)$ is the foreground redshift distribution
and $W(w)$ is the source lensing efficiency distribution given in equation (\ref{eq:W}). 

It is only when $\alpha\neq 1$
that we obtain magnification bias. We obtain a positive cross-correlation  only when $\alpha >1$
and anticorrelation when $\alpha <1$. Equations~(\ref{eq:mag-corr}) and (\ref{eq:xi}) show how magnification bias observations allow us to measure information about the amplitude, shape and evolution of the matter power spectrum, together with information about the bias and geometrical factors in the expanding background. 

From (Eq.~\ref{eq:xi}) we can also obtain the cosmic magnification power spectrum (\citealt{bartelmann01}):
\begin{equation}
\label{eq:Clgmu}
C_{\ell}^{g\mu} = \langle a_{\ell m}^g a_{\ell m}^{\mu*} \rangle = \int \frac{dk}{k} \Delta^2(k) W_{\ell}^g(k) W_{\ell}^{\mu}(k),
\end{equation}
where $W_{\ell}^g(k)$ has the same meaning as in Eq.~(\ref{eq:Clgg}), and $W_{\ell}^{\mu}(k)$ contains the prefactors and $w$ integral from equations \ref{eq:mag-corr} and \ref{eq:xi}.

%%%%%%%%%%%%%%%%%%%%%%%%%%%%%%%%%%%%%%%%%%%%%%%%%%%%%%%%%%%%%%%%%%
%%%%%%%%%%%%%%%%%%%%%%%%%%%%%%%%%%%%%%%%%%%%%%%%%%%%%%%%%%%%%%%%%%
%%%%%%%%%%%%%%%%%%%%%%	Cosmological Implications		%%%%%%%%%%%%%%%%%%%%%%
%%%%%%%%%%%%%%%%%%%%%%%%%%%%%%%%%%%%%%%%%%%%%%%%%%%%%%%%%%%%%%%%%%
%%%%%%%%%%%%%%%%%%%%%%%%%%%%%%%%%%%%%%%%%%%%%%%%%%%%%%%%%%%%%%%%%%

%\section{Cosmological Implications}

\section{Predictions for measurements with forthcoming surveys}

%%%%%%%%%%%%%%%%%%%%%%%%%%%%%%%%%%%%%%%%%%%%%%%%%%%%%%%%%%%%%%%%%%
%%%%%%%%%%%%%%%%%%%%%%%%%%%%%%%%%%%%%%%%%%%%%%%%%%%%%%%%%%%%%%%%%%
%%%%%%%%%%%%%%%%%%%%%%%%%%%%	ACF tests		%%%%%%%%%%%%%%%%%%%%%%%%%%%%
%%%%%%%%%%%%%%%%%%%%%%%%%%%%%%%%%%%%%%%%%%%%%%%%%%%%%%%%%%%%%%%%%%
%%%%%%%%%%%%%%%%%%%%%%%%%%%%%%%%%%%%%%%%%%%%%%%%%%%%%%%%%%%%%%%%%%

\subsection{Autocorrelation predictions}
We computed the predicted auto-correlation source power spectra for LOFAR, EMU and WODAN using Eq.~\ref{eq:Clgg}; the errors were assumed to follow:
\begin{equation}
\label{eq:err-clgg}
\sigma_{C_{\ell}^{gg}} = \hksqrt{\frac{2\left(C_{\ell}^{gg}+\frac{1}{\bar{n}}\right)^2}{(2\ell+1)f_{\rm sky}}} ,
\end{equation}
where $f_{\rm sky}$ is the sky coverage of the survey and $\bar{n}$ is the mean number of sources per steradian.
This assumes that systematics are sub-dominant, and there are no effect from the finite size of objects (i.e. we are not close to the confusion limit).

\begin{figure*}
\begin{center}
\epsfig{file=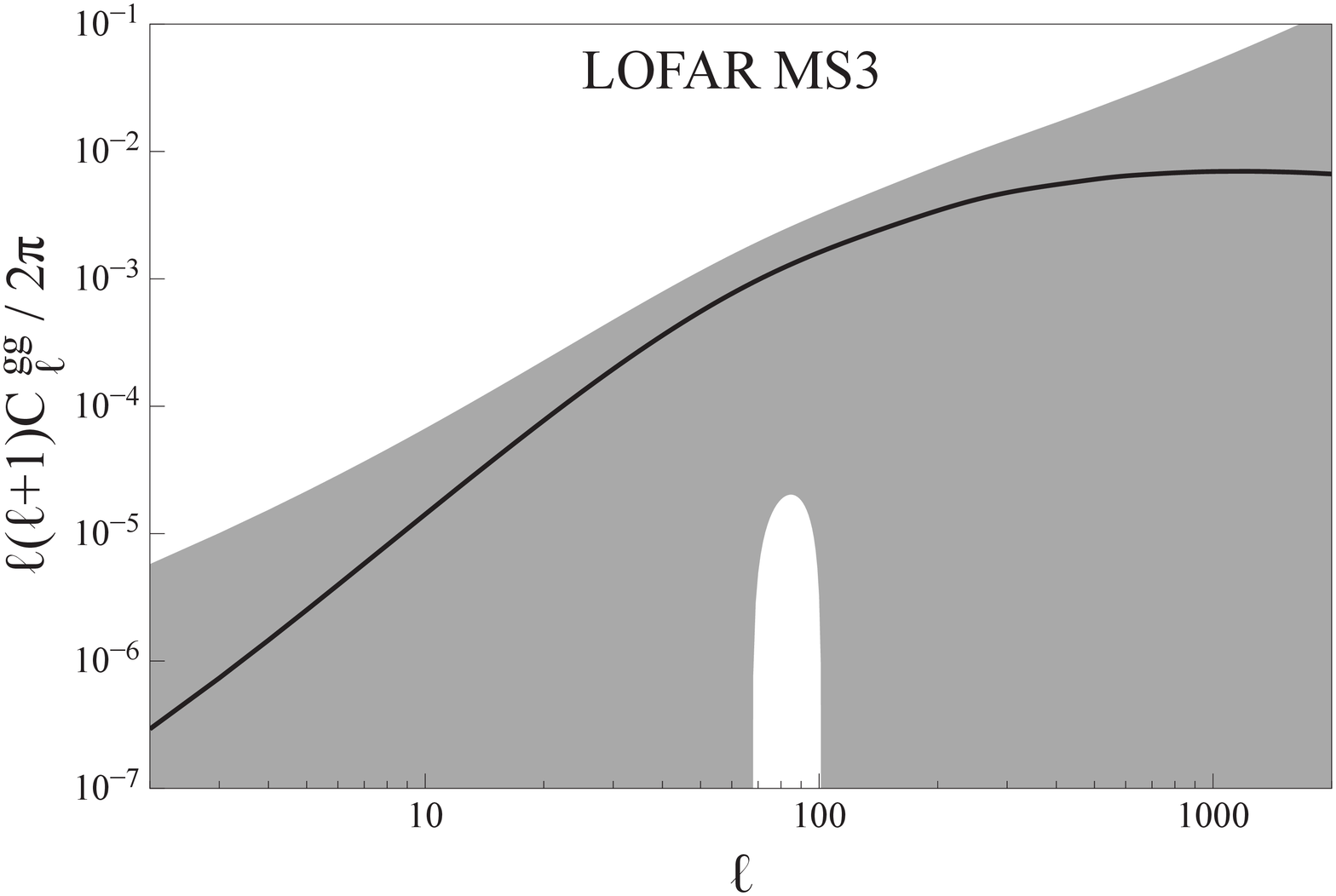, width=0.49\linewidth}
\epsfig{file=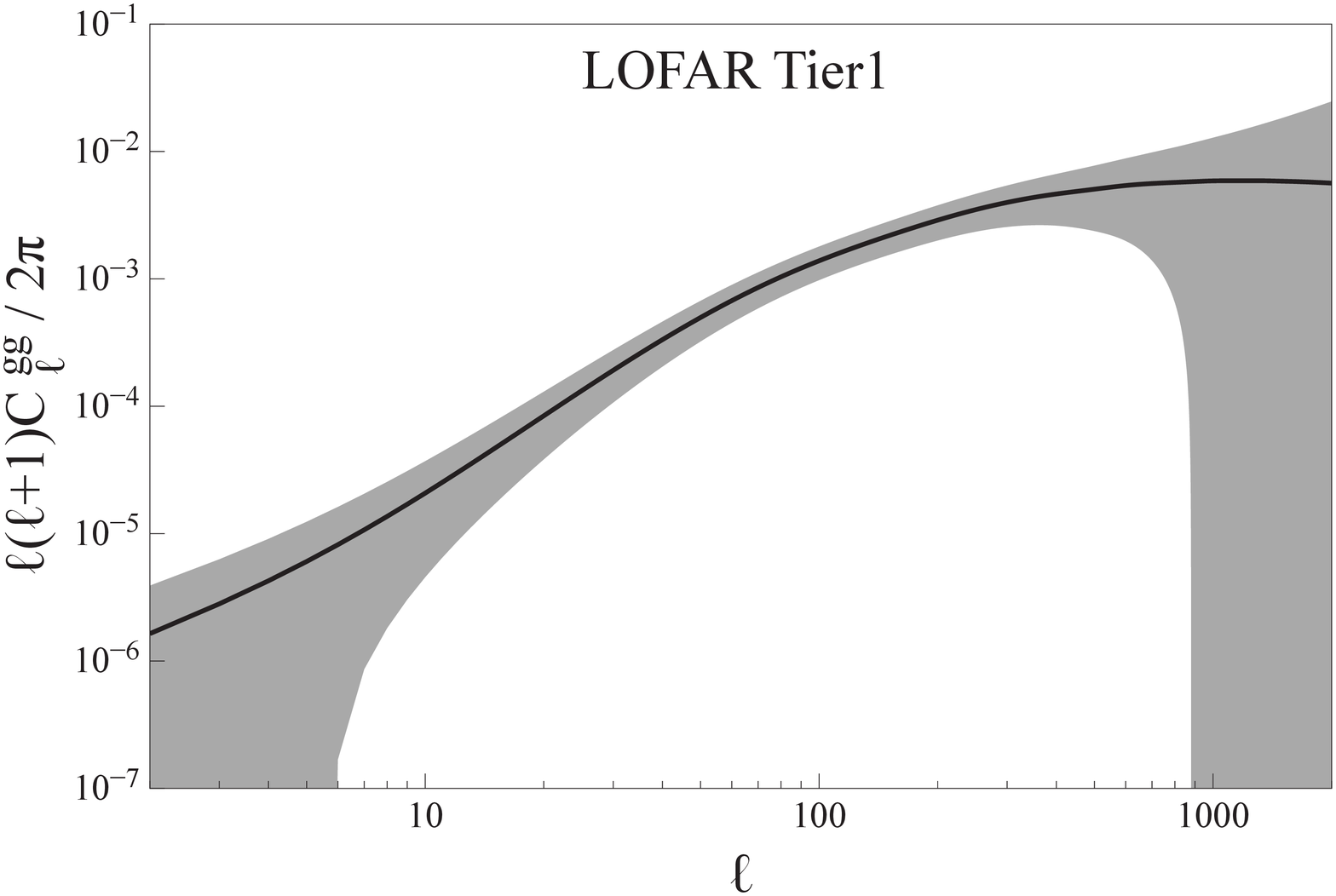, width=0.49\linewidth}
\epsfig{file=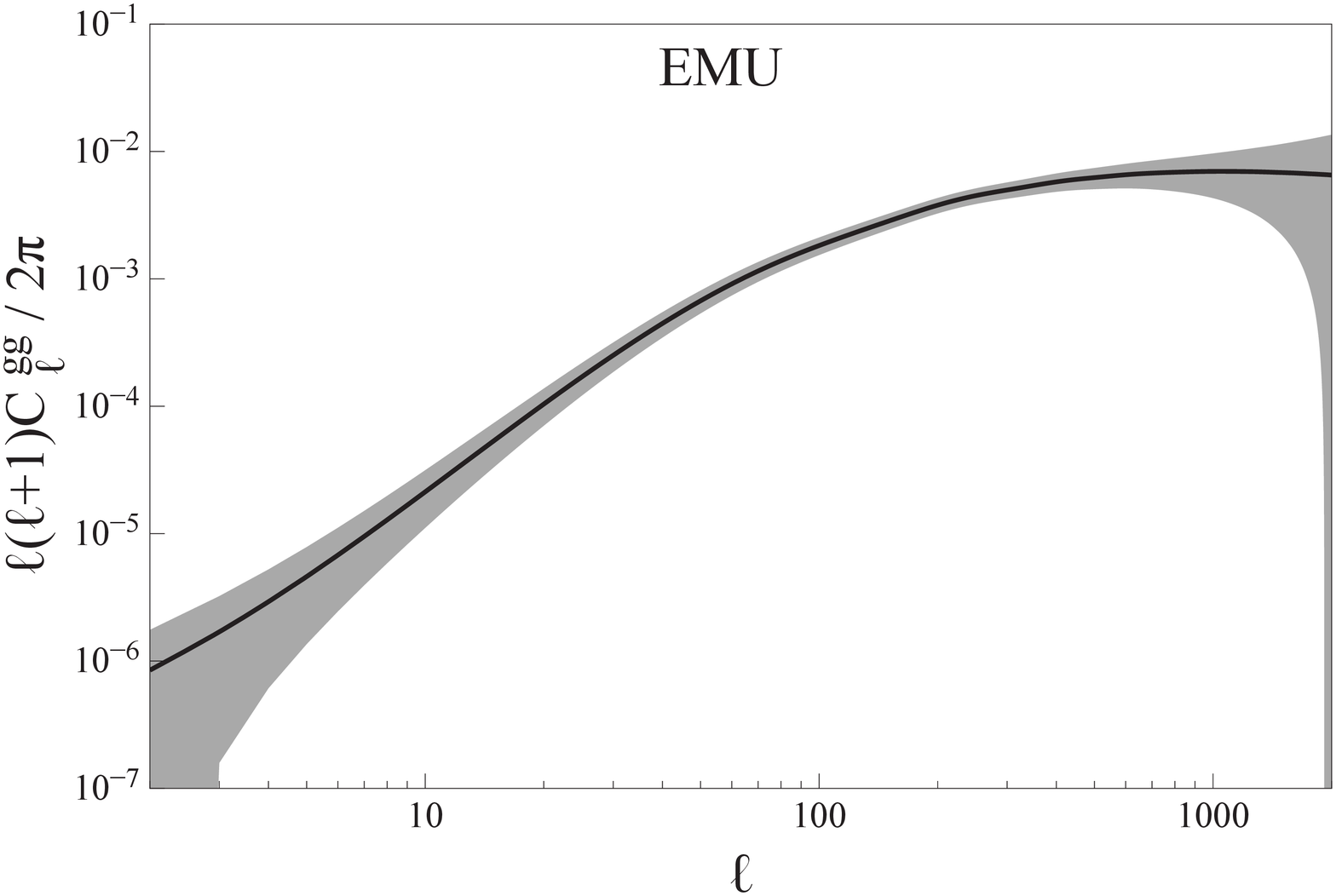, width=0.49\linewidth}
\epsfig{file=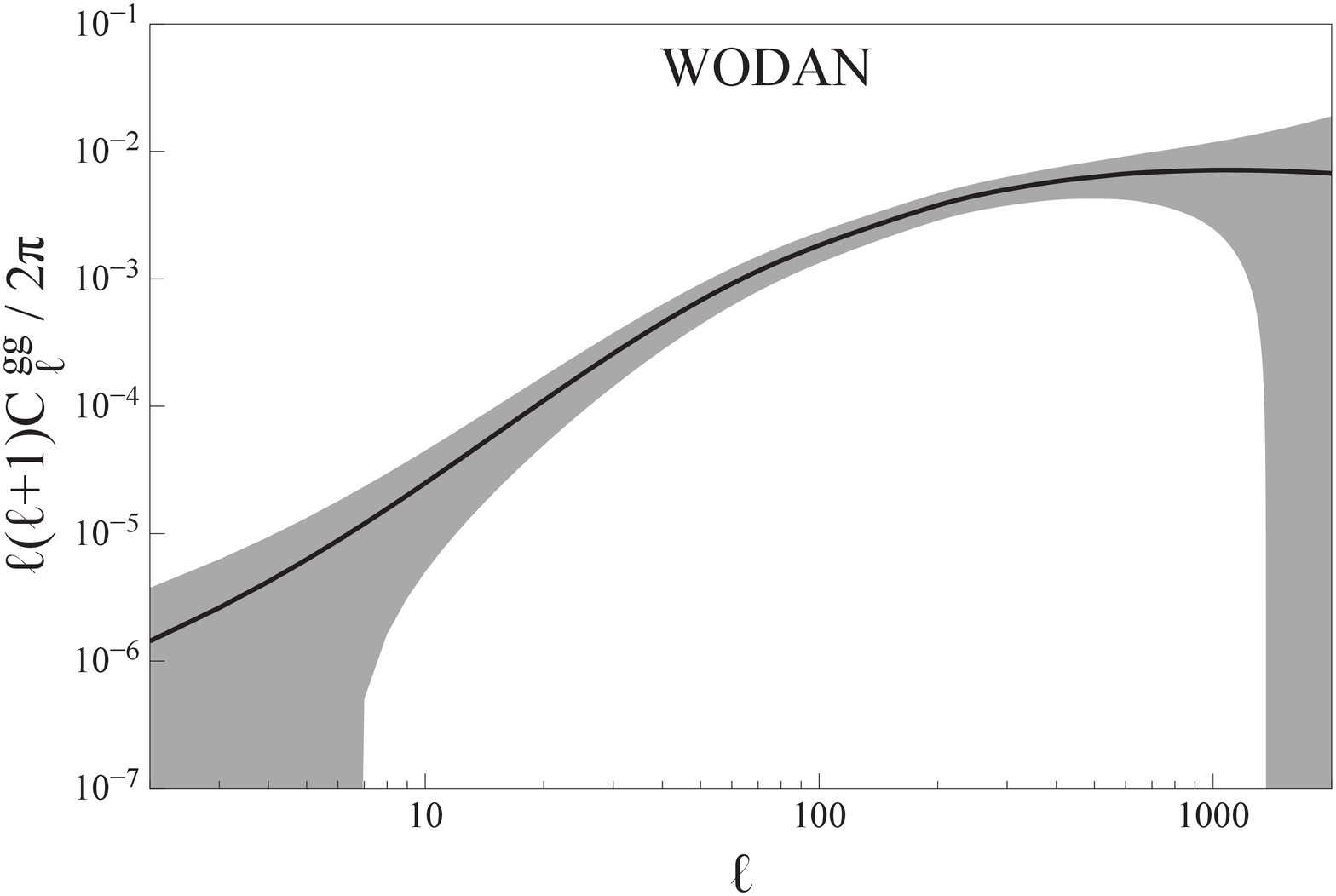, width=0.49\linewidth}
\caption{Source power spectra (Eq.~\ref{eq:Clgg}) of the combined source populations (black solid lines) for the different surveys, with 1-$\sigma$ errors (grey shaded regions), as in Eq.~\ref{eq:err-clgt}.}
\label{fig:ACF}
\end{center}
\end{figure*}

In Fig. \ref{fig:ACF} we show $C_{\ell}^{gg}$ for the combined source populations of the four different surveys considered. As shown, the errors on large and small scales are more pronounced since they are dominated by cosmic variance and shot noise, respectively; at intermediate scales, surveys with higher number density will provide the best measurements (in this case EMU). We will examine what can be learned cosmologically from these measurements in section 7. \\

\begin{figure}
\begin{center}
\epsfig{file=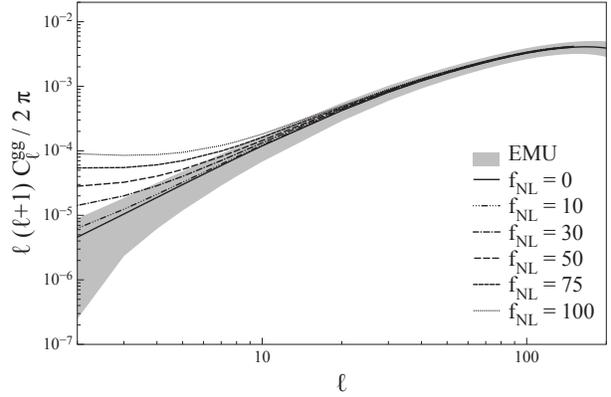, width=0.98\linewidth}
\caption{Source power spectrum of EMU radio sources for different values of the non-Gaussianity parameter $f_{\rm NL}$; shaded regions are errors for the EMU survey as in Eq.~\ref{eq:err-clgg}.}
\label{fig:ACF-NG}
\end{center}
\end{figure}

In Fig.~\ref{fig:ACF-NG} we plot the predicted source power spectrum of EMU radio sources for different values of the non-Gaussianity parameter $f_{\rm NL}$; the black solid line is the standard Gaussian prediction, the other lines being the prediction for non-Gaussian clustering, and the shaded area is 1-$\sigma$ errors (per mode) as in Eq.~\ref{eq:err-clgg}.
The presence of the non-Gaussian bias of Eq.~\ref{eq:ng-bias} enhances the clustering at large scales, thus increasing the amplitude of the autocorrelation function at those scales. 
A $\chi^2$ analysis shows that EMU should be able to distinguish (at 1-$\sigma$ level) a $f_{\rm NL}$ of 8 from a purely Gaussian model; 
it is worth noting that the current limit on $f_{\rm NL}$ from WMAP 7-year data is $f_{\rm NL}^{\rm LSS} = 42\pm27$ at 68\% CL (\citealt{komatsu11}\footnote{Note that in the CMB convention this is $32\pm21$}), and any detection of $f_{\rm NL} \gg 1$ would rule out all single scalar field inflation models \citep{komatsu10}.

It is intriguing to note that the observed autocorrelation function of the NVSS Survey has a shape that differs from the $\Lambda$CDM prediction at relatively large angular separation (\citealt{xia10}); the observed behaviour can be explained using a non-Gaussian correction (\citealt{xia10}), a peculiar bias model \citep{raccanelli08}, or some systematic errors not yet found in the NVSS Survey.
The degeneracy between models of bias and non-Gaussian corrections can be broken because their effect has a different redshift and scale dependence (the non-Gaussian correction is important only at large scales, because of the $1/k^2$ term in Eq.~\ref{eq:ng-bias}). 

It is also interesting to note that a similar excess power at large scales has been found in spectroscopic (\citealt{kazin10}, \citealt{samushia11}) and photometric (\citealt{thomas10}) data sets, although \citet{ross11} suggest that this is likely to be due to masking effects from stellar sources.
With the forthcoming all-sky radio surveys, measuring the angular autocorrelation will be an interesting check for this problem. 
On the other hand, the CMB shows a lack of correlation at angular scales $>$ 60 degrees (\citealt{bennett10}, \citealt{copi10}), discrepant with the concordance model of cosmology. The significance and origin of this is unclear. Certainly radio surveys that cover large fractions of the full sky will help to resolve this puzzle.

%%%%%%%%%%%%%%%%%%%%%%%%%%%%%%%%%%%%%%%%%%%%%%%%%%%%%%%%%%%%%%%%%%
%%%%%%%%%%%%%%%%%%%%%%%%%%%%%%%%%%%%%%%%%%%%%%%%%%%%%%%%%%%%%%%%%%
%%%%%%%%%%%%%%%%%%%%%%%%%%%%	  ISW tests		%%%%%%%%%%%%%%%%%%%%%%%%%%%%
%%%%%%%%%%%%%%%%%%%%%%%%%%%%%%%%%%%%%%%%%%%%%%%%%%%%%%%%%%%%%%%%%%
%%%%%%%%%%%%%%%%%%%%%%%%%%%%%%%%%%%%%%%%%%%%%%%%%%%%%%%%%%%%%%%%%%

\subsection{Cross-correlation predictions}
\label{sec:test-isw}
As we have seen in section \ref{sec:isw}, the cross-correlation between the CMB and the LSS depends on various factors from both the window functions in Eq.~(\ref{eq:ClgT}); it is influenced by the evolution of the gravitational potential (Eq.~\ref{eq:flT}) and by the clustering and bias of structures (Eq.~\ref{eq:flg}), and for this reason it has been used to test and constrain cosmological issues such as the evolution and clustering of structures (\citealt{raccanelli08}, \citealt{massardi10}, \citealt{schaefer09}), models of dark energy (\citealt{pogosian05}, \citealt{xia09}) and alternative models for the gravitational potential, such as the DGP, Unified Dark Matter cosmologies and Brans-Dicke theories (\citealt{giannantonio08b}, \citealt{bertacca11}, \citealt{defelice}).

The detection of the ISW effect via the cross-correlation of the LSS with the CMB is cosmic-variance limited, as it affects only the largest angular scales; therefore, the best measurement possible is a complete full sky survey with negligible shot noise. Regarding the CMB, the data provided by WMAP is already precise enough at low $\ell$, and the improvement that Planck will provide does not substantially affect the ISW detection significance.

\begin{figure*}
\begin{center}
\epsfig{file=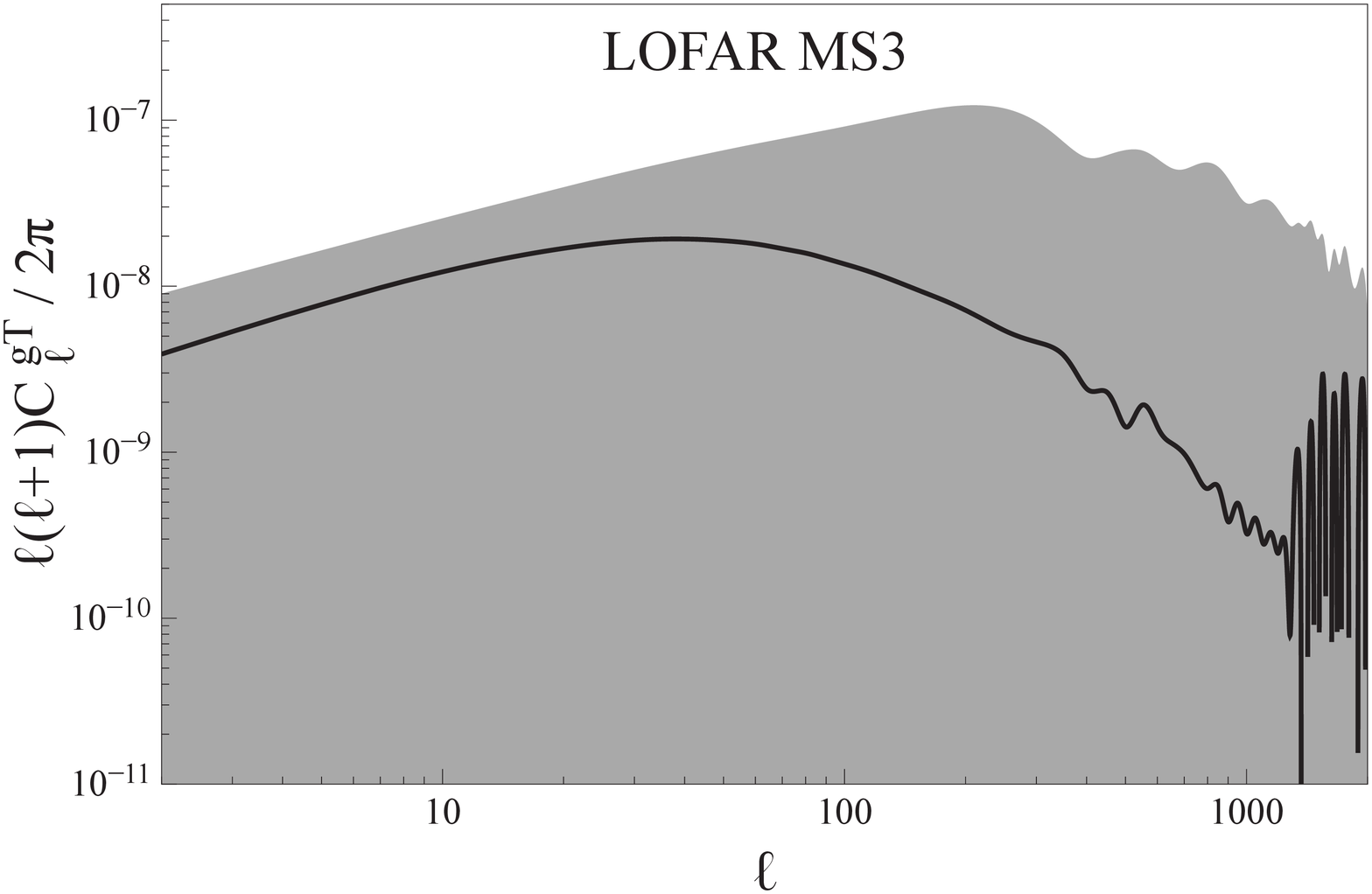, width=0.49 \linewidth}
\epsfig{file=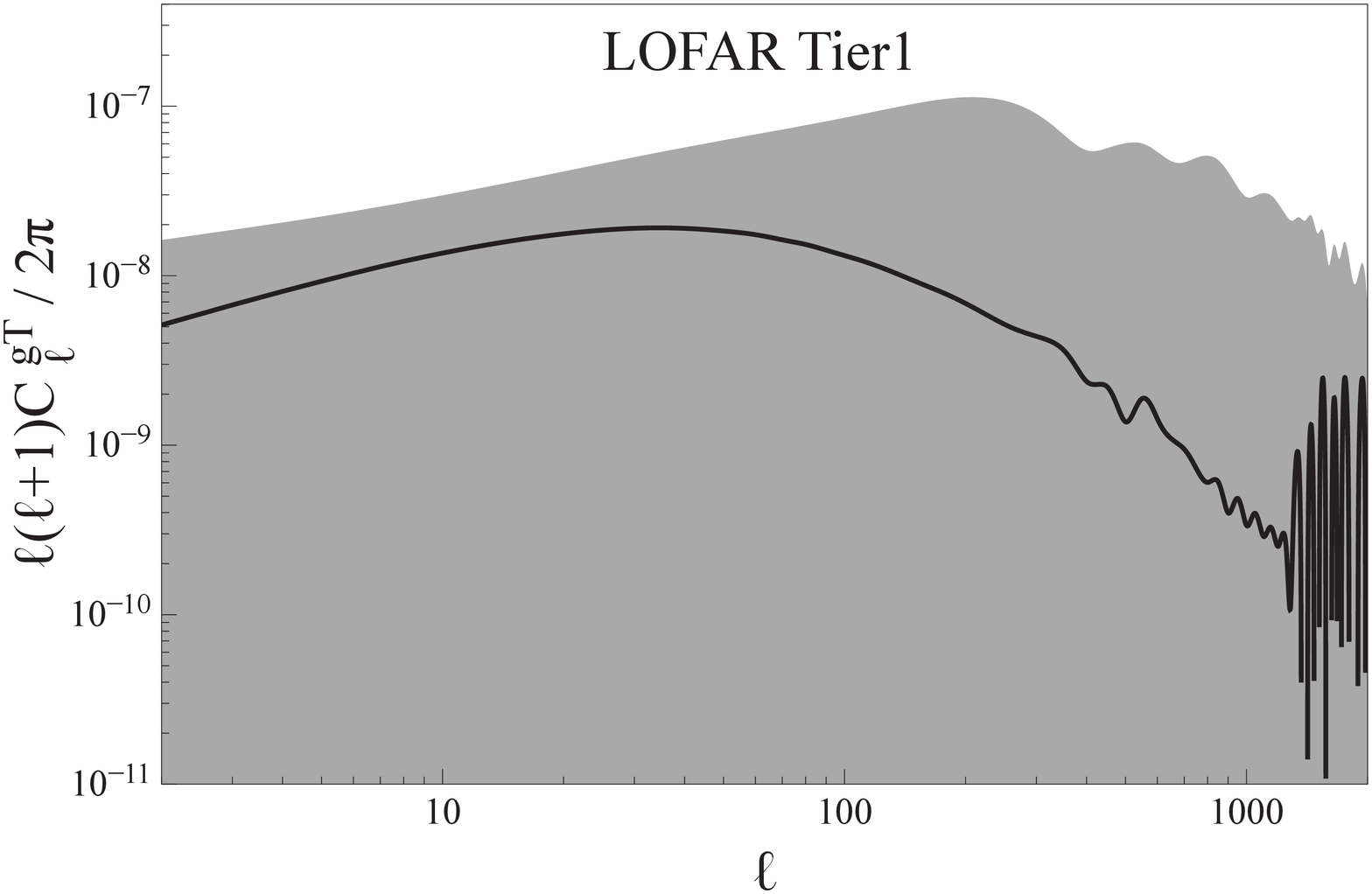, width=0.49 \linewidth}
\epsfig{file=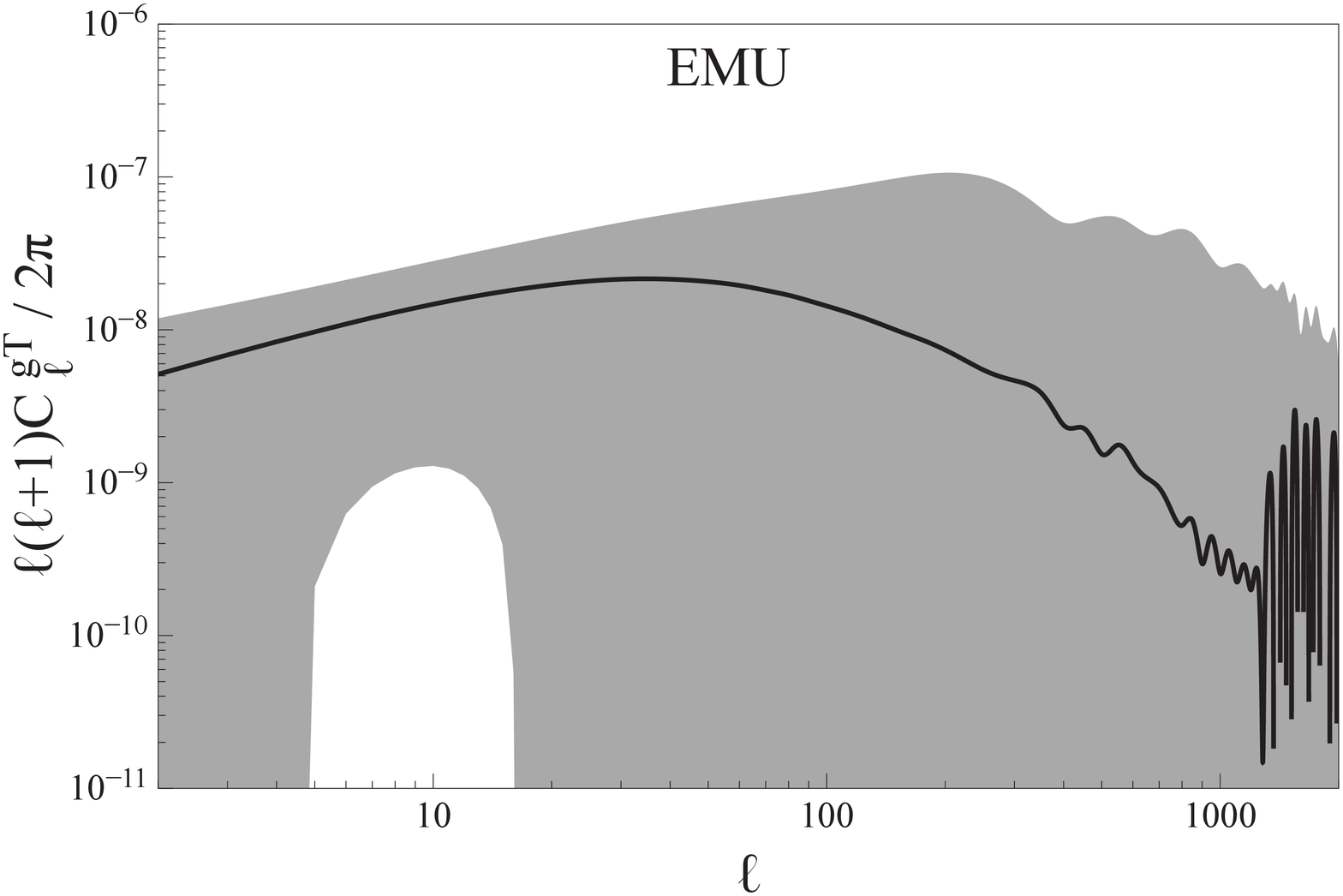, width=0.49 \linewidth}
\epsfig{file=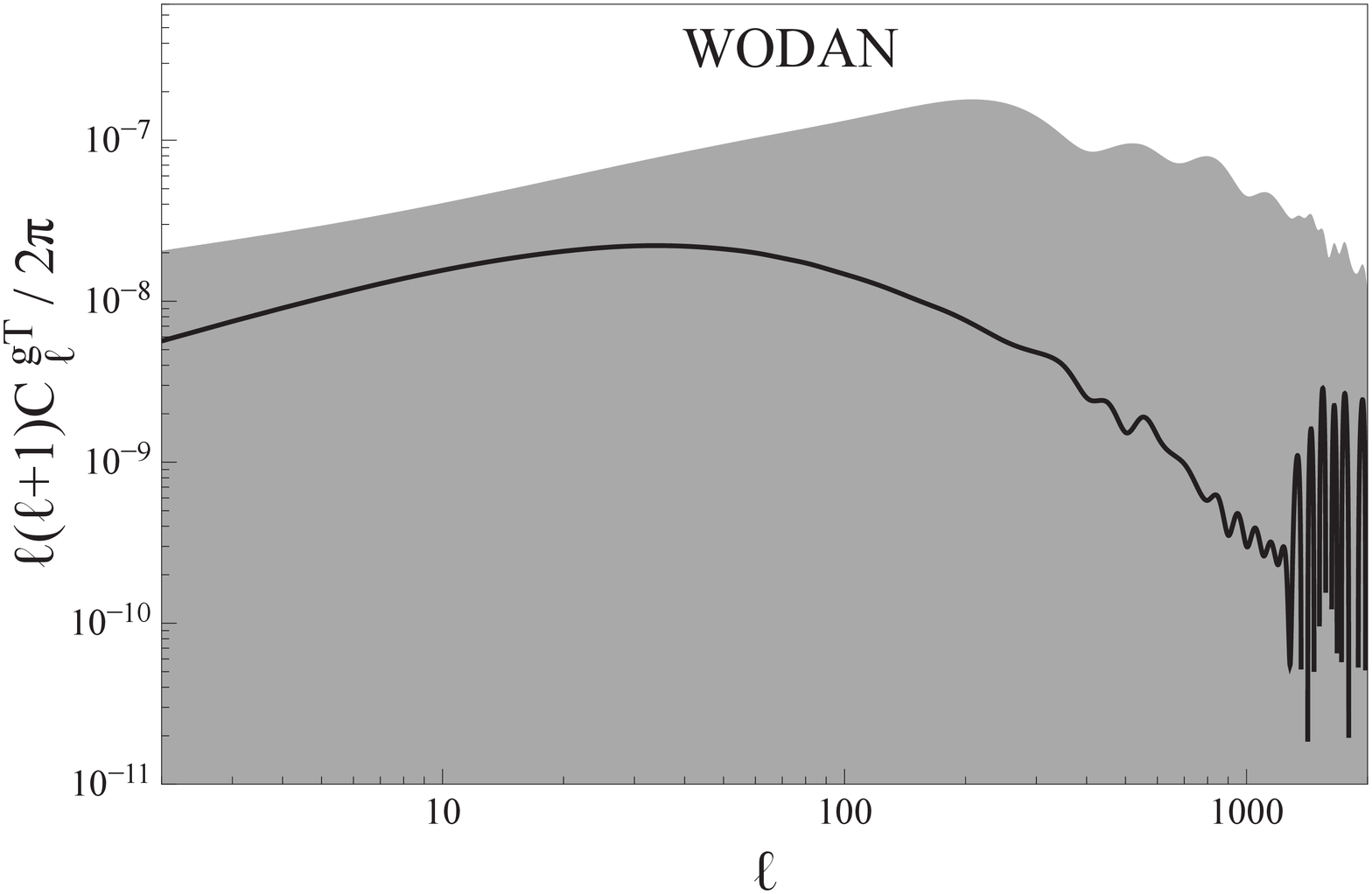, width=0.49 \linewidth}
\caption{Cross-correlations of radio sources with the CMB (Eq.~\ref{eq:cthetagT}). Solid lines are the theoretical $\Lambda$CDM prediction, the shaded area corresponds to cosmic variance errors, as in Eq.~\ref{eq:err-clgt}.}
\label{fig:ClgT}
\end{center}
\end{figure*}

In Figure~\ref{fig:ClgT} we show the predicted cross-correlations of the CMB with the combined radio source distributions.
Solid lines are standard $\Lambda$CDM+GR model, shaded regions are errors, calculated via (see e.g. \citealt{cabre07}):
\begin{equation}
\label{eq:err-clgt}
\sigma_{C_{\ell}^{gT}} = \hksqrt{\frac{\left(C_{\ell}^{gT}\right)^2 + C_{\ell}^{gg}C_{\ell}^{TT}}{(2\ell+1)f_{\rm sky}}},
\end{equation}
where $f_{\rm sky}$ is the sky coverage of the survey. 
The shot noise should be negligible for the cross-correlation of the CMB with LOFAR Tier1, EMU and WODAN on the scales of interest, given the high number density per $\ell$ mode. 

\begin{figure}
\begin{center}
\epsfig{file=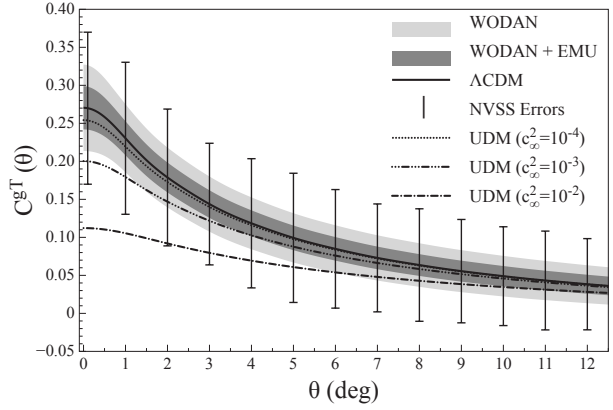, width=0.98\linewidth}
\caption{Cross-correlation of WODAN sources with the CMB. Black solid line is the $\Lambda$CDM prediction, black dashed line is the UDM prediction for $c_{\infty}^2=10^{-2}$ (see text for details); shaded regions are errors, light grey for the WODAN survey, dark grey for the EMU+WODAN combination; error bars are NVSS errors.}
\label{fig:CCF-comp-udm}
\end{center}
\end{figure}

% UDM Models testing
Fig.~\ref{fig:CCF-comp-udm} shows the predicted cross-correlation function, with cosmic variance errors, for the WODAN survey (light grey area), the WODAN+EMU combination (dark grey area) and the measured NVSS errors (error bars) as a comparison. Note the substantial improvement in ISW measurements provided by the all-sky survey, compared to  NVSS or WODAN alone.

\begin{figure}
\begin{center}
\epsfig{file=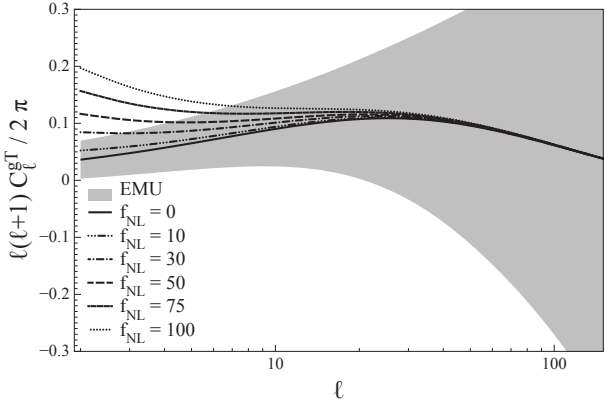, width=0.98\linewidth}
\caption{Cross-power spectrum of EMU radio sources with the CMB (Eq.~\ref{eq:ClgT}) for different values of the non-Gaussianity parameter $f_{\rm NL}$; shaded regions are errors for the EMU survey as in Eq.~\ref{eq:err-clgt}.}
\label{fig:ClgT-NG}
\end{center}
\end{figure}

The enhanced clustering due to non-Gaussianity would also modify the cross-correlation of galaxies with the CMB \citep{xia10b}, through the modified bias of Eq.~(\ref{eq:ng-bias}) in the galaxy window function (Eq.~\ref{eq:flg}). 
The effect is more significant on the largest scales, as shown in Fig.~\ref{fig:ClgT-NG}, which presents the cross-correlation of the CMB with EMU radio sources for different values of $f_{\rm NL}$. As in the auto-correlation case, we performed a $\chi^2$ analysis to predict what level of non-Gaussianity we should be able to detect; we used again the simulations of EMU data as its ISW detection should be the best of the surveys analysed (see Fig.~\ref{fig:ClgT}), and we found that these data would detect a $f_{\rm NL}$ of 11 at 1-$\sigma$ level.

As an initial example of cosmological constraints, in Fig.~\ref{fig:CCF-comp-udm},  the black dashed line is the predicted cross-correlation function for Unified Dark Matter scalar field cosmologies, where dark matter and dark energy are part of a single component. The key parameter in this model is the speed of sound (today) of the dark component, $c_{\infty}^2$, that has to be different from zero but small enough to let the dark component cluster (see \citealt{bertacca08, bertacca11} for details). Detecting a non-zero speed of sound would be an indication of a non-$\Lambda$CDM universe.

Using NVSS we are able to see differences from the $\Lambda$CDM case from $c_{\infty}^2=10^{-2}$, while the plot shows that
using the combined full sky EMU+WODAN will allow us to constrain values of the sound of speed of $c_{\infty}^2=10^{-4}$, using the ISW effect. Further details and forecasts on how well we will be able to test UDM cosmological models with these surveys will be part of a subsequent paper.

\begin{figure}
\begin{center}
\epsfig{file=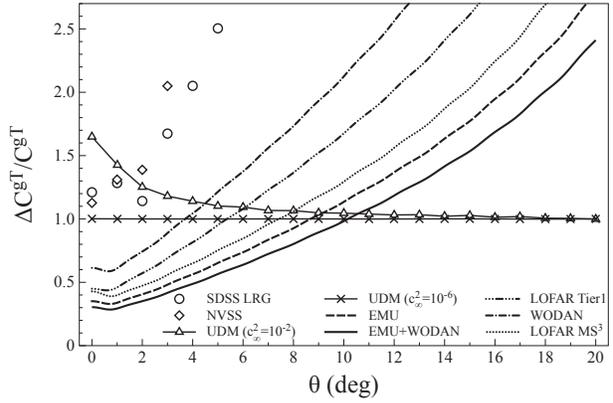, width=0.98\linewidth}
\caption{
Constraining power of cross-correlation CMB-radio sources for the different surveys; lines are radio surveys used in this paper, symbols are measurements from NVSS and SDSS, lines connecting symbols are thresholds to detect UDM models with a non-zero speed of sound (see text for details).}
\label{fig:DELTA-CCF-comp-udm}
\end{center}
\end{figure}

To predict the significance and constraining power of ISW measurements with the forthcoming radio surveys, in Fig. \ref{fig:DELTA-CCF-comp-udm} we define $\Delta C^{gT}/C^{gT}$ as the width of the entire 1-$\sigma$ constraint, i.e.:
\begin{equation}
\label{eq:deltaover}
\frac{\Delta C^{gT}}{C^{gT}} = \frac{[C^{gT}(\theta)+\sigma_{C^{gT}}]-[C^{gT}(\theta)-\sigma_{C^{gT}}]}{C^{gT}(\theta)} = 
\frac{2\sigma_{C^{gT}}}{C^{gT}},
\end{equation}
where $\sigma_{C^{gT}}$ is the error on the cross-correlation function in real space.
In Fig. \ref{fig:DELTA-CCF-comp-udm} we show $\Delta C^{gT}/C^{gT}$ (Eq.~\ref{eq:deltaover}) for the total surveys. 
We compare these with current measurements of $\Delta C^{gT}/C^{gT}$ from cross-correlations of NVSS and SDSS LRGs with WMAP maps, along with the threshold to actually distinguish between $\Lambda$CDM and other cosmological models.

As one can see, for small values of $\theta$ the constraining power is maximum, and all the surveys considered should have 
an increased discriminatory power;
in the case of whole-sky combined surveys, we obtain
that $\Delta C^{gT}/C^{gT}$ is less than half of that of NVSS and SDSS, for small $\theta$.

%%%%%%%%%%%%%%%%%%%%%%%%%%%%%%%%%%%%%%%%%%%%%%%%%%%%%%%%%%%%%%%%%%
%%%%%%%%%%%%%%%%%%%%%%%%%%%%%%%%%%%%%%%%%%%%%%%%%%%%%%%%%%%%%%%%%%
%%%%%%%%%%%%%%%%%%%%%%%%		Cosmic Magnification tests		%%%%%%%%%%%%%%%%%%%%%%
%%%%%%%%%%%%%%%%%%%%%%%%%%%%%%%%%%%%%%%%%%%%%%%%%%%%%%%%%%%%%%%%%%
%%%%%%%%%%%%%%%%%%%%%%%%%%%%%%%%%%%%%%%%%%%%%%%%%%%%%%%%%%%%%%%%%%

\subsection{Magnification Bias}
\label{sec:cosmag-pred}
We compute the power spectra for the magnification bias using equation (\ref{eq:Clgmu}); we consider the experiment where background radio sources from LOFAR, EMU and WODAN are cross-correlated with foreground galaxies from SDSS for the northern hemisphere and DES for the southern one. We note that we expect to have much better and wider data than the SDSS on the timescale of the radio surveys considered with the Pan-STARRs $3\pi$ sr survey \citep{kaiser10}, 
which when complete will provide imaging data to a depth between SDSS and DES so our analysis should be considered as conservative.
In the northern sky we use SDSS galaxies up to $z=0.35$ and radio sources as the background for higher redshift, while for the southern sky we use DES foregrounds up to $z=1$ and EMU sources as a background. To avoid the overlap between foreground and background galaxies, we remove the LOFAR, EMU and WODAN galaxies at $z<1$, i.e. we assume that via cross-matching between optical and radio bands, low z radio sources can be removed from our sample.  We assume that the bias for the foreground galaxies is unity since they are located at low redshifts. We follow \cite{scranton} to measure the weighted average power law slope $<\alpha-1>$, where $\alpha$ is given by Eq.~\ref{eq:alpha};
from our simulations, we obtain $-0.219$, $-0.147$, $0.1027$ and $0.121$, for LOFAR MS$^3$, LOFAR Tier 1, EMU and WODAN respectively.

\begin{figure*}
\begin{center}
\epsfig{file=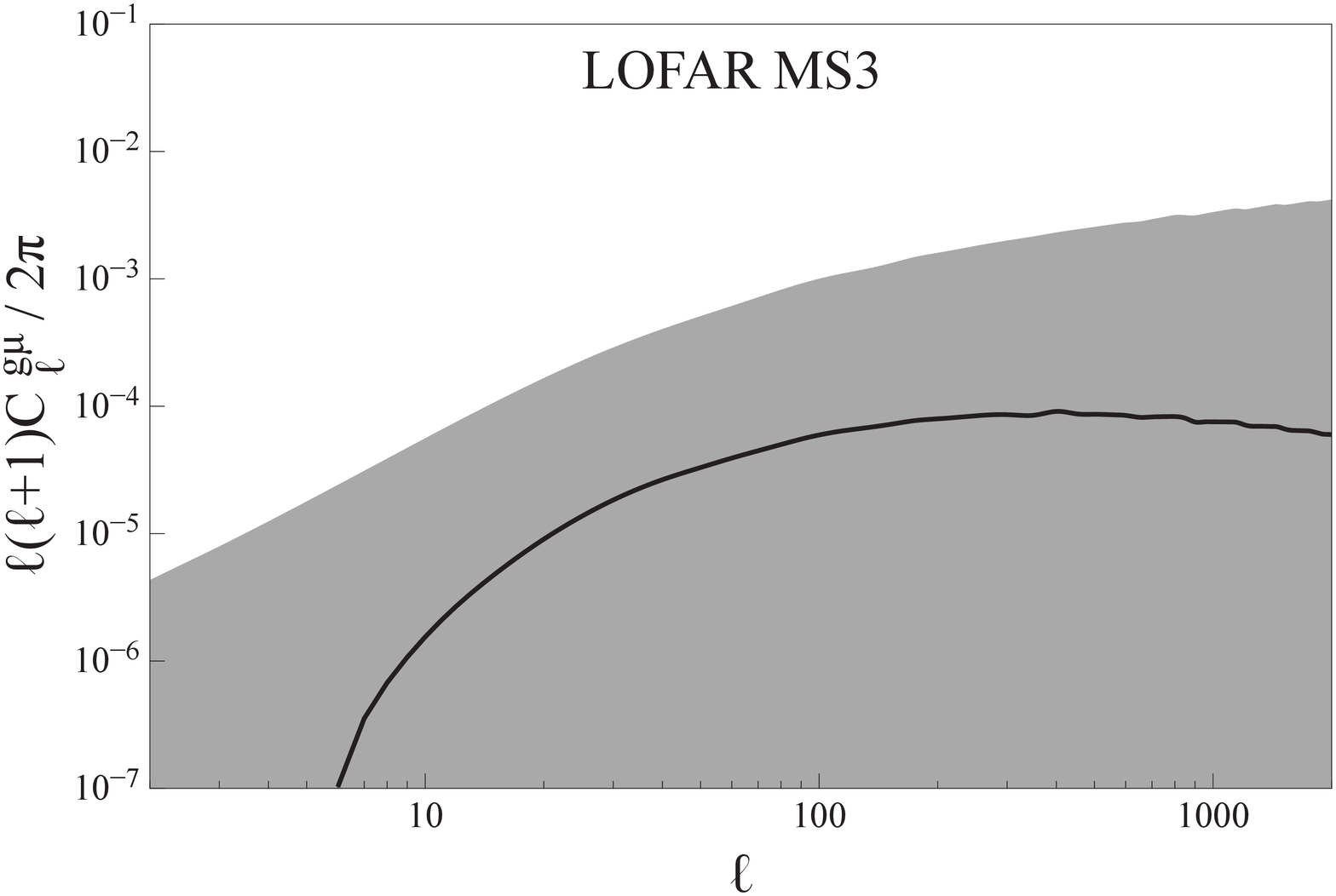, width=0.49\linewidth}
\epsfig{file=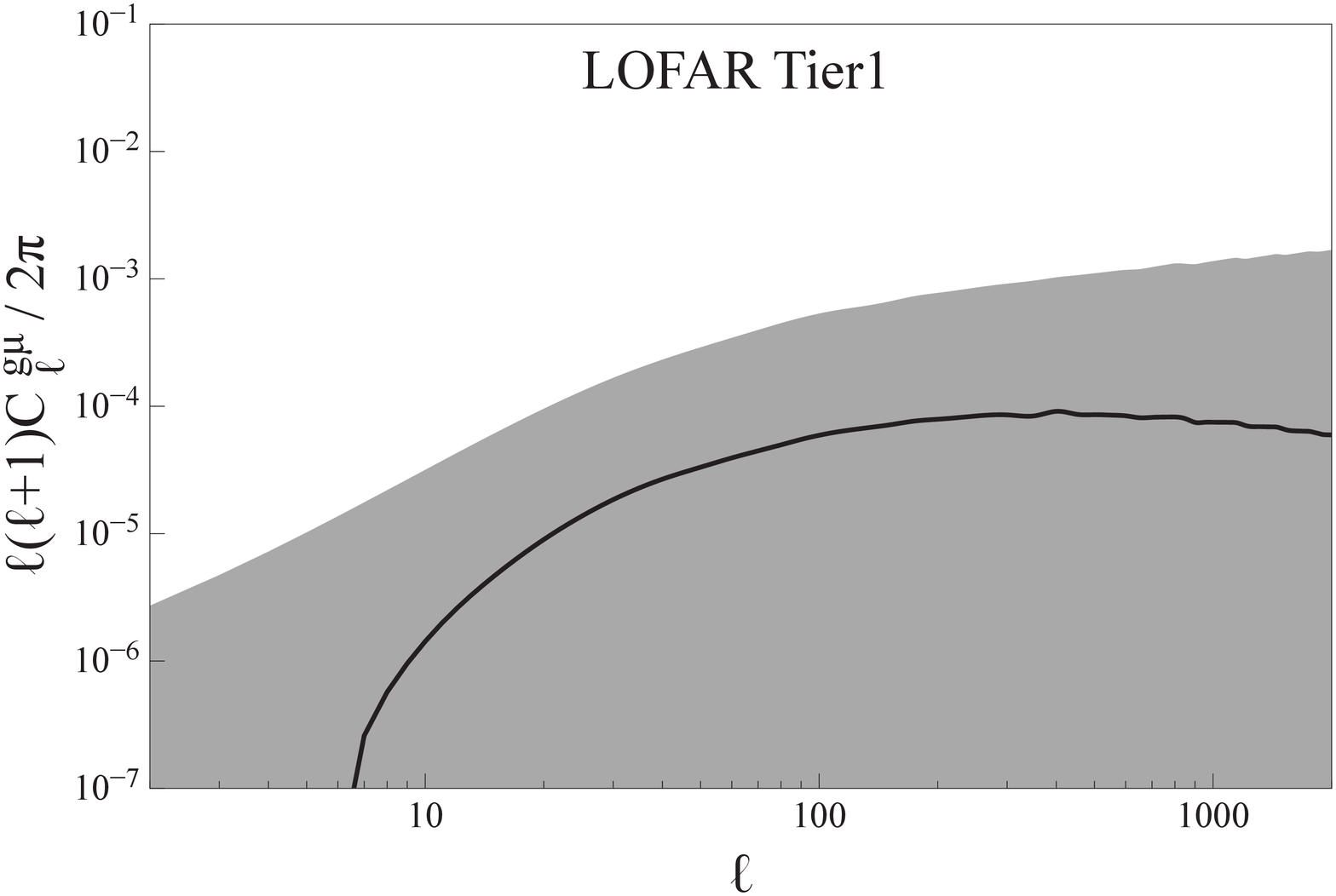, width=0.49\linewidth}
\epsfig{file=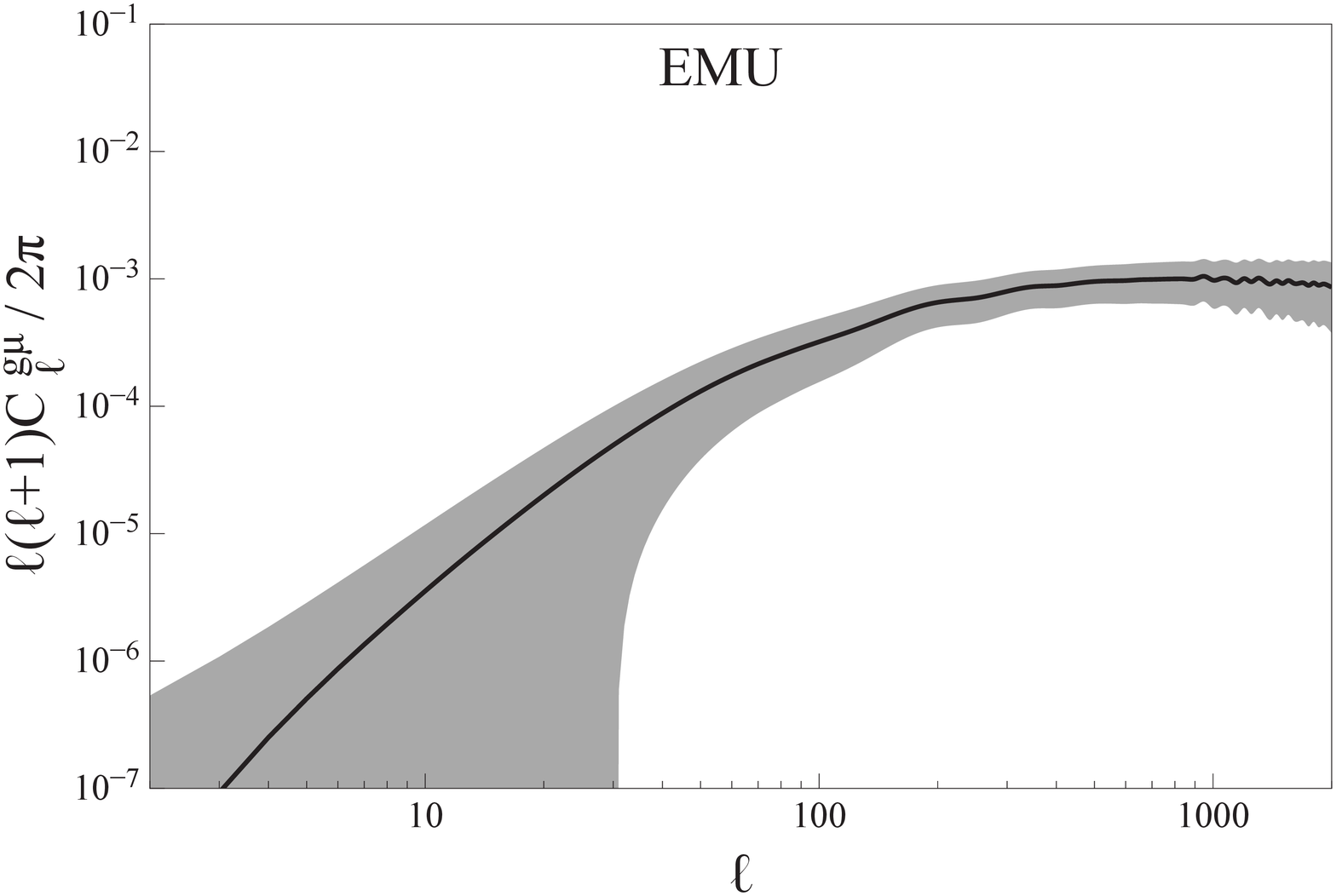, width=0.49\linewidth}
\epsfig{file=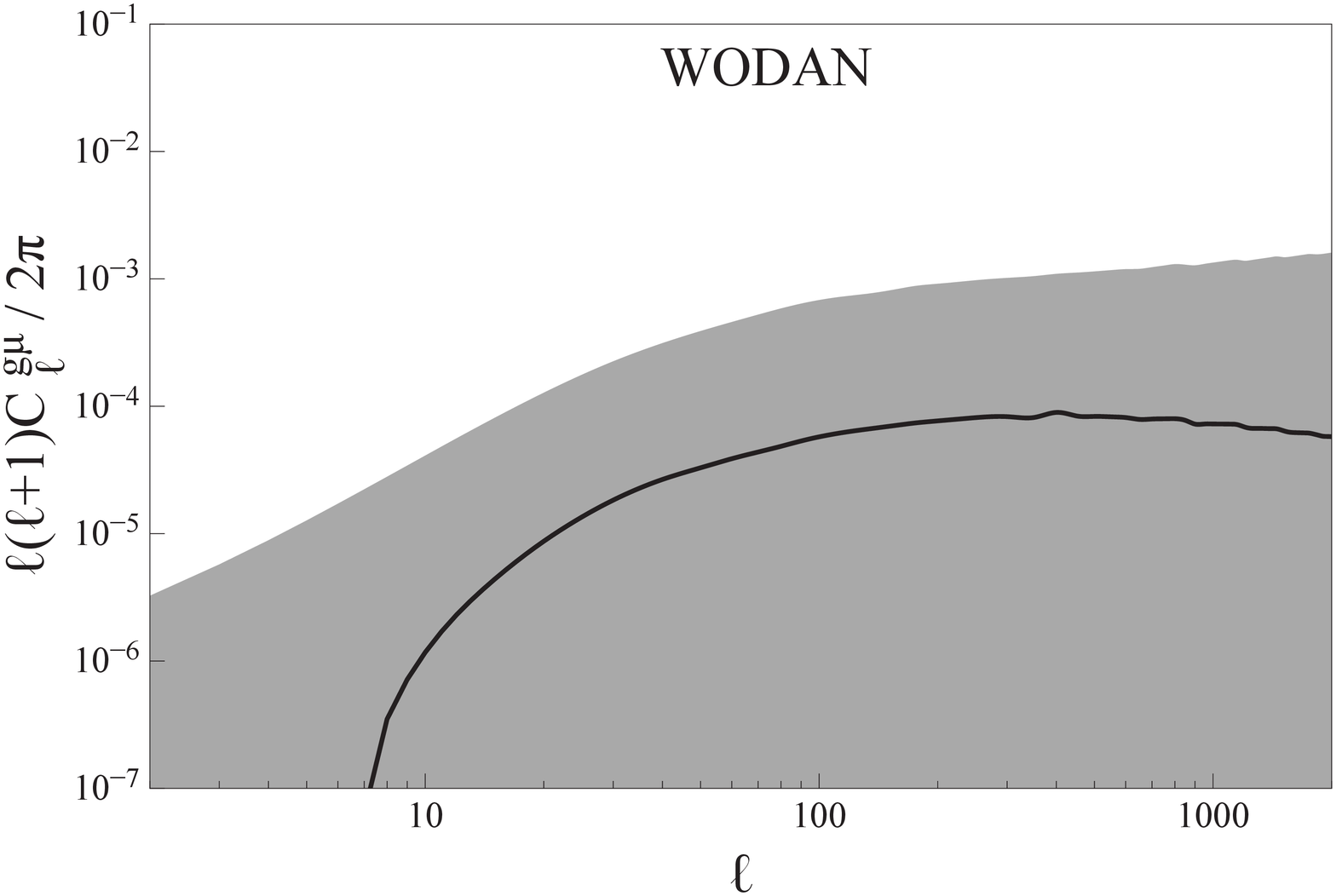, width=0.49\linewidth}
\caption{Magnification power spectra for LOFAR-SDSS, EMU-DES and WODAN-SDSS; black solid lines are theoretical predictions (Eq.~\ref{eq:Clgmu}), shaded regions are errors, as in Eq.~(\ref{eq:err-clgmu}).}
\label{fig:Clgmu}
\end{center}
\end{figure*}

In Fig.~\ref{fig:Clgmu} we show the cross-correlation of radio background with optical foreground sources power spectra, computing the errors according to \citet{zhang06}:
\begin{equation}
\label{eq:err-clgmu}
\sigma_{C_{\ell}^{g\mu}}=\hksqrt{\frac{C_{\mu{g}}^2+(C_g^b+C_{\rm shot}^b)(C_g^f+C_{\rm shot}^f)}{(2\ell+1) f_{\rm sky}}} ,
\end{equation}
where $f$ and $b$ denote the foreground and background sources, respectively, and ``shot" stands for the shot noise.

We can see that using this probe we have better constraining power when we have higher number density and magnification index. EMU will provide moderate constraints for this probe in combination with DES; we emphasise that a limiting factor for the northern surveys (LOFAR and WODAN) in our analysis is actually the optical data. We would therefore 
be able to tighten constraints for these surveys if Pan-STARRS data were used as a foreground instead of SDSS.

%%%%%%%%%%%%%%%%%%%%%%%%%%%%%%%%%%%%%%%%%%%%%%%%%%%%%%%%%%%%%%%%%
%%%%%%%%%%%%%%%%%%%%%%%%%%%%%%%%%%%%%%%%%%%%%%%%%%%%%%%%%%%%%%%%%
%%%%%%%%%%%%%%%%%%%%%%%%%	Cosmological Constraints	%%%%%%%%%%%%%%%%%%%%%%%%%
%%%%%%%%%%%%%%%%%%%%%%%%%%%%%%%%%%%%%%%%%%%%%%%%%%%%%%%%%%%%%%%%%
%%%%%%%%%%%%%%%%%%%%%%%%%%%%%%%%%%%%%%%%%%%%%%%%%%%%%%%%%%%%%%%%%

%\subsection{Cosmological Constraints}

\section{Cosmological model constraints}
\label{sec:cosmodel}
Having calculated the constraints on each cosmological probe, it is now possible to determine how the LOFAR, EMU and WODAN surveys can improve measurements of cosmological parameters.

Starting from Einstein's field equations:
\begin{equation}
\label{eq:einstein}
G_{\mu\nu}=\frac{8\pi G}{c^4} T_{\mu\nu},
\end{equation}
where $G_{\mu\nu}$ is the Einstein tensor, $T_{\mu\nu}$ is the energy-momentum tensor and $G$ is Newton's gravitational constant,
we study the improvements these surveys will bring to the measurement of parameters in the dynamical dark energy (DE) and modified gravity (MG) scenarios; 
we investigate these issues using Fisher Matrix techniques, following \citet{MGCAMB}.

%%%%%%%%%%%%%%%%%%%%%%%%%%%%%%%%%%%%%%%%%%%%%%%%%%%%%%%%%%%%%%%%%%
%%%%%%%%%%%%%%%%%%%%%%%%%%%%%%%%%%%%%%%%%%%%%%%%%%%%%%%%%%%%%%%%%%
%%%%%%%%%%%%%%%%%%%%%%%%%%%%	Dark Energy	%%%%%%%%%%%%%%%%%%%%%%%%%%%%
%%%%%%%%%%%%%%%%%%%%%%%%%%%%%%%%%%%%%%%%%%%%%%%%%%%%%%%%%%%%%%%%%%
%%%%%%%%%%%%%%%%%%%%%%%%%%%%%%%%%%%%%%%%%%%%%%%%%%%%%%%%%%%%%%%%%%

\subsection{Dynamical Dark Energy}
\label{sec:de}
In the cosmological framework of General Relativity, it has been necessary to modify Eq.~\ref{eq:einstein} to account for the observed acceleration of the expansion of the Universe;
the simplest modification is the introduction of a cosmological constant, as first suggested by \cite{zeldovich67}, that can be interpreted as vacuum energy; in this case one modifies the right hand side of Eq.~\ref{eq:einstein}:
\begin{equation}
\label{eq:vacuum}
G_{\mu\nu} = T_{\mu\nu} + T^{\rm vac}_{\mu\nu},
\end{equation}
where throughout this section we set $G=1/8\pi$ and $c=1$ for simplicity, and:
\begin{equation}
\label{eq:tvac}
T^{\rm vac}_{\mu\nu} = - \Lambda g_{\mu\nu}.
\end{equation}
The first step toward understanding the nature of dark energy is to clarify whether it is a simple cosmological constant or it originates from other sources that dynamically change in time. 
The dynamical models can be distinguished from the cosmological constant by considering the evolution of the equation of state of dark energy:
\begin{equation}
\label{eq:wde}
w = \frac{p}{\varrho},
\end{equation}
where $p$ and $\varrho$ are the pressure density and energy density of the fluid, respectively. In the cosmological constant model, $w=-1$, while for dynamical models $w=w(a)$.

To evaluate the potential of the considered radio surveys to constrain the dynamics of different models of dark energy, we adopt the following parametrisation for the DE equation-of-state (EoS) $w$ \citep{linder03}:
\begin{align}
w(a) = w_0 + w_a (1-a).
\end{align}
We use the best fit model $\{w_0,w_a\} = \{-0.89,-0.24\}$ from current data (see \citet{zhao10a}
for details) as the fiducial model, which is consistent with the prediction of the quintom model \citep{feng05}, 
and consistently include the dark energy perturbations in the calculation using the prescription proposed
in \citet{zhao05}.

%%%%%%%%%%%%%%%%%%%%%%%%%%%%%%%%%%%%%%%%%%%%%%%%%%%%%%%%%%%%%%%%%%
%%%%%%%%%%%%%%%%%%%%%%%%%%%%%%%%%%%%%%%%%%%%%%%%%%%%%%%%%%%%%%%%%%
%%%%%%%%%%%%%%%%%%%%%%%%%		Dark Gravity		%%%%%%%%%%%%%%%%%%%%%%%%%
%%%%%%%%%%%%%%%%%%%%%%%%%%%%%%%%%%%%%%%%%%%%%%%%%%%%%%%%%%%%%%%%%%
%%%%%%%%%%%%%%%%%%%%%%%%%%%%%%%%%%%%%%%%%%%%%%%%%%%%%%%%%%%%%%%%%%

\subsection{Modified Gravity}
\label{sec:dg}
An intriguing alternative to dark energy for the explanation of the accelerated expansion of the universe is the ``modified gravity" approach \citep{durrer08}, which states that gravity needs to be modified, i.e. weakened on large-scales; 
an attractive feature of modified gravity models is that one can alter the Einstein-Hilbert action so that accelerated solutions of the background of the Universe can be obtained without the need for a dark energy component.

In this case we modify the geometric side of Eq.~\ref{eq:einstein}:
\begin{equation}
\label{eq:darkgrav}
G_{\mu\nu} + G^{\rm dark}_{\mu\nu} = T_{\mu\nu},
\end{equation}

Modified gravity models can mimic the $\Lambda$CDM model in the sense that they include the background expansion, but in general they predict different dynamics for the growth of cosmic structures. Radio source number counts and ISW measurements directly probe structure formation, therefore they can constrain modified gravity scenarios. 

Here we follow \citet{zhao10b} and consider scalar metric perturbations around a FRW background for which the line element in the conformal Newtonian gauge is:
\be
\label{metric}
ds^2=-a^2(\tau)\l[\l(1+2\Psi\r)d\tau^2-\l(1-2\Phi\r) d\vec{x}^2\r] \ ,
\ee
where $\Phi$ and $\Psi$ are functions of time and space. 

We use the following parametrisation to describe the relations specifying how the metric perturbations relate to each other, and
how they are sourced by the perturbations of the energy-momentum
tensor:
\begin{align}
\label{gamma}
\frac{\Phi}{\Psi}&=\eta(a,k), \\
\label{parametrization-Poisson}
\Psi&=\frac{-4\pi G a^2 \mu(a,k) \varrho\Delta}{k^2} \ ,
\end{align}
where $\Delta$ is the gauge-invariant comoving density contrast defined as:
\begin{equation}
\Delta \equiv \delta + 3\f{aH}{k} v\, ;
\label{Def:Delta}
\end{equation}
$\eta(a,k)$ and $\mu(a,k)$ are two time- and scale-dependent functions encoding the modifications of gravity that can be written as:
\begin{equation}
\label{eq:etaak}
\eta(a,k) = \frac{1+\beta_1\lambda^2_1 k^2 a^s}{1+\lambda^2_1 k^2 a^s} \\
\end{equation}
\begin{equation}
\label{eq:muak}
\mu(a,k) = \frac{1+\beta_2\lambda^2_2 k^2 a^s}{1+\lambda^2_2 k^2 a^s} ,
\end{equation}
where $\lambda_i^2$ and $\beta_i$ are parameters and $a^s$ gives the time dependence of the deviation from GR; $\eta(a,k)=\mu(a,k)=1$ in GR, while in a modified gravity model $\mu$ and $\eta$ can
in general be functions of both time and scale (\citealt{bertschinger08, MGCAMB, linder11}).

Since we are interested in testing GR at late times, we will consider a simple approximation to Equation~\ref{eq:etaak} and~\ref{eq:muak} where
we assume $\mu(a,k)=\eta(a,k)=1$ at early times, with a transition to some other values at late times. This is natural in the existing models of modified gravity that aim to explain the late-time acceleration, where departures from GR occur at around the present day horizon scales. Also, the success in explaining the BBN and CMB physics relies on GR being valid at high redshifts.

To model the time evolution of $\mu$ and $\eta$ we use the hyperbolic
tangent function to describe the transition from unity to the constants
$\mu_0$ and $\eta_0$:
\begin{align}
\mu(z)=&\frac{1-\mu_0}{2}\Big(1+{\rm tanh}\frac{z-z_s}{\Delta{z}}\Big)+\mu_0~, \\
\eta(z)=&\frac{1-\eta_0}{2}\Big(1+{\rm tanh}\frac{z-z_s}{\Delta{z}}\Big)+\eta_0~. 
\end{align}
where $z_s$ denotes the threshold redshift where we start to modify gravity, and 
$\mu_0, \eta_0$ are free parameters; following \citet{zhao10b}, we fix the transition width $\Delta{z}$ to be $0.05$.

%%%%%%%%%%%%%%%%%%%%%%%%%%%%%%%%%%%%%%%%%%%%%%%%%%%%%%%%%%%%%%%%%%
%%%%%%%%%%%%%%%%%%%%%%%%%%%%%%%%%%%%%%%%%%%%%%%%%%%%%%%%%%%%%%%%%%
%%%%%%%%%%%%%%%%%%%%%%%%%		Fisher Matrix		%%%%%%%%%%%%%%%%%%%%%%%%%
%%%%%%%%%%%%%%%%%%%%%%%%%%%%%%%%%%%%%%%%%%%%%%%%%%%%%%%%%%%%%%%%%%
%%%%%%%%%%%%%%%%%%%%%%%%%%%%%%%%%%%%%%%%%%%%%%%%%%%%%%%%%%%%%%%%%%

\subsection{Observables and Fisher Matrices} 
\label{sec:fisher}
We use the observables including the LOFAR, EMU and WODAN radio source auto-correlation, $C_{\ell}^{gg}$, cross-correlation, $C_{\ell}^{gT}$, and magnification bias, $C_{\ell}^{g\mu}$, functions in a Fisher analysis.
To obtain the auto-correlation functions we consider radio source predicted distributions for LOFAR, EMU and WODAN;
for the ISW signal we cross-correlate the radio source distributions with the CMB,
while to obtain the magnification correlations we use SDSS DR7 and DES galaxy populations as foreground lenses for the northern and southern hemispheres, respectively.
The two-point functions we use can be generalised as:
\begin{equation}
C_\ell^{XY}= 4\pi \int \frac{dk}{k} \Delta^{2}(k) W_{\ell}^X(k) W_{\ell}^Y(k), 
\label{eq:cl-gen}
\end{equation} 
where $\Delta^{2}(k)$ is the power spectrum and $W_{\ell}^{X,Y}(k)$ denote angular window functions. Here $X,Y\in[T,g,{\mu}]$,
where $T,g$ and $\mu$ indicate the CMB temperature, radio source counts and magnification respectively.

Given the specifications of the proposed future surveys, the
Fisher matrix (\citealt{fisher35}, \citealt{tegmark97}) enables us to quickly estimate the
errors on the cosmological parameters around the fiducial values.
For Gaussian-distributed observables, such as
$C^{XY}_\ell$, the Fisher matrix is given by:
\begin{equation}
F_{\alpha\beta} =
f_{\rm sky} \sum_{\ell=\ell_{\rm min}}^{\ell_{\rm max}}\frac{2\ell +
1}{2} {\rm Tr}\left( \frac{\partial {\bf C_\ell}}{\partial p_\alpha}
{\bf \tilde{C}_\ell^{-1}}\frac{\partial {\bf C_\ell}}{\partial
p_\beta} {\bf \tilde{C}_\ell^{-1}} \right) \ , 
\label{eq:Fisher} 
\end{equation}
where $p_{\alpha(\beta)}$ is the $\alpha(\beta)$-th cosmological
parameter and ${\bf \tilde{C}_\ell}$ is the ``observed'' covariance
matrix with elements $\tilde{C}^{XY}_\ell$ that include
contributions from noise: 
\begin{equation}
\tilde{C}^{XY}_\ell=
C^{XY}_\ell+N^{XY}_\ell \ . 
\label{eq:NoiseAdd} 
\end{equation}
Eq.~(\ref{eq:Fisher}) assumes that all fields $X(\hat{\bf
n})$ are measured over contiguous regions covering a fraction
$f_{\rm sky}$ of the sky. The value of the lowest multipole can be
estimated from $\ell_{\rm min} \approx [\pi /(2f_{\rm sky})]$, where
the square brackets denote the rounded integer;
for the noise matrix $N^{XY}_\ell$ we use Eq.~(\ref{eq:err-clgg}, \ref{eq:err-clgt}, \ref{eq:err-clgmu}).

To perform the Fisher analysis, we first parametrize our cosmology using: 
\begin{equation}
\label{eq:paratriz} 
{\bf P} \equiv (\omega_{b}, \omega_{c},
\Theta_{s}, \tau, n_s, A_s, \aleph, \beth) ,
\end{equation}
where
$\omega_{b}\equiv\Omega_{b}h^{2}$ and
$\omega_{c}\equiv\Omega_{c}h^{2}$ are the physical baryon and cold
dark matter densities relative to the critical density respectively,
$\Theta_{s}$ is the ratio (multiplied by 100) of the sound
horizon to the angular diameter distance at decoupling, $\tau$
denotes the optical depth to re-ionization, $n_s$ and $A_s$ are the
primordial power spectrum index and amplitude, respectively, and $\aleph \in [w_0, \eta_0]$, $\beth \in [w_a, \mu_0]$ are the parameters we want to measure.
We assume a flat Universe and an effective dark energy equation of state $w=-1$ throughout the expansion history; we also combine the latest supernovae Ia luminosity distance from the UNION2 sample \citep{amanullah10} to tighten the constraints.  

Finally, given the uncertainties in the measurement of the bias and the redshift distribution for radio surveys, we marginalise over the amplitude of the product $b(z) \times N(z)$.
We note that the models we use are constrained by the total radio source counts and our current knowledge of the evolution of the sub-populations. The main uncertainties in these distributions is in the high-redsift ($z>1$) evolution of the FRI radio galaxies (see e.g. \citealt{clewley04}; \citealt{sadler07}), however rapid progress on pinning down the evolution of these source should be made over the next few years by combining deep multi-wavelength survey data with deep radio continuum data (e.g. \citealt{smolcic09}; \citealt{mcalpine11}).
The final results also depend on the shape of this product, which is not precisely known; however, we verified that modifications at the level of a few percent in the peak position, amplitude or width do not significantly affect our results. 
A complete analysis of the impact of this uncertainty on the measurement of cosmological parameters is beyond the scope of this paper and it is left for future work. 
Of course, a careful treatment of this issue will be required in the real data analyses, as we mention in Section~\ref{sec:uncert}.

We use {\tt MGCAMB} \citep{MGCAMB}\footnote{http://userweb.port.ac.uk/~zhaog/MGCAMB.html} to calculate the observables in modified gravity for LOFAR, EMU and WODAN, and use Eq.~(\ref{eq:Fisher}) to calculate the Fisher matrices using the preferred model of current data as a fiducial model; following \citet{MGCAMB} and \citet{zhao10b}, we assume as fiducial $\{w_0,w_a\} = \{-0.89,-0.24\}$ for the dynamical dark energy parameters, and $\{\eta_0,\mu_0\} = \{1.3,0.87\}$ for the modified gravity parameters.

%%%%%%%%%%%%%%%%%%%%%%%%%%%%%%%%%%%%%%%%%%%%%%%%%%%%%%%%%%%%%%%%%%
%%%%%%%%%%%%%%%%%%%%%%%%%%%%%%%%%%%%%%%%%%%%%%%%%%%%%%%%%%%%%%%%%%
%%%%%%%%%%%%%%%%%%%%%%%%%		Results		%%%%%%%%%%%%%%%%%%%%%%%%%%%%
%%%%%%%%%%%%%%%%%%%%%%%%%%%%%%%%%%%%%%%%%%%%%%%%%%%%%%%%%%%%%%%%%%
%%%%%%%%%%%%%%%%%%%%%%%%%%%%%%%%%%%%%%%%%%%%%%%%%%%%%%%%%%%%%%%%%%

\section{Results}
\label{sec:results}
\begin{figure*}
\begin{center}
\epsfig{file= 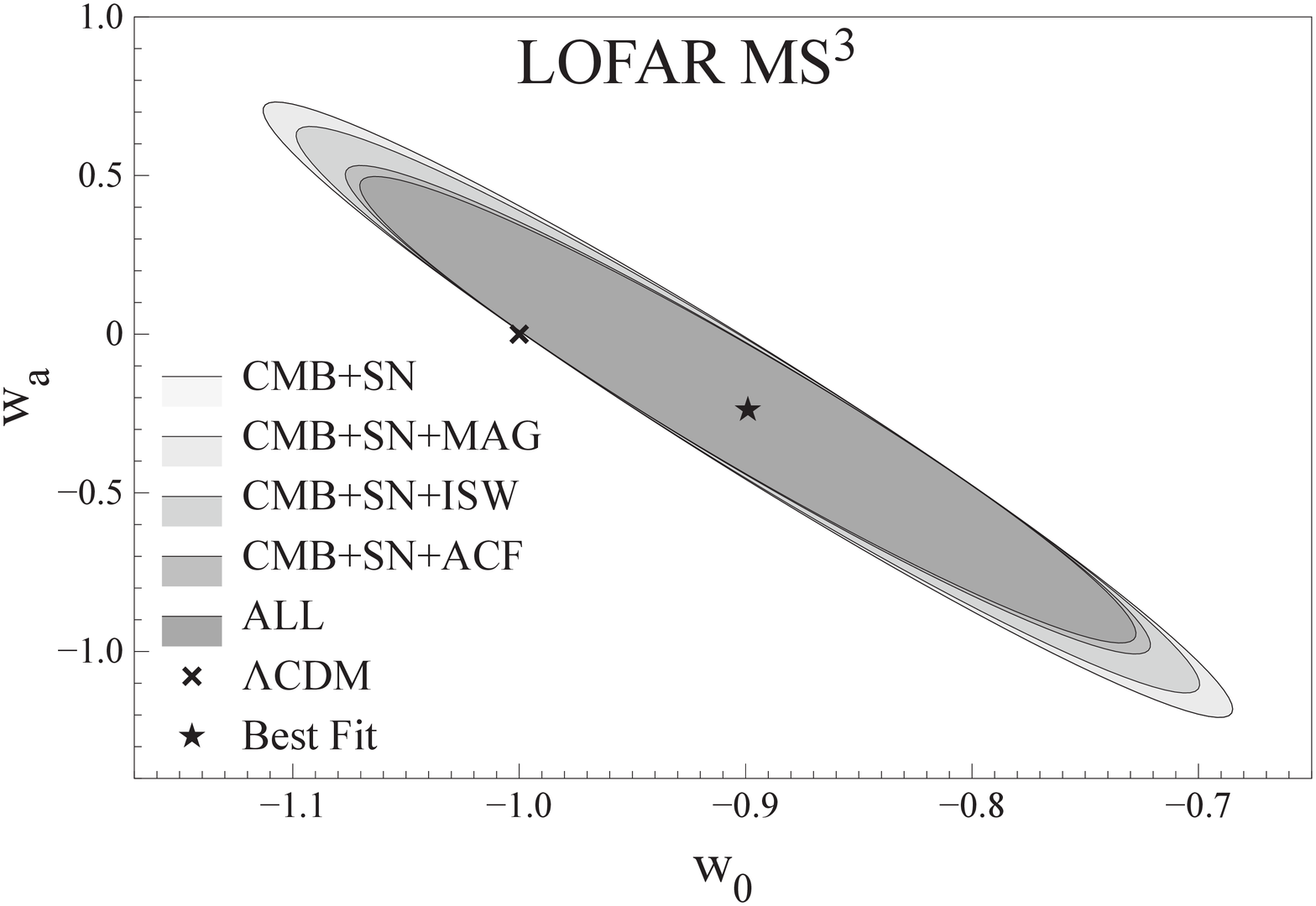, width=0.49\linewidth}
\epsfig{file= 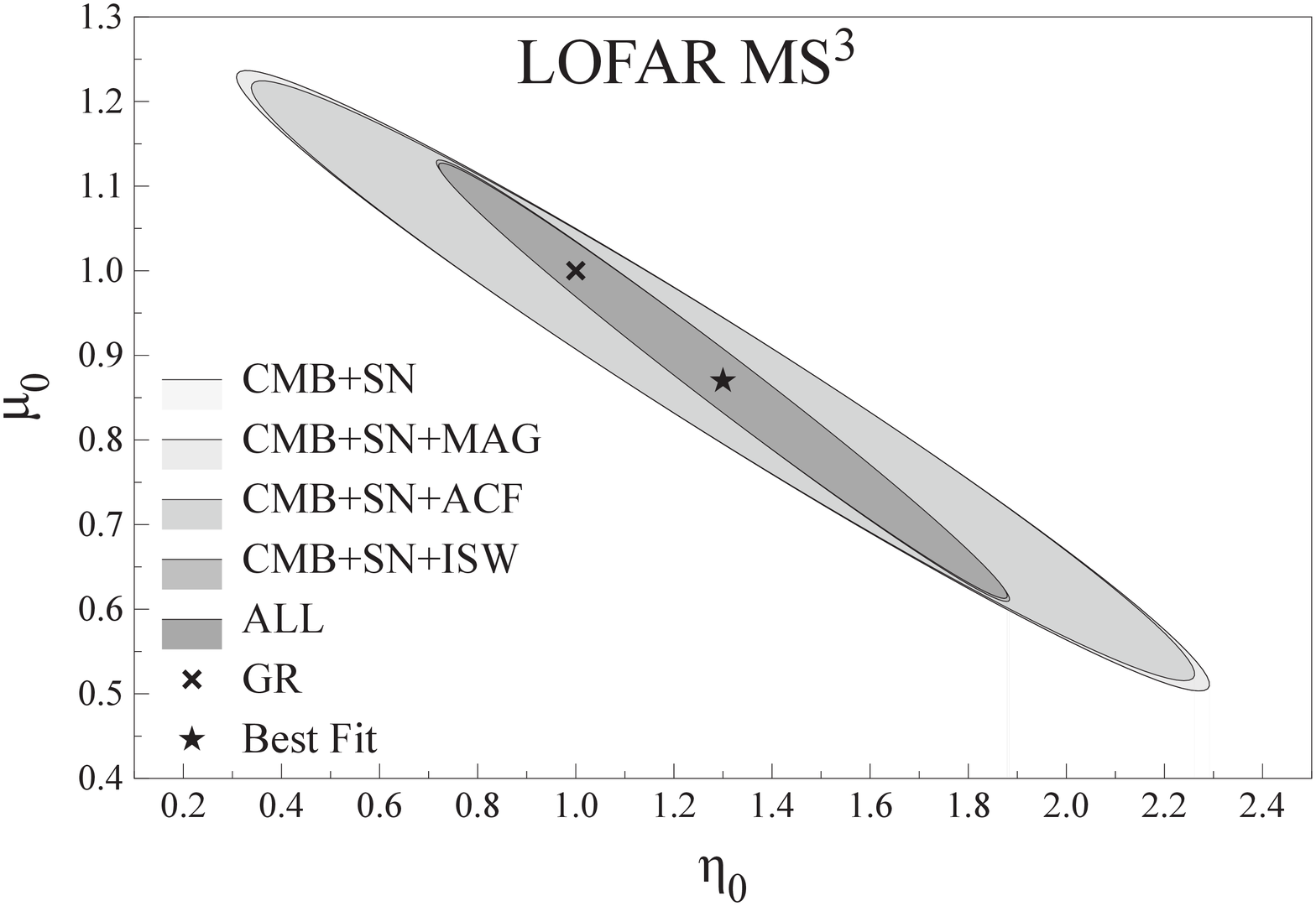, width=0.49\linewidth}
\caption{Forecast of constraints for dark energy (left) and modified gravity (right) parameters, for the LOFAR MS$^3$ survey. Ellipses show constraints for different combinations of probes (see text for details).}
\label{fig:forecast-lofms3}
\end{center}
\end{figure*}

\begin{figure*}
\begin{center}
\epsfig{file= 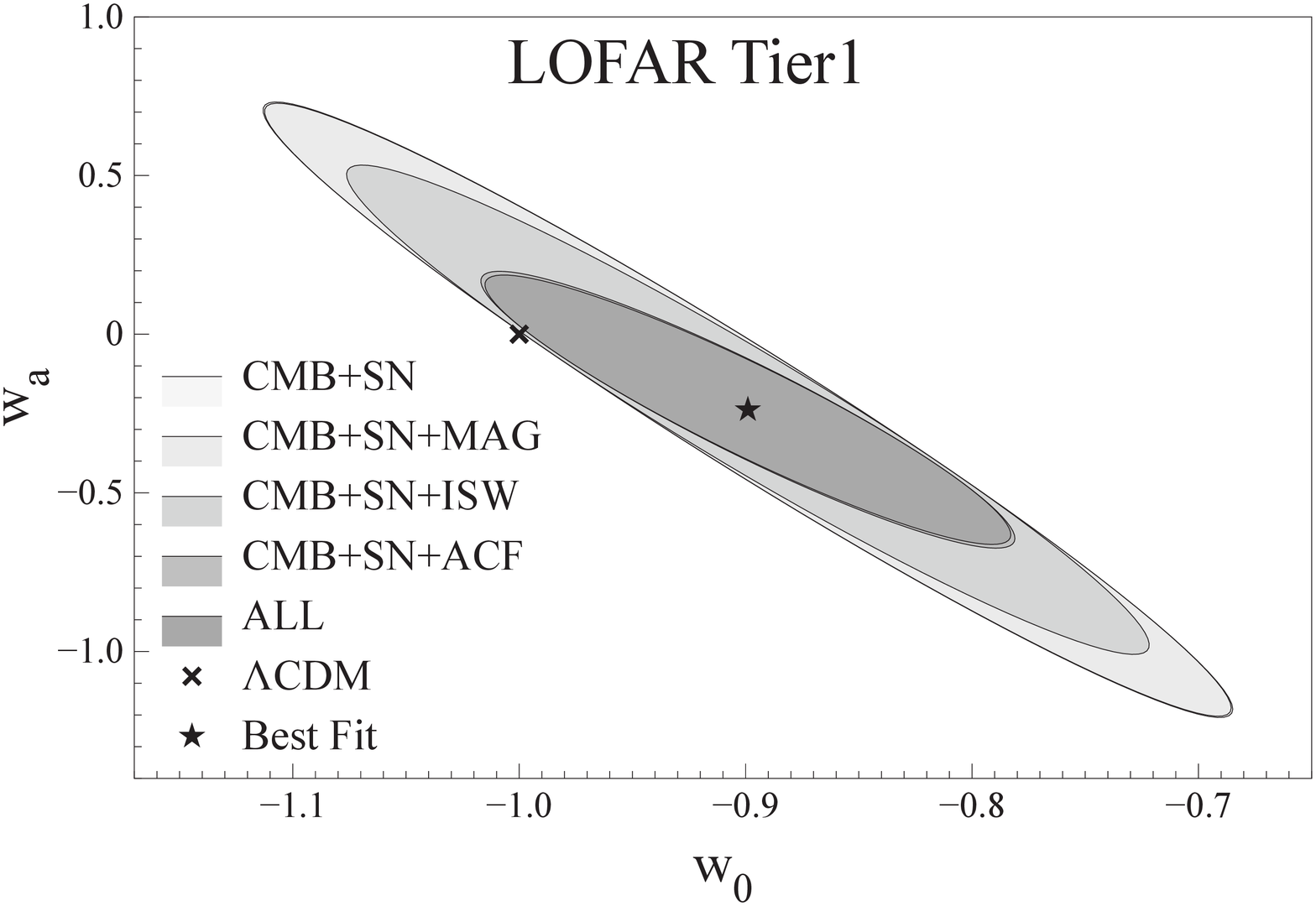, width=0.49\linewidth}
\epsfig{file= 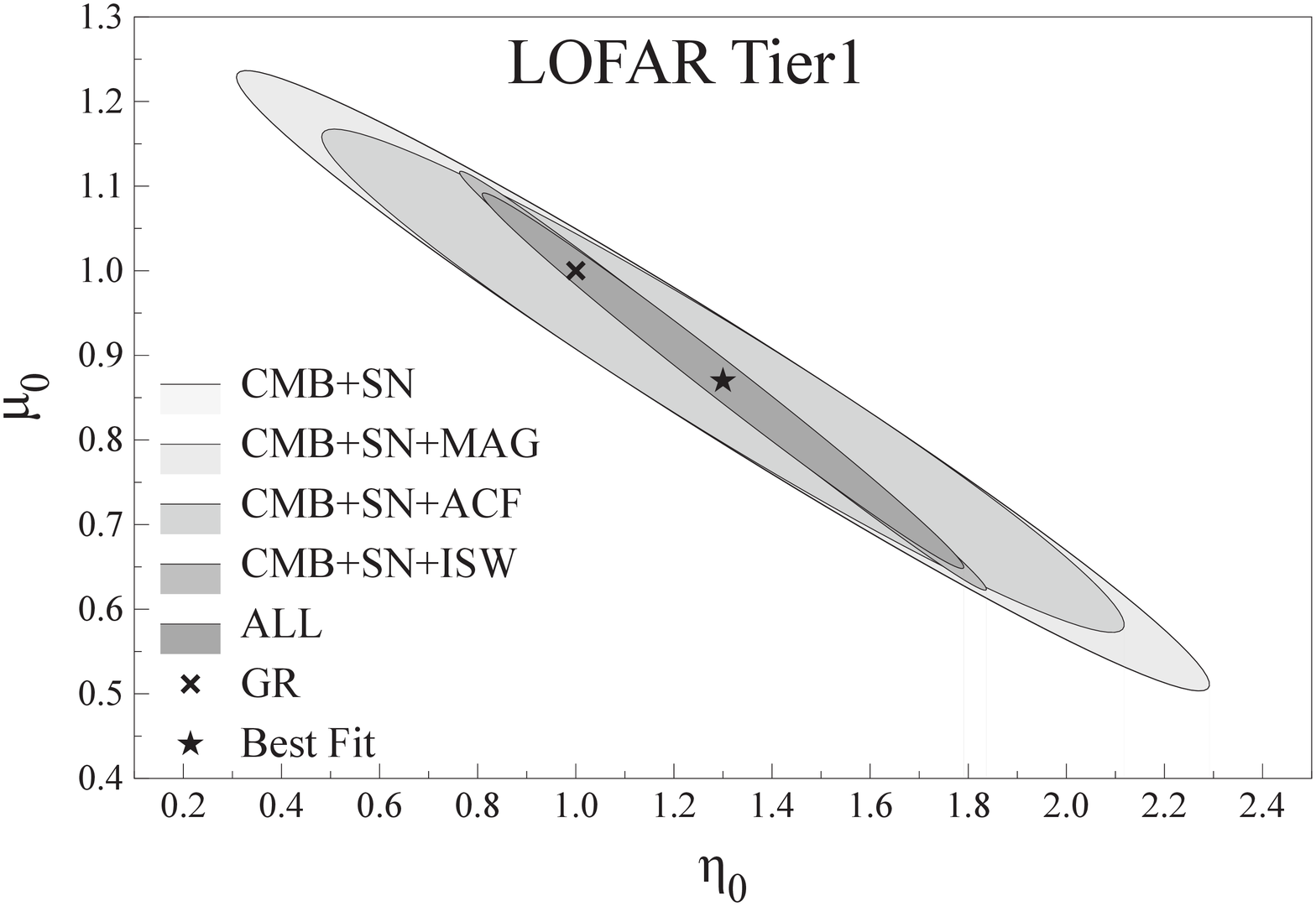, width=0.49\linewidth}
\caption{Forecast of constraints for dark energy (left) and modified gravity (right) parameters, for the LOFAR Tier 1 survey. Ellipses show constraints for different combinations of probes (see text for details).}
\label{fig:forecast-loft1}
\end{center}
\end{figure*}

\begin{figure*}
\begin{center}
\epsfig{file= 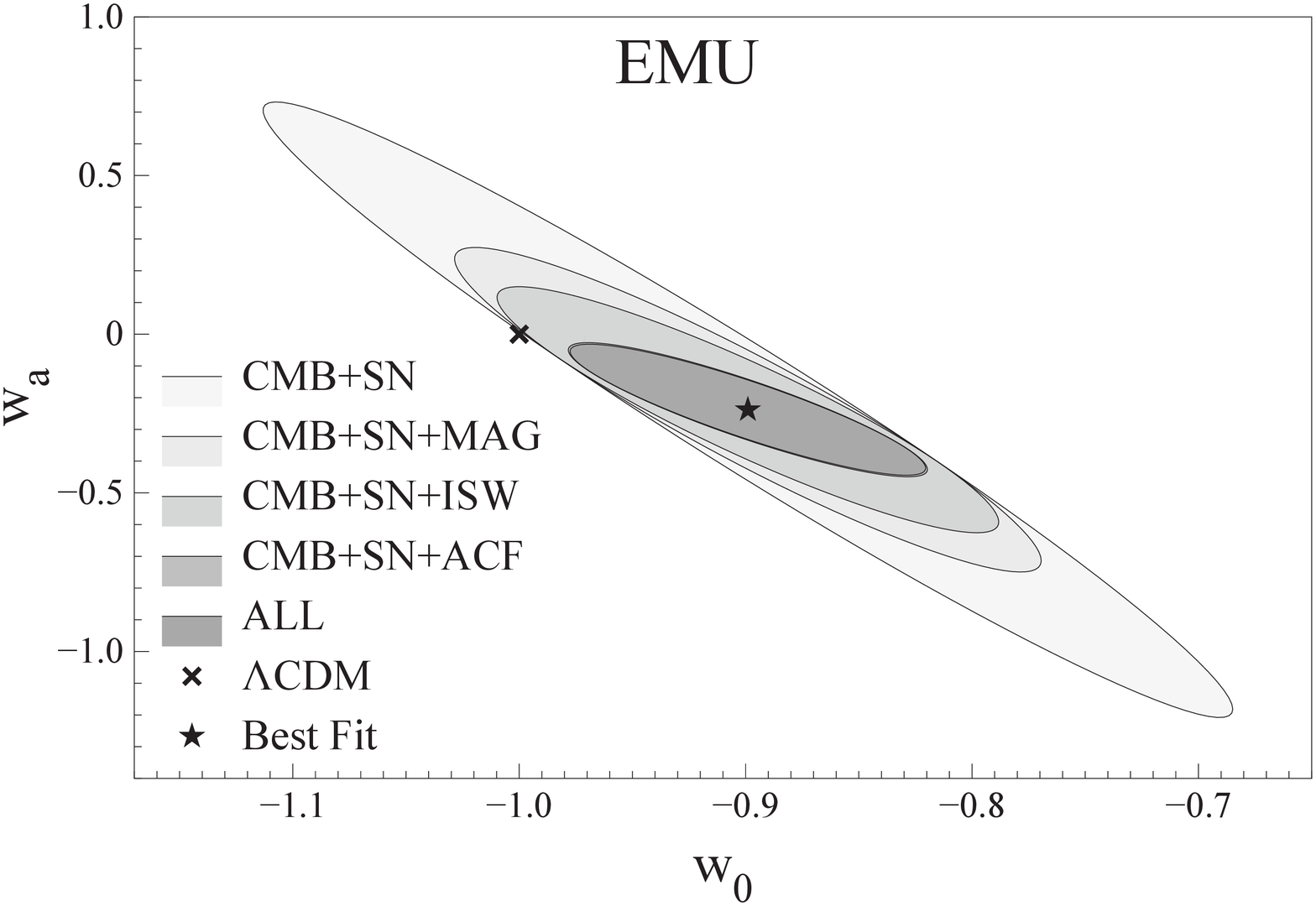, width=0.49\linewidth}
\epsfig{file= 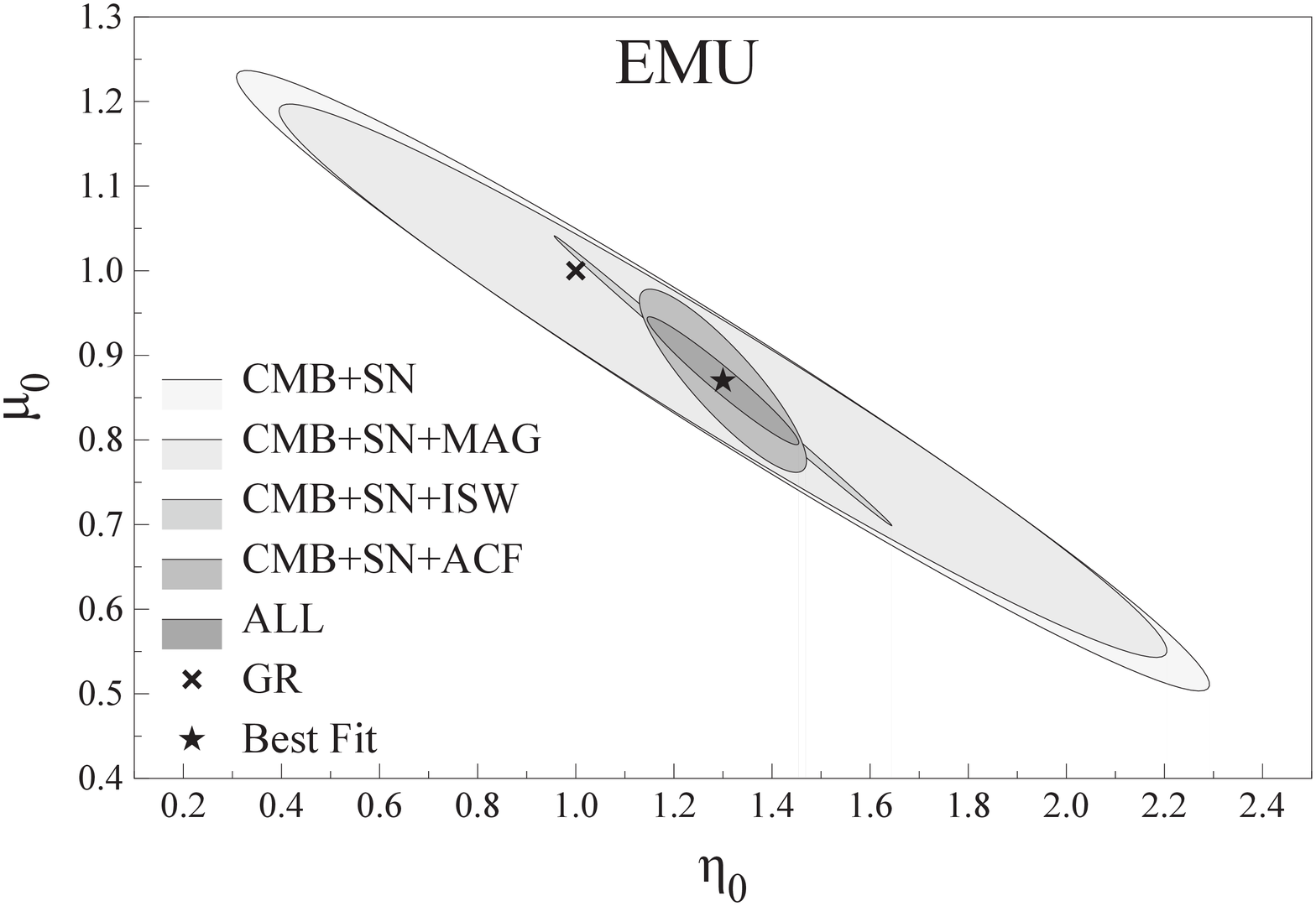, width=0.49\linewidth}
\caption{Forecast of constraints for dark energy (left) and modified gravity (right) parameters, for the EMU survey. Ellipses show constraints for different combinations of probes (see text for details).}
\label{fig:forecast-emu}
\end{center}
\end{figure*}

\begin{figure*}
\begin{center}
\epsfig{file= 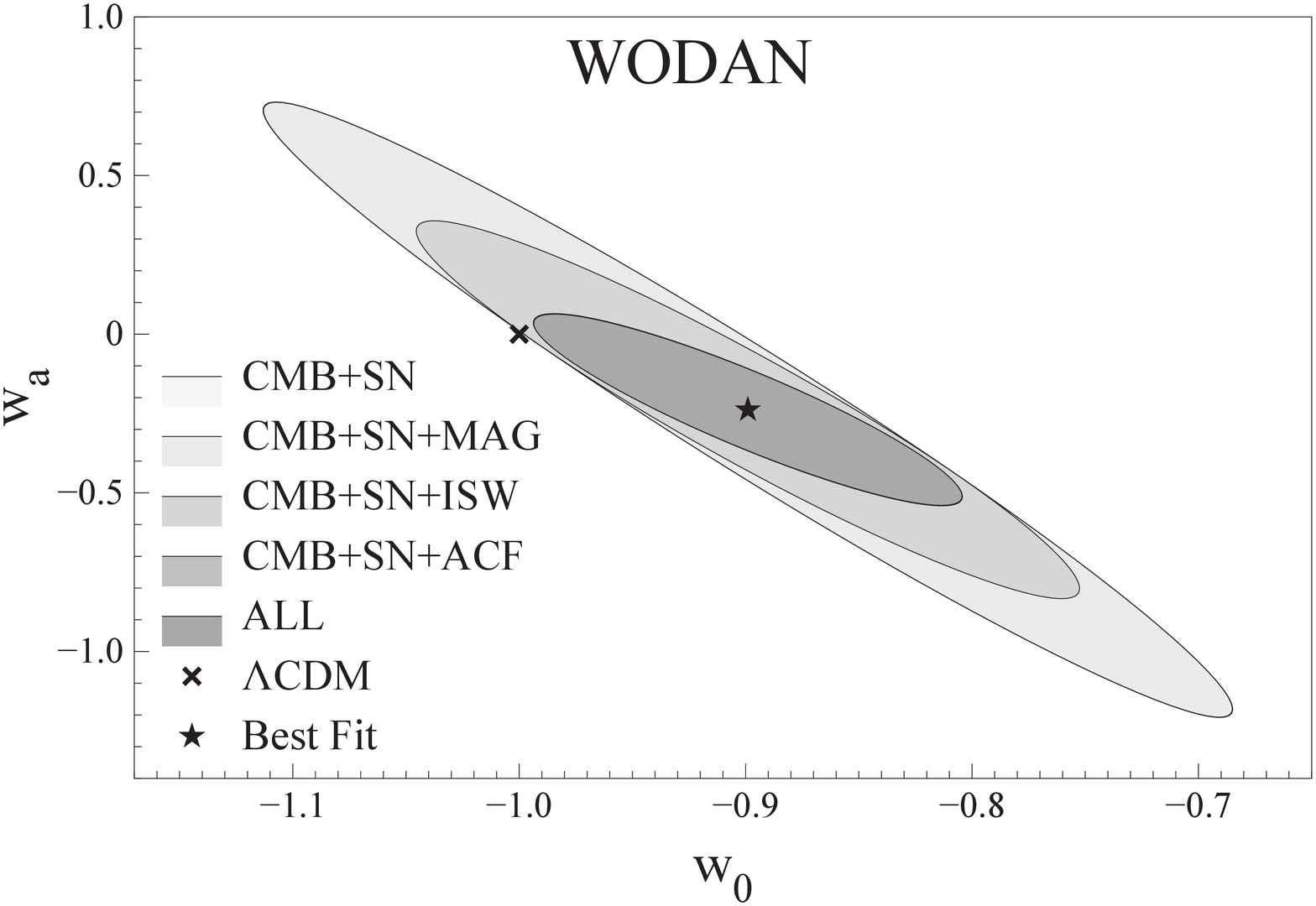, width=0.49\linewidth}
\epsfig{file= 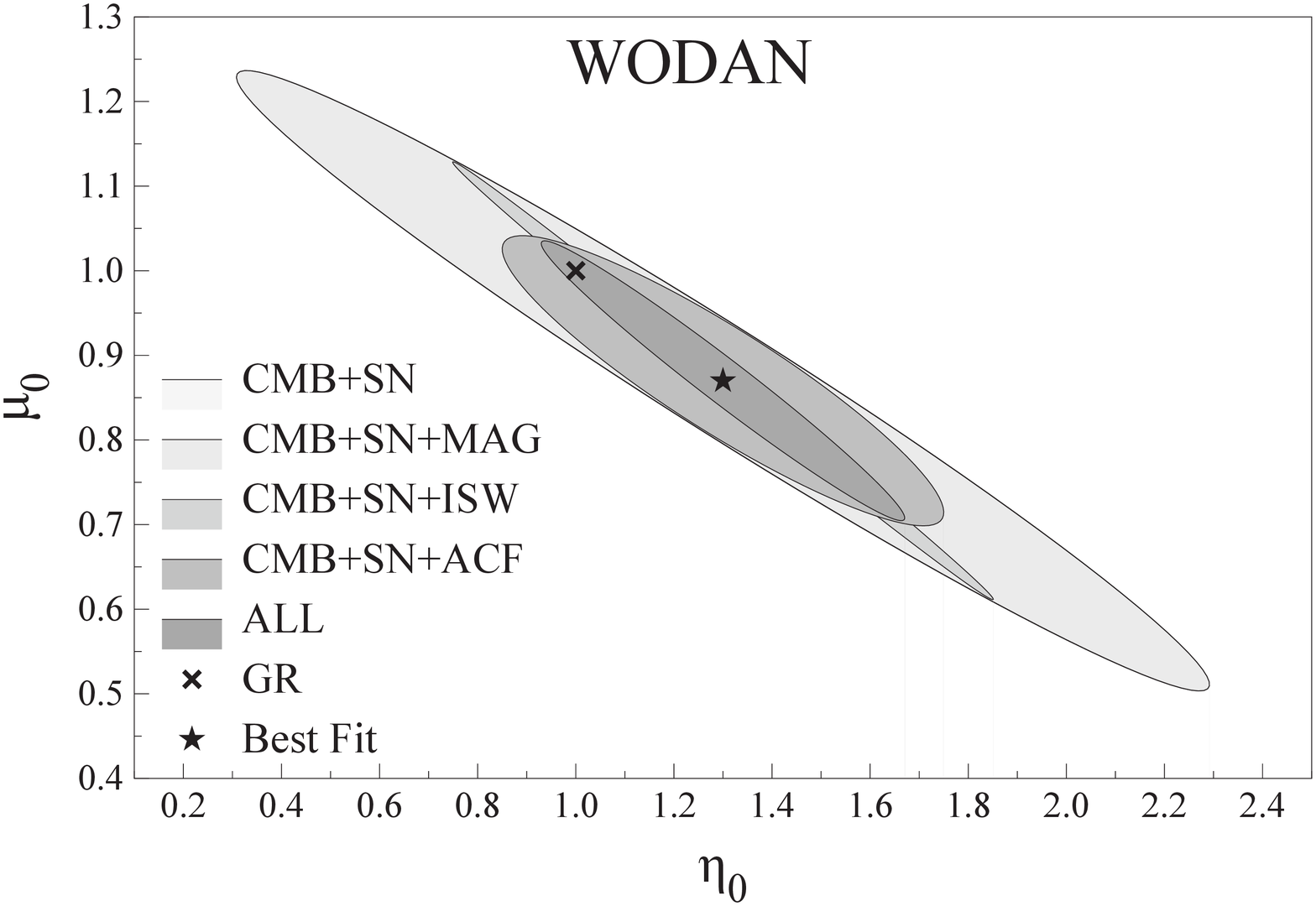, width=0.49\linewidth}
\caption{Forecast of constraints for dark energy (left) and modified gravity (right) parameters, for the WODAN survey. Ellipses show constraints for different combinations of probes (see text for details).}
\label{fig:forecast-wod}
\end{center}
\end{figure*}

The results of our forecasts are shown in Fig.~\ref{fig:forecast-lofms3}, \ref{fig:forecast-loft1}, \ref{fig:forecast-emu}, and \ref{fig:forecast-wod}; 
we plot the limits it will be possible to obtain, using the surveys considered, in the measurements of the dynamical dark energy and modified gravity parameters. 
To highlight the constraining ability of different observables, we show the contours for different data combinations:
lighter grey areas are limits from the Planck CMB\footnote{http://www.rssd.esa.int/index.php?project=planck} plus Supernov\ae Ia measurements \citep{amanullah10}, while darker grey areas are improvements we will have adding
the auto-correlation of radio sources,
the radio sources-CMB cross-correlation (ISW), 
the foreground galaxy-background radio source cross-correlation (Cosmic Magnification)
and a combination of all the measurements; the crosses refer to the standard model (cosmological constant and General Relativity), stars indicate the current best fit from a combination of probes using WMAP, SDSS and CFHTLS (see \citealt{zhao10b} for details).

We can see that the precision in the measurements of cosmological parameters will be significantly increased by the addition of the probes considered; in particular we note that the ISW effect is more powerful in testing models for gravity than models of dark energy.
If it turns out that gravity needs to be modified, the ISW effect measured with radio surveys will be a powerful probe to measure the modified gravity parameters,
the physical reason being that if $\mu_0$ transits from 1 to another value at low $z$, indicating a deviation from GR, the growth will be enhanced; this will change $(\dot{\Phi}+\dot{\Psi})$ significantly, hence generating a large ISW signal (Eq.~\ref{eq:deltat-isw}).

Analyses of clustering and magnification bias also tighten the constraints on gravity; 
the magnification signal measures information about the power spectrum of ($\Phi+\Psi$), which is largely controlled by $\eta_0$, and also tests $\mu_0$ via growth of structures.
The ACF is also sensitive to $\mu_0$ for this reason.
These probes will also be useful in measuring parameters of the dark energy component, if GR is correct even at the largest scales. This is because $w(a)$ changes the growth in a very smooth way; so while it does not generate a large ISW signal, it does change ($\Phi+\Psi$) power integrated along the line of sight, so can be noticed by magnification bias and the projected ACF.

Looking at the different surveys, we can see that they will allow precise measurements of cosmological parameters.
LOFAR Tier1, EMU and WODAN should all be able to increase the precision in the dark energy and modified gravity measurements, compared with that predicted for CMB+SNe, by a significant amount;
it is also interesting to note that adding measurements from LOFAR MS$^3$, which is the least powerful of the surveys we considered (due to the lower number density of sources), will already decrease the errors in the measurements on modified gravity parameters with respect to the CMB+SNIa ones.

Finally, in Fig.~\ref{fig:tot-constraints} we show the constraints on the parameters of dynamical dark energy and modified gravity that will be possible to obtain using the combination of EMU and WODAN, and we compare them to the current best measurements available \citep{zhao10b}. We see that there is substantial improvement, which we quantify in Table~\ref{tab:constraints}; this reports limits on the measurements of the four parameters for the different techniques using the single surveys and the EMU+WODAN combination.

%%%%%%%%%%%%%%%%%%%%%%%%%%%%%%%%%%%%%%%%%%%%%%%%%%%%%%%%%%%%
%%%%%%%%%%%%%%%%%%%%%%		PLOTS		%%%%%%%%%%%%%%%%%%%%%%%%
%%%%%%%%%%%%%%%%%%%%%%%%%%%%%%%%%%%%%%%%%%%%%%%%%%%%%%%%%%%%

%%%%%%%%%%%%%%%%%%%%%%%%%%%%%%%%%%%%%%%%%%%%%%%%%%%%%%%%%%%
%%%%%%%%%%%%%%%%%%%%%%%%%%%%%%%%%%%%%%%%%%%%%%%%%%%%%%%%%%%
%%%%%%%%%%%%%%%%%%%%%%%%%%%%%%%%%%%%%%%%%%%%%%%%%%%%%%%%%%%

%%%%%%%%%%%%%%%%%%%%%%%%%%%%%%%%%%%%%%%%%%%%%%%%%%%%%%%%%%%%%%%%%%
%%%%%%%%%%%%%%%%%%%%%%%%%%%%%%%%%%%%%%%%%%%%%%%%%%%%%%%%%%%%%%%%%%
%%%%%%%%%%%%%%%%%%%%%%	Discussion and Conclusions		%%%%%%%%%%%%%%%%%%%%%%
%%%%%%%%%%%%%%%%%%%%%%%%%%%%%%%%%%%%%%%%%%%%%%%%%%%%%%%%%%%%%%%%%%
%%%%%%%%%%%%%%%%%%%%%%%%%%%%%%%%%%%%%%%%%%%%%%%%%%%%%%%%%%%%%%%%%%

\section{Discussion and Conclusions}
In this paper we have presented a forecast of the cosmological measurements that will be possible with data from the forthcoming LOFAR, EMU and WODAN radio surveys.
We have used the correlation spectra of radio sources: the auto-correlation, the correlations with the CMB and with foreground galaxies, alone and in combination, to predict measurements of cosmological parameters.

As mentioned in Section \ref{sec:nz}, the EMU and WODAN surveys can be combined in order to obtain a complete full sky catalogue and so the largest possible sky coverage, and Fig.~\ref{fig:tot-constraints} shows the improvements this combination will produce in constraining cosmological parameters. 
However, this combination and all the measurements we highlighted will require a very careful treatment of observational data and systematic errors; future work will concentrate on detailed analysis of these issues.

%%%%%%%%%%%%%%%%%%%%%%%%%%%%%%%%%%%%%%%%%%%%%%%%%%%%%%%%%%%%%%%%%%
%%%%%%%%%%%%%%%%%%%%%%%%%%%%%%%%%%%%%%%%%%%%%%%%%%%%%%%%%%%%%%%%%%
%%%%%%%%%%%%%%%%%%%%		Implications for Survey Design		%%%%%%%%%%%%%%%%%%%%%%
%%%%%%%%%%%%%%%%%%%%%%%%%%%%%%%%%%%%%%%%%%%%%%%%%%%%%%%%%%%%%%%%%%
%%%%%%%%%%%%%%%%%%%%%%%%%%%%%%%%%%%%%%%%%%%%%%%%%%%%%%%%%%%%%%%%%%

\subsection{Implications for Survey Design}
\label{sec:uncert}
The tests described in this paper will be very sensitive to systematic errors. For example, to measure magnification bias requires that the background source samples are uniformly surveyed (or that the threshold variation and completeness are well understood) over large areas, placing a  stringent requirement on the flux calibration of the surveys. Systematics such as these lead to a number of requirements on the surveys; here we make some initial comments about the nature of these requirements:
\begin{itemize}
\item Uniformity and completeness. It is important that the tests described in this paper are either conducted on a uniform sample, or one where fluctuations are well understood. For instance a uniform sample could be created by imposing a flux-density cut which is sufficiently above the sensitivity limit at the most insensitive part of the survey, so that there are few spurious sources, and so that sources are not being lost to systematic effects. Detailed simulations will be necessary to check the impact of the flux threshold given the consequent non-uniform signal-to-noise.\\
\item Calibration accuracy of individual surveys. Most surveys typically aim for a 1\% calibration accuracy; it will be important to try to maintain this level, given the need for uniformity described above, and the problems arising if these calibration errors occur systematically and not randomly across the field. \\
%At lofar frequencies, the system temperatures are mainly
%determined by the (Galactic) sky, which varies for position to
%position, so one has to correct for that.
%
\item Dynamic Range. If a strong radio source causes low-level artefacts, then that will affect the claimed number of faint (and therefore typically distant) galaxies, resulting in a spurious correlation between low-redshift and high-redshift galaxies; to first order one will see this as an increase in rms map noise towards bright sources. Any of the cosmic measurements need to take this into account, possibly through masking the affected area, with the consequence of reducing the sky-coverage slightly.\\
\item Cross-calibration of different surveys. It would be useful for all of the surveys to overlap in some regions of the sky to ensure an accurate absolute flux scale.\\
\item Large scale gradients, especially in the declination direction, are virtually unavoidable due to changing UV coverage as a function of declination and increased system temperatures for low elevation observations. These need to be carefully corrected.\\
\item Bias and redshift distribution uncertainties; this is a well known issue for both the galaxy-galaxy and galaxy-CMB temperature spectra and for the redshift distribution only for cosmic magnification. To take this uncertainty into account, we marginalised over the amplitude of $b(z) \times N(z)$ (see Section~\ref{sec:fisher} for more details). A reliable measurement of redshift and bias for the radio continuum population will allow us to improve the constraining power of the techniques considered in this paper. This is the subject of a future paper \citep{lindsay}.
\end {itemize}

%%%%%%%%%%%%%%%%%%%%%%%%%%%%%%%%%%%%%%%%%%%%%%%%%%%%%%%%%%%%%%%%%%
%%%%%%%%%%%%%%%%%%%%%%%%%%%%%%%%%%%%%%%%%%%%%%%%%%%%%%%%%%%%%%%%%%
%%%%%%%%%%%%%%%%%%%%%%		Additional Measurements		%%%%%%%%%%%%%%%%%%%%%%
%%%%%%%%%%%%%%%%%%%%%%%%%%%%%%%%%%%%%%%%%%%%%%%%%%%%%%%%%%%%%%%%%%
%%%%%%%%%%%%%%%%%%%%%%%%%%%%%%%%%%%%%%%%%%%%%%%%%%%%%%%%%%%%%%%%%%

\subsection{Additional Measurements}
In addition to the techniques presented in this paper, LOFAR, EMU and WODAN data will enable several other cosmological analyses, which will be useful to test and improve our models. As examples, we briefly mention two interesting possibilities: the measurement of a dipole anisotropy and a study of the CMB Cold Spot.

The measurement of a dipole anisotropy in the distribution of radio sources can be used to test the distribution of matter at different distances and constrain our local motion with respect to the comoving cosmic rest frame. 

The dipole anisotropy in the cosmic microwave background has been detected with good precision, so an accurate measurement of the dipole anisotropy in the large scale mass distribution at lower redshift will allow a test of the homogeneity of the matter distribution in the universe: if there is agreement between the dipole in the CMB and the dipole of galaxies, this will suggest a large scale homogeneity; while a discrepancy between the CMB and nearby dipole would cast  doubt on the general assumption of isotropy and homogeneity of the Universe on large scales.

It is valuable to have radio sky surveys at different frequencies (such as LOFAR and WODAN), as the amplitude of the radio dipole is not only a function of our peculiar velocity, but also of the spectral index of radio emission \citep{ellis84}.

A detection of the dipole anisotropy in the radio source distribution has been reported using NVSS \citep{blake02}, but the significance of this measurement depends strongly on the number of sources; the surveys considered here will provide an impressive improvement in the precision of the dipole anisotropy measurement, being able to move from an uncertainty of $\sim$15 degrees in dipole direction of \citep{blake02}, to an improved accuracy of $\sim$2 deg, at 1-$\sigma$ level \citep{crawford09}.

Using the radio source distribution, it will also be possible to perform a number count analysis in order to search for a void in the direction of the Cold Spot (\citealt{cruz05}) in the Cosmic Microwave Background.
Several models have been proposed in order to explain this
anomaly, e.g. voids (\citealt{inoue06}, \citealt{rudnick07}), second order gravitational effects (\citealt{tomita08}) or a
brane-world model (\citealt{cembranos08}); \citet{cruz07} showed 
through a Bayesian statistical analysis that the cosmic texture explanation 
is favoured over the Rees-Sciama effect \citep{rees68} due to a void or the Sunyaev-Zel'dovich effect \citep{zel69} caused by a cluster.
Radio and optical data have been used to test the void hypothesis (\citealt{granett10}, \citealt{bremer10}),
trying to find a gap in the number density in the direction of the Cold Spot; no gap was found, however a further analysis using the EMU survey will be helpful, because the larger number density of sources at high redshifts will provide much better S/N for a potential void. Such an anomalously large void will also leave an imprint on ISW measurements \citep{granett08}, and that again can be examined using EMU data.

%\subsection{Prospects with other Surveys ***}

%%%%%%%%%%%%%%%%%%%%%%%%%%%%%%%%%%%%%%%%%%%%%%%%%%%%%%%%%%%%%%%%%%
%%%%%%%%%%%%%%%%%%%%%%%%%%%%%%%%%%%%%%%%%%%%%%%%%%%%%%%%%%%%%%%%%%
%%%%%%%%%%%%%%%%%%%%%%%%%		Conclusions		%%%%%%%%%%%%%%%%%%%%%%%%%
%%%%%%%%%%%%%%%%%%%%%%%%%%%%%%%%%%%%%%%%%%%%%%%%%%%%%%%%%%%%%%%%%%
%%%%%%%%%%%%%%%%%%%%%%%%%%%%%%%%%%%%%%%%%%%%%%%%%%%%%%%%%%%%%%%%%%

\subsection{Conclusions}
In this paper we have shown the potential of SKA pathfinder-generation radio surveys to provide competitive cosmological measurements able to test cosmological models and constrain parameters describing fundamental physics models.

Using simulated catalogues, we have predicted which measurements we will obtain with the source auto-correlation, the cross-correlation between sources and the CMB, the magnification bias, and a joint analysis together with the CMB power spectrum and Supernovae Ia.

We have shown examples of the constraining power in testing cosmological models alternative to the $\Lambda$CDM+GR model, looking for modifications coming from non-Gaussianity, alternative models for dark energy or modifications to the theory of gravity.
We have assumed that the surveys will achieve their target dataset and treatment of systematic errors, but have tried to be conservative in our analyses (e.g. marginalising over the amplitude of correlation power spectra, and using objects detected at 10$\sigma$ signal-to-noise threshold). \\

\begin{figure}
\begin{center}
\epsfig{file=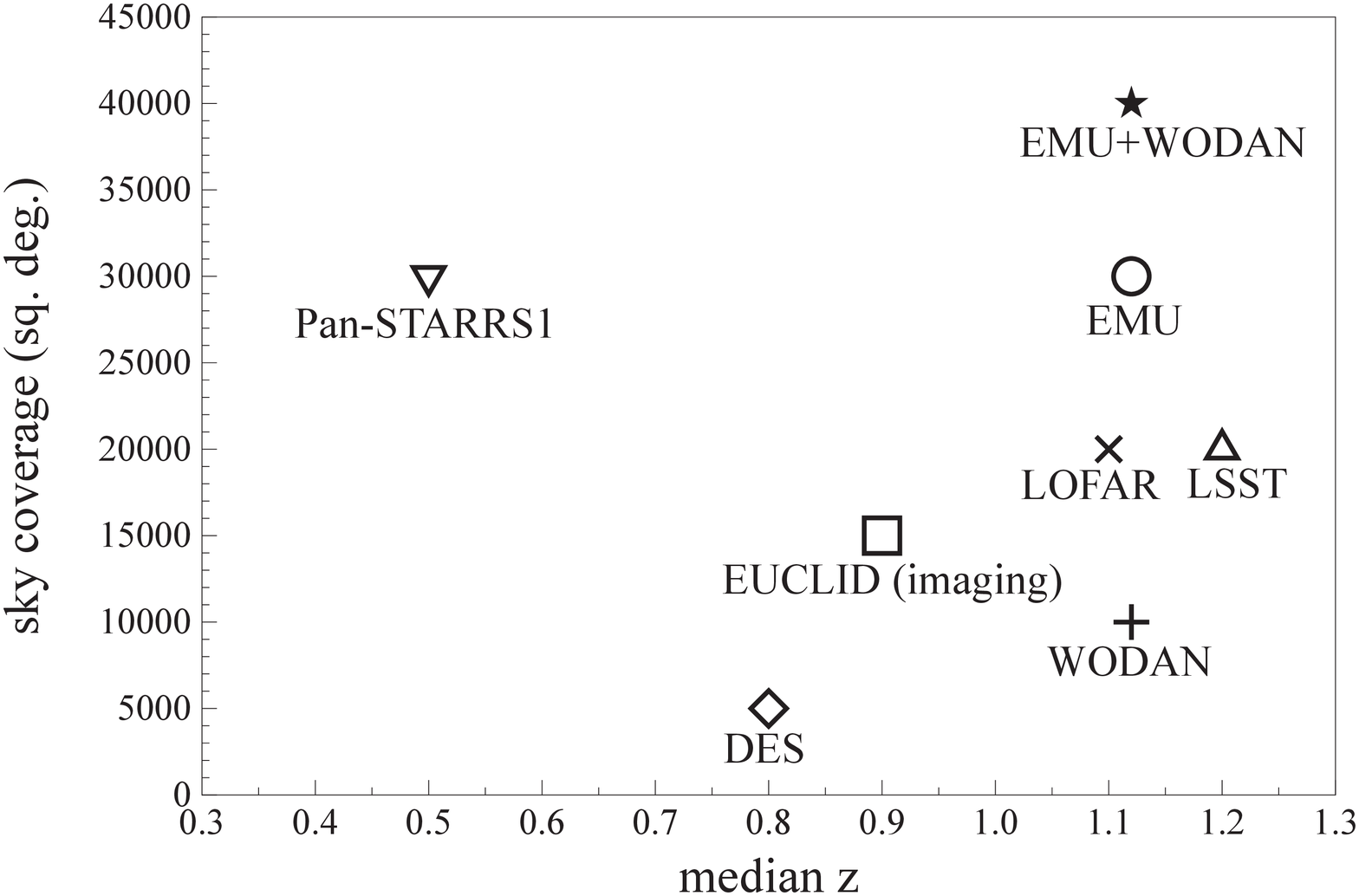, width=0.98\linewidth}
\caption{Comparison of median redshift and sky coverage of selected future imaging surveys.}
\label{fig:surveys}
\end{center}
\end{figure}

There are a number of other galaxy surveys at different wavelengths, which aim to measure cosmological parameters, and which are already collecting data or are being actively prepared for.
The radio surveys discussed in this paper are complementary to these surveys, because of the difference in area, redshift and number density covered, and so they will be able to provide useful information using some specific probes (i.e. ISW and Cosmic Magnification, as their constraining power is increased for larger sky coverage and higher redshifts).
In the period before SKA,
3D redshift surveys such as BOSS will provide more information on the power spectrum on intermediate scales and at low redshifts;
photometric surveys such as Pan-STARRS1\footnote{parameters from \citealt{baugh08}.} and
DES will also span a different part of the parameter space, as they will observe a larger number of objects, but at a lower median redshift and, in some cases, a smaller region of the sky.
Radio surveys cover larger volumes, and so provide more large-scale information; thus they will be complementary to these other surveys.
Next generation experiments such as Euclid and LSST will improve the quality of available data, but for some aspects the radio surveys of the current generation are still competitive, as can be seen from Fig.~\ref{fig:surveys}.
In the radio, NVSS has been used to perform cosmological analyses (e.g. \citealt{raccanelli08}, \cite{xia10}), and the surveys we considered here will have higher median redshift and number of objects observed, so they should improve the precision of the cosmological measurements available.

\begin{figure*}
\begin{center}
\epsfig{file= 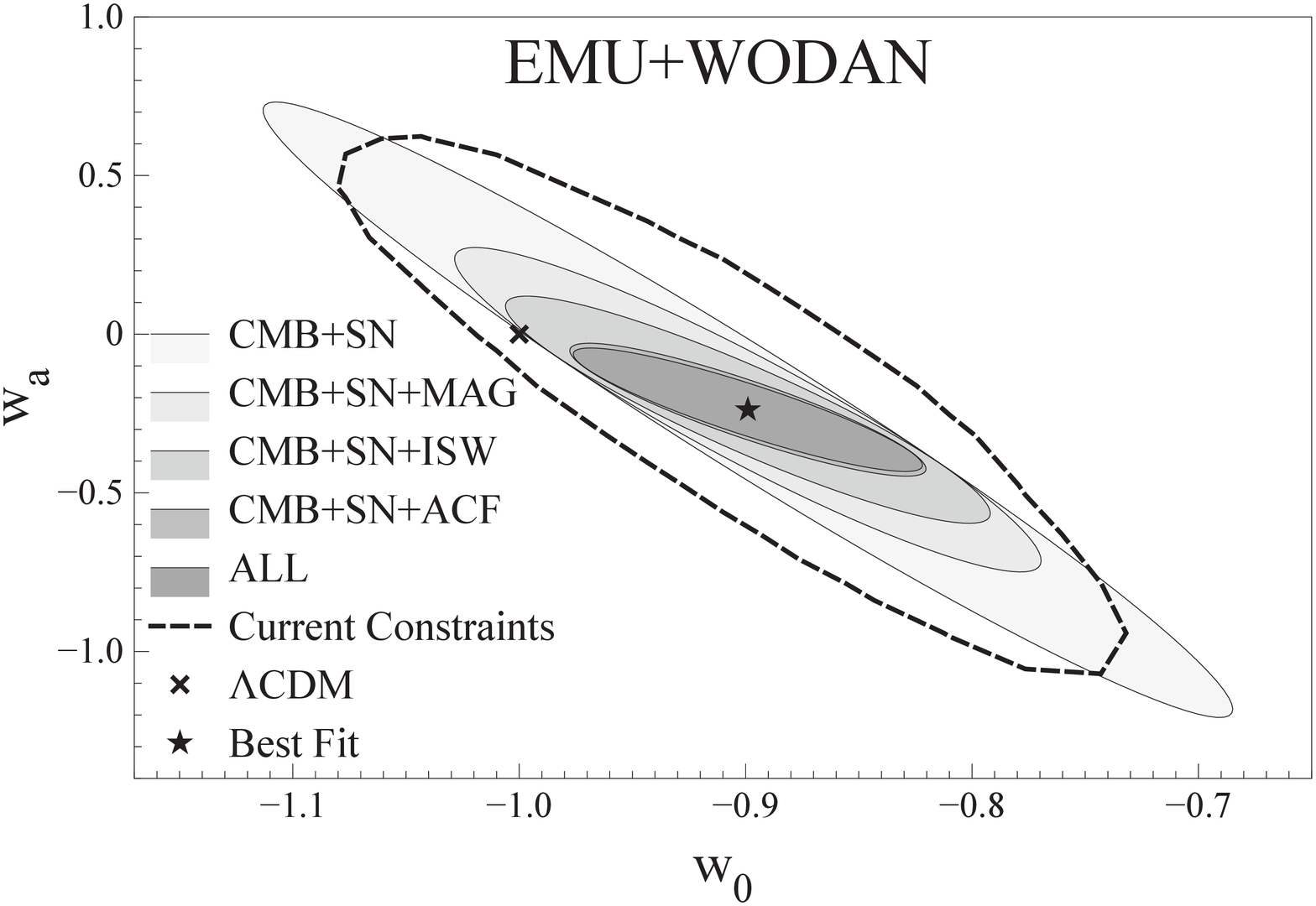, width=0.49\linewidth}
\epsfig{file= 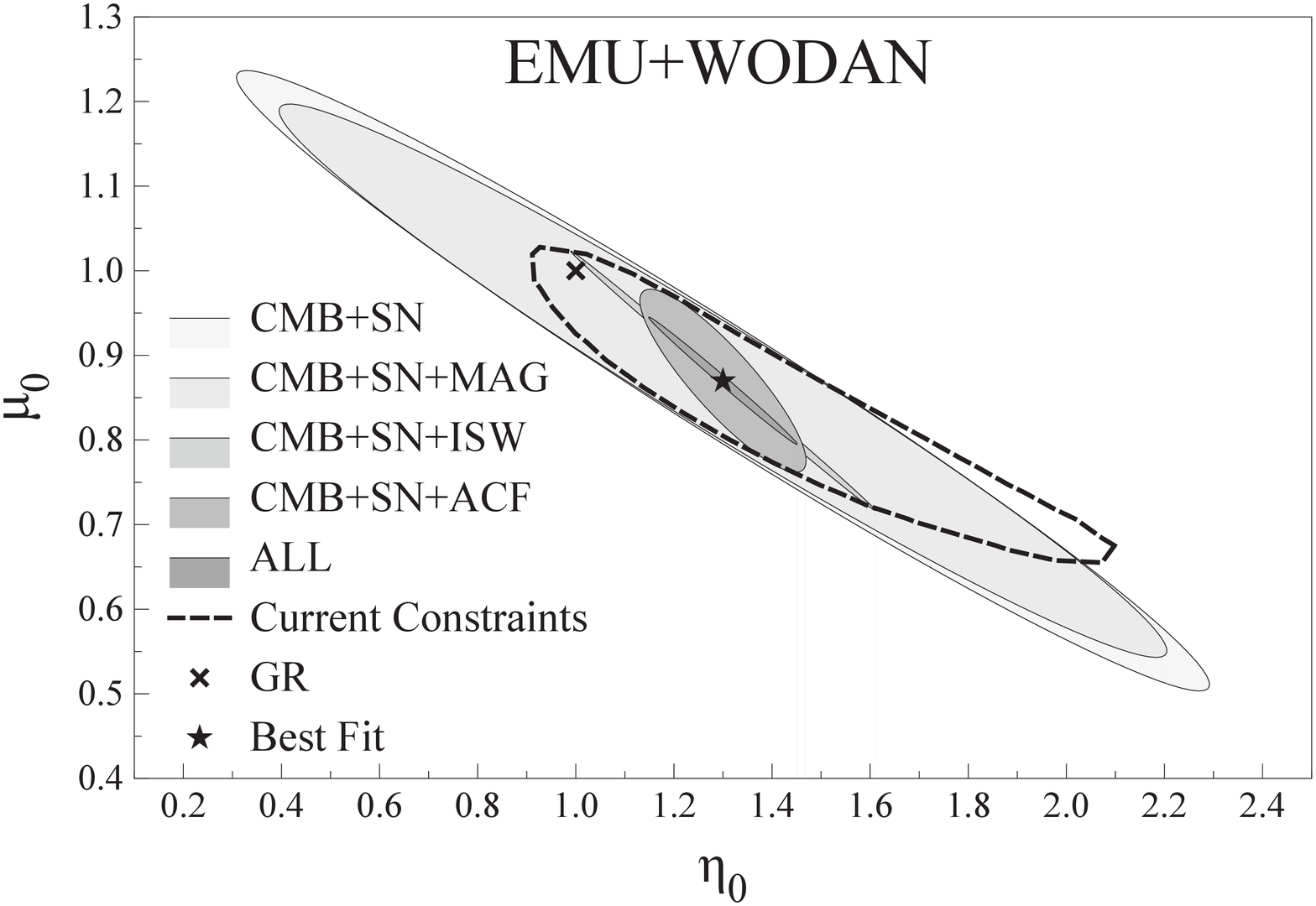, width=0.49\linewidth}
\caption{Forecast of constraints on dynamical dark energy (left) and modified gravity (right) parameters with the EMU+WODAN combination, for different combinations of probes (grey shaded areas), compared with current measurements (solid dashed lines).}
\label{fig:tot-constraints}
\end{center}
\end{figure*}

Our results show that the unprecedented combination of sky coverage, redshift range and sensitivity will enable high-precision measurements, competitive with current surveys in a conservative scenario. Examining Fig.~\ref{fig:tot-constraints} and Table~\ref{tab:constraints}, it is clear that the measurements that LOFAR, EMU and WODAN could provide are potentially decisive in ruling out a large part of the cosmological parameter space for dark energy and modified gravity models.

%%%%%%%%%%%%%%%%%%%%%%%%%%%%%%%%%%%%%%%%%%%%%%%%%%%%%%%%%%
%%%%%%%%%%%%%%%%%%%%%%	TOTAL PLOTS		%%%%%%%%%%%%%%%%%%%%%%
%%%%%%%%%%%%%%%%%%%%%%%%%%%%%%%%%%%%%%%%%%%%%%%%%%%%%%%%%%

%%%%%%%%%%%%%%%%%%%%%%%%%%%%%%%%%%%%%%%%%%%%%%%%%%%%%%%%%%%%
%%%%%%%%%%%%%%%%%%%%%		TOTAL TABLE		%%%%%%%%%%%%%%%%%%%%%%%%
%%%%%%%%%%%%%%%%%%%%%%%%%%%%%%%%%%%%%%%%%%%%%%%%%%%%%%%%%%%%

\begin{center}
  \begin{table}
	\begin{tabular}{ |l|l|l|l|l| }
	  \hline
	  \large{\textbf{Probe}} & \large{$\sigma_{w_0}$} & \large{\textbf{$\sigma_{w_a}$}} & \large{\textbf{$\sigma_{\eta_0}$}} & \large{\textbf{$\sigma_{\mu_0}$}} \\
	  \hline
	  CMB + SNe & 0.14 & 0.64 & 0.66 & 0.24  \\
	  \hline
	  CMB + SNe + LOFAR MS$^3$ ISW & 0.13 & 0.59 & 0.38 & 0.18 \\
	  \hline
	  CMB + SNe + LOFAR MS$^3$ ACF & 0.12 & 0.51 & 0.64 & 0.23 \\
	  \hline
	  CMB + SNe + LOFAR MS$^3$ MAG & 0.14 & 0.64 & 0.66 & 0.24 \\
	  \hline
	  CMB + SNe + LOFAR MS$^3$ ALL & 0.14 & 0.49 & 0.38 & 0.17 \\
	  \hline
	  CMB + SNe + LOFAR Tier1 ISW & 0.11 & 0.51 & 0.35 & 0.16 \\
	  \hline
	  CMB + SNe + LOFAR Tier1 ACF & 0.07 & 0.29 & 0.54 & 0.19 \\
	  \hline
	  CMB + SNe + LOFAR Tier1 MAG & 0.14 & 0.64 & 0.66 & 0.24 \\
	  \hline
	  CMB + SNe + LOFAR Tier1 ALL & 0.07 & 0.28 & 0.32 & 0.14 \\
	  \hline
	  CMB + SNe + EMU ISW & 0.07 & 0.25 & 0.22 & 0.11 \\
	  \hline
	  CMB + SNe + EMU ACF & 0.05 & 0.14 & 0.11 & 0.07 \\
	  \hline
	  CMB + SNe + EMU MAG & 0.08 & 0.34 & 0.60 & 0.21 \\
	  \hline
	  CMB + SNe + EMU ALL & 0.05 & 0.13 & 0.10 & 0.05 \\
	  \hline
	  CMB + SNe + WODAN ISW & 0.09 & 0.39 & 0.36 & 0.16 \\
	  \hline
	  CMB + SNe + WODAN ACF & 0.06 & 0.20 & 0.30 & 0.11 \\
	  \hline
	  CMB + SNe + WODAN MAG & 0.14 & 0.64 & 0.66 & 0.24 \\
	  \hline
	  CMB + SNe + WODAN ALL & 0.06 & 0.20 & 0.24 & 0.11 \\
	  \hline
	  CMB + SNe + EMU+WODAN ISW & 0.07 & 0.23 & 0.20 & 0.10 \\
	  \hline
	  CMB + SNe + EMU+WODAN ACF & 0.05 & 0.13 & 0.11 & 0.07 \\
	  \hline
	  CMB + SNe + EMU+WODAN MAG & 0.08 & 0.34 & 0.60 & 0.21 \\
	  \hline
	  CMB + SNe + EMU+WODAN ALL & 0.05 & 0.12 & 0.10 & 0.05 \\
	  \hline
	\end{tabular}
\caption{ Errors on measurements of dark energy and modified gravity parameters for the different surveys and probe combinations (current best measurements: $\{w_0,w_a\} = \{-0.89 \pm 0.11,-0.24 \pm 0.56\}$ , \citep{zhao10a}, $\{\eta_0,\mu_0\} = \{1.3 \pm 0.35,0.87 \pm 0.12\}$), \citep{zhao10b}.}
\label{tab:constraints}
   \end{table}
\end{center}

%%%%%%%%%%%%%%%%%%%%%%%%%%%%%%%%%%%%%%%%%%%%%%%%%%%%%%%%
%%%%%%%%%%%%%%%%%%%%%%%%%%%%%%%%%%%%%%%%%%%%%%%%%%%%%%%%
%%%%%%%%%%%%%%%%%%%%%%%%%%%%%%%%%%%%%%%%%%%%%%%%%%%%%%%%

%%%%%%%%%%%%%%%%%%%%%%%%%%%%%%%%%%%%%%%%%%%%%%%%%%%%%%%%%%%%%%%%%%
%%%%%%%%%%%%%%%%%%%%%%%%%%%%%%%%%%%%%%%%%%%%%%%%%%%%%%%%%%%%%%%%%%
%%%%%%%%%%%%%%%%%%%%%%%%%	Acknowledgments		%%%%%%%%%%%%%%%%%%%%%%%%%
%%%%%%%%%%%%%%%%%%%%%%%%%%%%%%%%%%%%%%%%%%%%%%%%%%%%%%%%%%%%%%%%%%
%%%%%%%%%%%%%%%%%%%%%%%%%%%%%%%%%%%%%%%%%%%%%%%%%%%%%%%%%%%%%%%%%%

%\newpage

\section*{Acknowledgments}
AR is grateful for the support from a UK Science and Technology
Facilities Research Council (STFC) PhD studentship. 
GBZ is supported by STFC grant ST/H002774/1.
DB acknowledges the support of an RCUK fellowship.
MJJ acknowledges the support of an RCUK fellowship.
WJP is grateful for support from the European Research Council, the
Leverhulme Trust and STFC.
DJS acknowleges financial support from Deutsche Forschungsgemeinschaft (DFG).
We thank Daniele Bertacca, Emma Beynon, Annalisa Bonafede, Rob Crittenden, Olivier Dor\'{e}, Tommaso Giannantonio, Ben Hoyle, Minh Huynh, Joseph Lazio, Alejo Martinez-Sansigre, Sabino Matarrese, Matthias Rubart, Hana Schumacher, Charles Shapiro, David Wands and Jun-Qing Xia for valuable discussions.
Numerical computations were carried out on the SCIAMA High Performance Compute (HPC) cluster which is supported by the ICG, SEPNet and the University of Portsmouth.

\end{document}